\documentclass[longauth]{aa}  
\usepackage{graphicx}
\usepackage{xcolor}
\usepackage[varg]{txfonts}
\usepackage{natbib}
\usepackage{amssymb}
\usepackage[rightcaption]{sidecap}
\sidecaptionvpos{figure}{b}
\usepackage{floatrow}

\pdfoutput=1
\usepackage{hyperref}
\begin{document}

   \title{The Galaxy Activity, Torus and Outflow Survey (GATOS)}
   \subtitle {I.~ALMA images of dusty molecular tori in Seyfert galaxies}

 \author{S.~Garc\'{\i}a-Burillo\inst{1}
	   \and	
	   A.~Alonso-Herrero\inst{2}	
	   \and
	   C.~Ramos Almeida\inst{3, 4}
	    \and
	   O.~Gonz\'alez-Mart\'{\i}n\inst{5}	
	   \and
	    F.~Combes\inst{6}
	   \and	   
	   A.~Usero \inst{1}
	   \and
	   S. H\"onig\inst{7}
	   \and
	   M.~Querejeta\inst{1}
	    \and
	   E.~K.~S.~Hicks\inst{8}	 
	   \and
	   L.~K.~Hunt\inst{9}
    	   \and
	   D.~Rosario\inst{10}
	   \and
	   R.~Davies\inst{11}
	   \and
	   P.~G.~Boorman\inst{12}
	   \and 
	   A.~J.~Bunker	\inst{13}   
	   \and
	   L.~Burtscher\inst{14}
	   \and
	   L.~Colina\inst{15}
           \and
	   T.~D\'{\i}az-Santos\inst{16}
	   \and
	   P.~Gandhi\inst{7}
	   \and
	   I.~Garc\'{\i}a-Bernete\inst{13}
	   \and
	   B.~Garc\'{\i}a-Lorenzo\inst{3, 4}
	   \and
	   K.~Ichikawa\inst{17}
	   \and
	   M.~Imanishi\inst{18, 19}
	   \and
	   T.~Izumi\inst{18,19}
	   \and
	   A.~Labiano\inst{15}
	   \and
	   N.~A.~Levenson\inst{20}
	   \and
	   E.~L\'opez-Rodr\'{\i}guez\inst{21, 22}
	   \and
	   C.~Packham\inst{23}
	   \and
	   M.~Pereira-Santaella\inst{15}
	   \and
	   C. Ricci\inst{24, 25}
	   \and
	   D.~Rigopoulou\inst{13}
	   \and
	   D.~Rouan\inst{26}
	   \and
	   T.~Shimizu\inst{11}
	   \and
	   M.~Stalevski\inst{27, 28}
	   \and
	   K.~Wada\inst{29, 30, 31}
	   \and
	   D.~Williamson\inst{7}	   	   	   	   	    
	    }
   \institute{
     Observatorio Astron\'omico Nacional (OAN-IGN)-Observatorio de Madrid, Alfonso XII, 3, 28014-Madrid, Spain 			  \email{s.gburillo@oan.es} 		  
        \and
    Centro de Astrobiolog\'{\i}a (CSIC-INTA), ESAC Campus, 28692 Villanueva de la Ca\~nada, Madrid, Spain  
       \and  
   Instituto de Astrof\'{\i}sica de Canarias, Calle V\'{\i}a L\'actea, s/n, E-38205 La Laguna, Tenerife, Spain
       \and
   Departamento de Astrof\'{\i}sica, Universidad de La Laguna, E-38205, La Laguna, Tenerife, Spain
      \and    
   Instituto de Radioastronom\'{\i}a y Astrof\'{\i}sica (IRyA-UNAM), 3-72(Xangari), 8701, Morelia, Mexico      
     \and  
     LERMA, Observatoire de Paris, Coll\`ege de France, PSL University, CNRS, Sorbonne University,  Paris   
   \and
   Department of Physics \& Astronomy, University of Southampton, Hampshire S17 1BJ, Southampton, UK
   \and
     Department of Physics \& Astronomy, University of Alaska Anchorage, Anchorage, AK 99508-4664, USA
   \and
   INAF - Osservatorio Astrofisico di Arcetri, Largo Enrico Fermi 5, 50125 Firenze, Italy   
   \and
    Centre for Extragalactic Astronomy, Department of Physics, Durham University, South Road, Durham DH1 3LE, UK
   \and   
    Max-Planck-Institut f\"ur Extraterrestrische Physik, Garching, Germany   
   \and   
   Astronomical Institute, Academy of Sciences, Bo\^cn\'{\i} II 1401, CZ-14131 Prague, Czechia
  \and
   Department of Physics, University of Oxford, Denys Wilkinson Building, Keble Road, Oxford, OX13RH, U.K.
  \and
   Leiden Observatory, PO Box 9513, 2300 RA Leiden, The Netherlands   
   \and
   Centro de Astrobiolog\'{\i}a (CSIC-INTA),    Carretera    de    Ajalvir,    28850    Torrej\'on    de    Ardoz,    Madrid,    Spain   
   \and
     Institute of Astrophysics, Foundation for Research and Technology-Hellas, GR-71110, Heraklion, Greece
    \and
   Astronomical Institute, Tohoku University 6-3 Aramaki, Aoba-ku, Sendai, 980-8578 Japan   
   \and
   National Astronomical Observatory of Japan, National Institutes of Natural Sciences (NINS), 2-21-1 Osawa, Mitaka, Tokyo 181-8588, Japan
   \and
   Department of Astronomy, School of Science, The Graduate University for Advanced Studies, SOKENDAI, Mitaka, Tokyo 181-8588, Japan
  \and
     Space Telescope Science Institute, Baltimore, MD 21218, USA
    \and
      Kavli Institute for Particle Astrophysics \& Cosmology (KIPAC), Stanford University, Stanford,CA 94305, USA
    \and
    SOFIA Science Center, NASA Ames Research Center, Moffett Field, CA 94035, USA
    \and
    The University of Texas at San Antonio, One UTSA Circle, San Antonio, TX 78249, USA    
    \and
      N\'ucleo de Astronom\'ia de la Facultad de Ingenier\'ia, Universidad Diego Portales, Av. Ej\'ercito Libertador 441, Santiago, Chile
    \and
      Kavli Institute for Astronomy and Astrophysics, Peking University, Beijing 100871, China
       \and
     LESIA, Observatoire de Paris, PSL Research University, CNRS, Sorbonne Universit\'es, UPMC Univ. Paris 06, Univ. Paris Diderot, Sorbonne Paris Cit\'e, 5 place Jules Janssen, 92190 Meudon, France  
     \and  
    Astronomical Observatory, Volgina 7, 11060 Belgrade, Serbia
     \and
    Sterrenkundig Observatorium, Universiteit Ghent, Krijgslaan 281-S9, Ghent, B-9000, Belgium
    \and
     Kagoshima University, Graduate School of Science and Engineering, Kagoshima 890-0065, Japan
    \and
     Ehime University, Research Center for Space and Cosmic Evolution, Matsuyama 790-8577, Japan
    \and
    Hokkaido University, Faculty of Science, Sapporo 060-0810, Japan
}
    
    \date{Received: April, 2021; Accepted:--, --}

 
  \abstract{We present the first results of the Galaxy Activity, Torus and Outflow Survey (GATOS), a project aimed at understanding the properties of the dusty molecular tori and their connection to the host galaxy in nearby Seyfert galaxies. Our project expands the range of Active Galactic Nuclei (AGN) luminosities and Eddington ratios covered by previous  surveys of Seyferts conducted by the Atacama Large Millimeter Array (ALMA) and allows us to study the gas feeding and feedback cycle in a combined sample of 19 Seyferts. We used ALMA to obtain new images of the emission of molecular gas and dust using the CO(3--2) and HCO$^+$(4--3) lines as well as their underlying continuum emission at 870~$\mu$m with high spatial resolutions ($0.1\arcsec\sim7-13$~pc) in the circumnuclear disks  (CND) of 10 nearby ($D<28$~Mpc) Seyfert galaxies selected from an ultra-hard X-ray survey.  Our new ALMA observations detect 870~$\mu$m continuum and CO line emission from spatially resolved disks located around the AGN in all the sources. The bulk of the 870~$\mu$m continuum flux can be accounted for by thermal emission from dust in the majority of the targets. For most of the sources the disks show a preponderant orientation perpendicular to the AGN wind axes, as expected for dusty molecular tori.  The median diameters and molecular gas masses of the tori are $\sim42$~pc, and $\sim6\times10^5$~M$_{\sun}$, respectively. We also detected the emission of the 4--3 line of HCO$^+$ in four GATOS targets. 
 The order of magnitude differences found in the CO/HCO$^+$ ratios within our combined sample point to a very different density radial stratification inside the dusty molecular tori of these Seyferts.  We find a positive correlation between the line-of-sight gas column densities responsible for the absorption of X-rays and  the  molecular gas column densities derived from CO   towards the AGN in our sources. Furthermore, the median values of both  column densities are similar. This suggests that the neutral gas line-of-sight column densities of the dusty molecular tori imaged by ALMA contribute significantly to the obscuration of X-rays.  The radial distributions of molecular gas in the CND of our combined sample  show signs of nuclear-scale molecular gas deficits. We also detect molecular outflows in the sources that show the most extreme nuclear-scale gas deficits in our sample. These observations find for the first time supporting evidence that the imprint of AGN feedback is more extreme in higher luminosity and/or higher Eddington ratio Seyfert galaxies.}

  \maketitle
%

\section{Introduction}

The fueling of super-massive black holes (SMBHs), a common element of the spheroidal components of galaxies, explains the onset of nuclear activity.
High-resolution observations of molecular gas in nearby Active Galactic Nuclei (AGN) have played a key role in addressing
the question of AGN fueling \citep[e.g. see reviews by][]{Gar12, Sto19}. These observations have also revealed  that  gas accretion onto SMBHs and their hosts can be regulated through the launching of molecular outflows in galaxies \citep[e.g.][]{flu19,Lut20,vei20}.  Molecular outflows have been mapped across a wide range of spatial scales in different galaxy populations. In the particular case of nearby AGN, these outflows have been imaged on nuclear scales in dusty molecular tori and their surroundings ($\sim$tens of pc), but also on the larger scales typical of circumnuclear disks (CND) ($\sim$hundreds of pc) \citep{GB14, Mor15, Gal16, Aal17,  Alo18, Alo19, Imp19, GB19, Dom20}. These different  manifestations of the outflow phenomenon constitute a key ingredient to understanding the co-evolution of galaxies and SMBHs.

The simplest version of the unified model of AGN  explains their  observational properties as only due to different lines of sight toward the central engine.  In type 2 AGN the obscuring material blocks our view of the Broad Line Region (BLR) and only emission  
from the narrow line region (NLR) is seen \citep[e.g.][]{Ant85, Ant93, Kro88, Urr95}. Type 1 AGN have an unobscured view and thus emission from both the BLR and NLR  is observed.   
Different observational properties led to an initial description of the obscuring material as
an optically and geometrically thick torus \citep[e.g.][]{Pie92, Pie93}, although
other geometries were also considered, including flared, tapered and warped disks \citep[e.g.][]{San89,Efs95a}. 
More recently, the simplest version of unifying theories has been debated due to the large observed spread in properties of the toroidal obscuration in AGN \citep{Ram09b, Ram11, Alo11, Eli12, Mat16, Gar19}. The application of the torus
paradigm has also been questioned in the early universe, where conditions may be significantly different from those observed locally \citep{Net15}.
 
In the canonical scheme, the torus is expected to be located  between the BLR and the NLR and extend to parsec scales. More complex models 
were developed  in recent decades as the community started  to probe the morphology, content, and radial extent of the circumnuclear gas in active galaxies \citep[see e.g. the reviews by][]{Bia12, Ram17}.
The first torus models were static and, for computational reasons, the dust was distributed homogeneously \citep{Pie92, Pie93, Gra94, Efs95a}.
The torus size was allowed to range from compact (a few parsecs) to a few hundred parsecs,  depending on the model.
With the introduction of the so-called clumpy torus model \citep{Nen02, Nen08a, Nen08b}, obscuration depends not only on the inclination but also on the number of clouds along the line of sight, i.e., on the covering factor. 
Clumpy torus models received observational support from the measurement of significant variations in the absorbing column density derived from X-ray observations on timescales ranging from days to years \citep[e.g.][]{Ris02, Mar14}.

The detection of extended
mid-infrared emission along the torus polar direction, initially in NGC~1068 and later on in other
nearby AGN, prompted the inclusion of a polar component in the models \citep[see][]{Efs95b, Gal15, Hoe17, Sta17}. The mid-infrared polar component has been observationally resolved on
parsec scales with interferometry  \citep{Tri09, Hoe12, Bur13, Hoe13, Lop14, Tri14, Lop16} and extends out to a few hundred
parsecs as revealed by ground-based imaging facilities \citep{Rad03, Pac05, Gal05, Roc06, Asm16, GarB16, Asm19}. Polar dust might also provide a natural explanation to the apparent isotropy of the infrared emission observed in AGN, irrespective of obscuring column density \citep[e.g.][]{Gan09, Lev09, Asm15, Full19}.

 The static ``torus'' and disk-wind models are effectively the starting point for more realistic dynamical models  to explain the formation,  maintenance and the eventual disappearance of the obscuring material around AGN. On small scales the
height of the obscuring material is proposed to be maintained by the AGN radiation pressure \citep{Pie92}, infrared radiation pressure \citep[e.g.][]{Kro07, Wil19, Hoe19, Ven20, Taz20}, magnetic support
\citep{Emm92, Eli06, Cha17, Vol18, Kud20, Lop20}, and/or stellar feedback \citep{Wad02}. Radiation hydrodynamical simulations including some of these physical processes  predict the launch of outflows. These can extend to scales of tens of parsecs mostly along the polar direction
 \citep{Wad12, Wad15, Wad16, Cha16, Cha17, Will20}. These simulations also produce outflow components along the equatorial direction of the nuclear torus/disk.
The new paradigm is now a ``dynamical torus'' that has  the inflowing \citep[but also outflowing, see][]{Ven20} reservoir of fueling material along the equatorial plane. The boosted acceleration due to infrared radiation pressure near the inner walls of the torus  can lift up material from the equatorial plane and launch a dusty and molecular wind. Under certain conditions part of this material can rain back down onto the equatorial plane 
\citep[see][]{Wad16}  or loop back towards the inner region on ballistic orbits \citep[see][]{Will20}. In this new scenario for the torus,  the equatorial and polar components both contribute to the obscuration of the central engine and the nature of the obscuring material around AGN is complex \citep[see reviews by][]{Ram17, Hoe19}.

Atacama Large Millimeter Array (ALMA) observations with physical resolutions of a few parsecs to $\sim 10-20\,$pc are providing strong support for this scenario. The first ALMA detection of an AGN torus was for the Seyfert galaxy NGC 1068 \citep{GB16,Gal16,Ima18,GB19,Imp19,Ima20}. The measured torus diameter (ranging from 7 to up to 30\,pc) from a variety of molecular transitions and dust continuum emission is  a factor of a few
larger than found by the mid-IR observations, which are tracing warmer emission originating in a more compact region. More recently, the K-band observations done with the VLT/GRAVITY instrument have imaged the inner dust sublimation region inside the  torus of NGC~1068 \citep{Gravity20}.  
The NGC~1068 torus is not by any means unique. Relatively large (diameters of up to 20-50 pc) and massive ($\sim 10^5-10^7M_\odot$) tori/disks are detected in cold molecular gas in local Seyferts,  low-luminosity AGN, and compact obscured nuclei \citep{Aal17, Alo18, Izu18, Sal18, Alo19, Aal19, Com19, Aud19, Aal20, Aud20}. The nuclear tori/disks detected with ALMA are not found in isolation at the centers of AGN but are well connected with molecular gas in the galaxy disk and can be decoupled both morphologically and kinematically from their host galaxies. Rather than a simple compact rotating structure, the torus also appears to be highly turbulent \citep{GB16} and possibly outflowing \citep{Gal16, Alo18, GB19, Imp19, Aal20}. Theoretical arguments and dynamical simulations suggest that these phenomena could be associated with the ``outflowing torus'' scenario described above.

This is the first paper in a series aimed at understanding the properties of the dusty molecular tori\footnote{Hereafter we use the term dusty molecular `torus' to denote the compact disk detected both in dust continuum and molecular line emission around the central engine, and which tends to show an equatorial geometry relative to the AGN wind/jet axis. The detailed internal morphology of this dusty molecular disk feature is nevertheless still to be determined.} and the connection to their host galaxies in nearby Seyfert sources. This is one of the main goals of our Galaxy Activity, Torus, and Outflow Survey (GATOS). In the second paper of this series (paper II: Alonso-Herrero et al. 2021 in prep.; hereafter AH21), we will study the relation between the torus and polar dust emissions  and compare them with predictions from disk+wind models.  To select our Seyfert galaxies, we used the 70 month {\it Swift}/BAT $14-195\,$~keV all-sky catalog \citep{Bau13}
 to draw our parent sample from. This catalog provides a sample that is unbiased of obscuration/absorption, even up to column densities of $N_{\rm H} \simeq 10^{24}\,{\rm cm}^{-2}$ as well as  providing
 a nearly complete selection for nearby AGN at $L_{\rm AGN}$(14-150 keV)$> 10^{42}\,{\rm erg s}^{-1}$. Optical and IR AGN surveys, on the other hand, can be incomplete for nearby galaxies 
 and biased against AGN with strong star-formation activity. We imaged with ALMA the emission of molecular gas and dust in the CND of our targets  using the CO(3--2) and HCO$^+$(4--3) lines and their underlying continuum emission, with spatial resolutions $\simeq 0\farcs1$~(7-13~pc). The use of these transitions allows us to probe simultaneously a range of physical conditions  ($n$(H$_2$)~$\simeq$ a few $10^4$-a few$10^6$cm$^{-3}$, $T_{K}\geq20-50$~K) that are well suited to probing the bulk of the gas reservoirs in the typical CND environments of AGN, which  are known to host predominantly dense and hot molecular gas \citep{GB14, Vit14, GB16, Gal16, Ima18, Imp19, Ima20}. As an additional advantage of the use of the 3--2 line of CO, the analysis of the CO spectral line energy distribution of the  radiation-driven fountain model of the torus discussed by \citet{Wad18} showed that the CO--to--H$_2$ conversion factor has a comparatively weaker dependence on the intensity for mid--J CO lines. Furthermore, the obtention of 351~GHz continuum images in our targets is a prerequisite to deriving the dust content of tori by keeping low the contribution from other mechanisms alien to thermal dust emission, which can be prevalent at lower frequencies.

 The paper is organized as follows. We present in Sect.~\ref{GATOS} the GATOS sample. Sect.~\ref{observations} describes the ALMA observations  and ancillary data used in this work. Sect.~\ref{continuum} describes the decomposition of the submillimeter continuum images of our Seyfert galaxies and derives the sizes, orientations and masses of their dusty disks. A description of the molecular gas distributions derived from the CO(3--2) line maps is included in Sect.~\ref{co}. We  compare the images and line ratios derived  from the CO(3--2) and HCO$^+$(4--3) lines in Sect.~\ref{faces}. Sect.~\ref{Xrays} discusses the relation between the dusty molecular tori and X-ray emission in our Seyferts. Sect.~\ref{gas-radial} compares the radial distribution of molecular gas in NUGA and GATOS Seyferts and describes a scenario accounting for the imprint of AGN feedback.
The main conclusions of this work are summarized in Sect.~\ref{summary}.

\section{The sample} \label{GATOS}

 We initially selected a volume-limited sample from the Swift/BAT catalog of AGN with
distances of $10-40\,$Mpc in the southern hemisphere (declinations below $+20^{\circ}$) for our ALMA observations. We set the lower distance limit in order not to include low luminosity AGN since
these are targeted by other ALMA programs (see below). The upper limit to the distance allows for a reasonably sized sample of 31 local AGN and
sufficient spatial resolution to isolate the dusty molecular tori with ALMA. All our targets have modeled 14-195\,keV observations from which the X-ray column densities ($N_{\rm H}$) and intrinsic (absorption 
corrected) hard and  ultra hard X-ray luminosities in the $2-10\,$keV and  $14-150\,$keV  bands, respectively, are derived \citep{Ric17a}.  

\begin{figure*}
   \centering
    \includegraphics[width=0.55\textwidth]{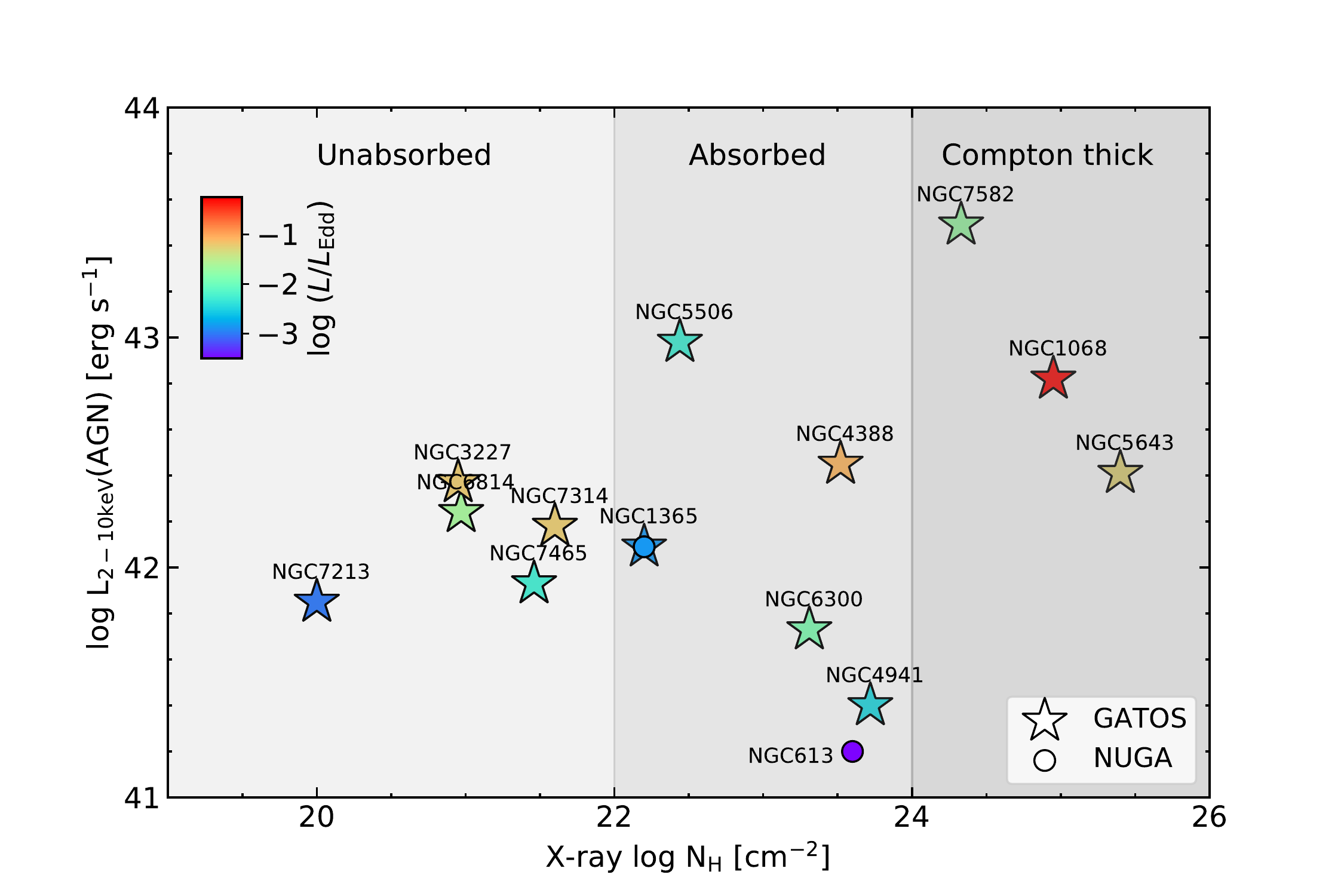}
    \includegraphics[width=0.43\textwidth]{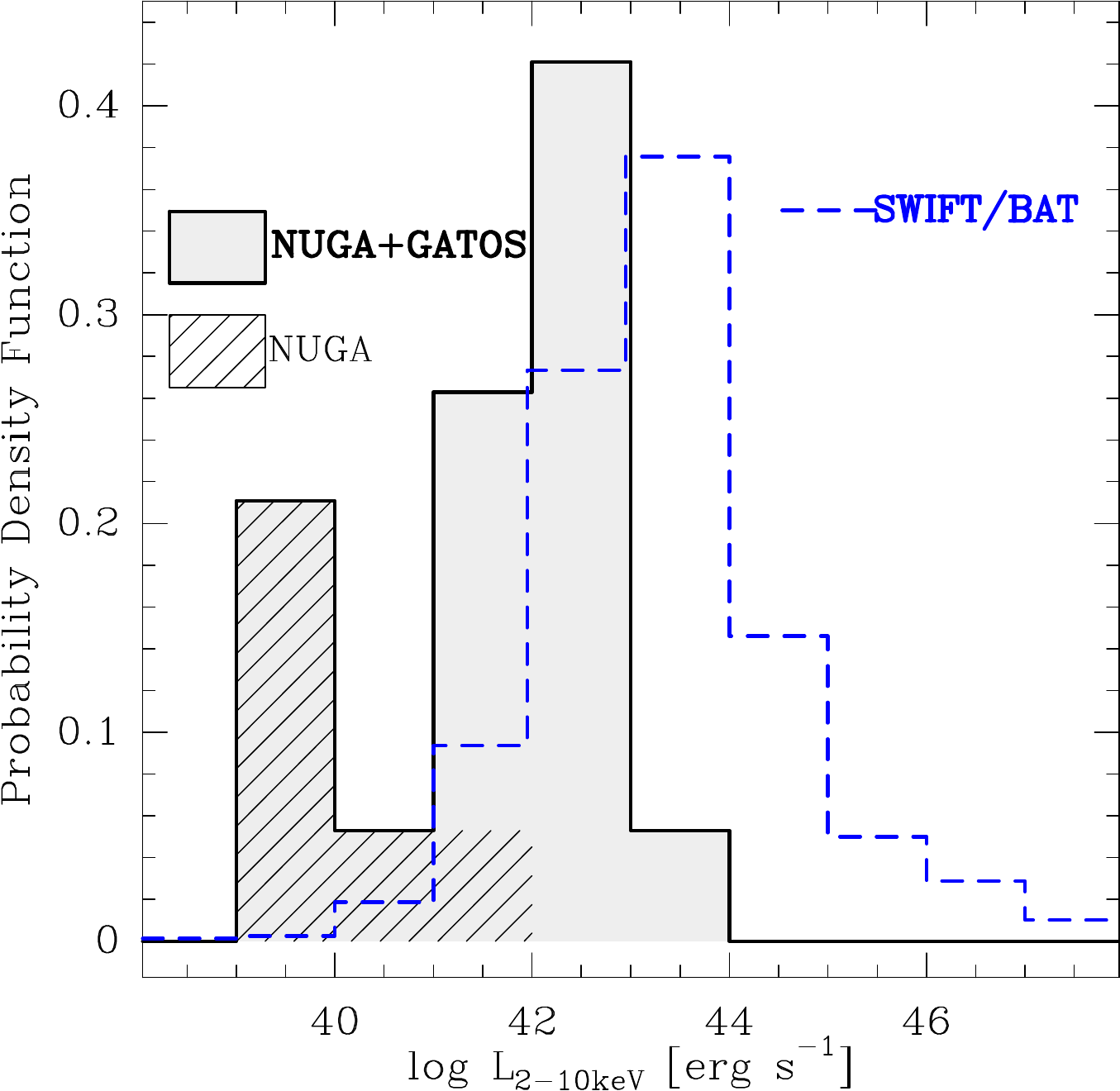}
     \caption{{\it Left panel:} Comparison of the distribution of intrinsic AGN luminosities  measured in the 2-10~keV X-ray band (corrected for absorption)  and the column densities, $N_{\rm H}$,  derived from hard X-rays, as measured
     by Swift-BAT in the 14-150~keV band for the  combined GATOS sample listed in Table~\ref{Tab1}  (star markers) \citep{Ric17a}. We also include the data of  NGC~613 from the NUGA sample \citep{Com19} (circle), which also has NGC~1365 as a target in common with GATOS. Symbols are color-coded to reflect the Eddington ratios estimated by \citet{Kos17}. {\it Right panel:} Comparison of the probability density function (PDF) of  AGN luminosities in the 2-10~keV band  for 
 the combined sample of GATOS and NUGA (gray-filled histogram) and the  Swift-BAT sample of  \citet{Ric17a} (blue histogram). We identify the distribution of NUGA targets by the hatched area. This figure illustrates the complementarity of NUGA and GATOS samples in
 covering a significant ($\sim$4.5~dex) range of intrinsic AGN luminosities with the combined sample of 19 nearby Seyferts used in this work.}  
   \label{sample}
    \end{figure*}

We finally  selected a subset of 10 targets within the GATOS sample defined above for the new ALMA Cycle 6 observations presented in this paper. This is hereafter referred to as the GATOS core sample. We limited the distances to $<28\,$Mpc  to be able to spatially resolve tori as small as $\sim10$\,pc with an angular 
resolution of $\sim0\farcs1$ and the luminosities to  $L_{\rm AGN}$(14-150 keV) $\ge 10^{42}\,{\rm erg s}^{-1}$ to complement ongoing ALMA surveys of nearby 
Seyferts such as the Nuclei of Galaxies (NUGA) survey  \citep{Com19, Aud19}. The NUGA sample includes seven nearby low-luminosity Seyfert galaxies:  NGC\,613, NGC\,1326,  NGC\,1365, NGC\,1433 NGC\,1566, NGC\,1672, and NGC\,1808. With the exception of NGC~1365, which is also originally  part of the GATOS sample, the sources of NUGA have significantly lower Eddington ratios ($\sim$10$^{-6.5}$--$10^{-3.5}$) and AGN luminosities ($L_{\rm AGN}$(2-10~keV)$~\sim10^{39.1-41.3}$erg~s$^{-1}$) compared to the GATOS core sample \citep[see][and Table~\ref{Tab1}]{Com19}.  Hereafter, when we refer to NUGA targets we will leave out NGC~1365, which satisfies the selection criterion used in the definition of GATOS sources. The sample used in this paper, listed in Table~\ref{Tab1}, contains three additional GATOS targets that have already been observed by ALMA in previous cycles in band~7 with a frequency coverage,  a spatial resolution and sensitivity similar to those of the core sample, namely, NGC~1068 \citep{GB19}, NGC~1365 \citep{Com19}, and NGC~3227  \citep{Alo19}. As can be seen from Fig.~\ref{sample}, the selection of GATOS sources 
 listed in Table~\ref{Tab1} probes a range of $\sim$2~dex in AGN luminosities ($L_{\rm AGN}$(2-10~keV)$~\sim10^{41.4-43.5}$erg~s$^{-1}$) and $\sim$2.7~dex in Eddington ratios ($\sim$10$^{-3.0}$--$10^{-0.3}$), encompassing both unabsorbed ($N_{\rm H} <$10$^{22}$~cm$^{-2}$, $\sim$type 1) and absorbed ($N_{\rm H}>$10$^{22}$cm$^{-2}$, $\sim$type 2) AGN including Compton-thick objects ($N_{\rm H}>$10$^{24}$cm$^{-2}$). Furthermore, the combination of GATOS and NUGA samples, contributing 13 and 6 targets, respectively, allows us to compile a list of 19 suitable nearby AGN. Our combined sample spans a range of $\sim$6~dex in Eddington ratios  and   $\sim$4.5~dex in AGN luminosities. Since our combined sample covers a relatively small volume ($D<28\,$Mpc), it probes preferentially lower luminosities with respect to the overall {\it Swift}/BAT sample, as illustrated by the probability density functions (PDF) shown in Fig.\,\ref{sample}. One of the benefits of the Swift/BAT  selection criteria is that it allows us to recover a large fraction of the Compton-thick AGN population in the luminosity range we probe \citep[see Fig. 3b of][]{Ric15}. We plan to extend our sample with ALMA to higher luminosities in a forthcoming study. The main difference between the Swift/BAT and the NUGA+GATOS samples shown in Fig.\,\ref{sample} can be explained by the fact that the all-sky Swift/BAT survey is flux-limited \citep{Bau13}. These differences are exacerbated when the PDF of NUGA and Swift/BAT targets are compared, as NUGA preferentially selects nearby \citep[$D<17\,$Mpc;][]{Com19} low-luminosity AGN.

Appendix~\ref{App1} provides a short description of some of the main properties derived from previous observations obtained at optical, near infrared, radio centimeter and X-ray wavelengths for the galaxies of the GATOS core sample. In particular, we adopt the orientation (major axis) and extent (opening angle) of the ionized winds in the galaxies of the core sample derived from the references listed in Appendix~\ref{App1}. Furthermore, the parameters of the ionized winds in the three additional GATOS targets listed in Table~\ref{Tab1} are adopted  from \citet{Das06} for NGC~1068 \citep[see also][]{Bar14, May17, Min19}, \citet{Ven18} and \citet{Min19} for NGC~1365, and \citet{Alo19} for NGC~3227.

\begin{table*}[ht!]
\caption{Main properties of the sample used in this paper}
\centering
\resizebox{1\textwidth}{!}{ 
\begin{tabular}{lccccccccccc} 
\hline
\noalign{\smallskip} 
  Name        & $\alpha_{\rm 2000}^{[a]}$ & $\delta_{\rm 2000}^{[a]}$ & Dist$^{[b]}$  & Hubble type$^{[b]}$ & AGN type$^{[b]}$  & PA$_{\rm phot}$$^{[c]}$ & i$_{\rm phot}$$^{[c]}$ & log$_{\rm10}$$L_{\rm 14-150~keV}^{[d]}$  &   log$_{\rm10}$$L_{\rm 2-10~keV}^{[d]}$  & log$_{\rm10}$$\frac{L_{\rm AGN}}{L_{\rm Edd}}^{[e]}$  & log$_{\rm10}$$N(H)_{\rm Xabs}^{[d]}$ \\
              	   &       $^h$ $^m$ $^s$                       &          $^{\circ}$ $\arcmin$ $\arcsec$                   & Mpc   &         --        &    --            &  $^{\circ}$  &  $^{\circ}$ &  erg s$^{-1}$        &       erg s$^{-1}$  &           ---                              &    cm$^{-2}$      \\  
	   \hline
	   \hline
 \noalign{\smallskip} 	   	   
  NGC6300     & 17:16:59.473   &   -62:49:13.98 & 14.0  & SB(rs)b  &  Sy2    &  120  & 53   & 42.3       & 41.7       & -1.9      &   23.3   \\
  NGC5643     & 14:32:40.778   &  -44:10:28.60  & 16.9  &  SAB(rs)c &  Sy2   & 98 & 30  & 43.0       & 42.4 &  -1.3      &   25.4  \\
  NGC7314     &  22:35:46.230 &   -26:03:00.90  & 17.4  &  SAB(rs)bc &  Sy1.9, S1h &  3 & 70  &  42.2   & 42.2    &  -1.2      &   21.6   \\
  NGC4388     &  12:25:46.820  &   +12:39:43.45   & 18.1  &  SA(s)b &  Sy1.9, S1h &  91& 90 &  43.0     & 42.5  &  -1.1      &   23.5  \\
  NGC4941     &  13:04:13.143  &    -05:33:05.83   & 20.5  &  (R)SAB(r)ab &  Sy2    &  22 & 37  &  42.0    & 41.4  &  -2.4      &   23.7   \\
  NGC7213     &   22:09:16.260 &    -47:09:59.95 & 22.0  &  SA(s)a &  Sy1.5, radio-source & 124  & 39 & 42.3  &   41.9  &  -3.0      &   20.0  \\
  NGC7582     &    23:18:23.621 &   -42:22:14.06 & 22.5  &   (R')SB(s)ab &  Sy2 , S1i   &  156 & 68  & 43.2   & 43.5 &  -1.7      &   24.3    \\
  NGC6814     &    19:42:40.576 &   -10:19:25.50 & 22.8  &  SAB(rs)bc &  Sy1.5 &  108 & 52 & 42.6    & 42.2  &  -1.6      &   21.0   \\
  NGC5506     &   14:13:14.901  &   -03:12:27.22 & 26.4  &   Sa peculiar &  Sy1.9, S1i & 89 & 90 & 43.2  &  43.0   &  -2.3      &   22.4   \\
  NGC7465     &   23:02:00.952  &  +15:57:53.55  & 27.2  &  (R')SB(s)0 &  Sy2, S3    & 162  & 64 & 42.0   & 41.9   &  -2.2      &   21.5   \\
 \noalign{\smallskip}
  \hline
   \noalign{\smallskip}
  NGC1068     &   02:42:40.771  &  -00:00:47.84  & 14.0  & (R)SA(rs)b  & Sy2   &  73 & 35 & 42.7    & 42.8  &  -0.3      &  25.0  \\
  NGC1365     &    03:33:36.458 &   -36:08:26.37   &  18.3  &  (R')SBb(s)b   & S1.8    & 23 & 63 & 42.3  &  42.1   & -2.8       &  22.2  \\
  NGC3227     &   10:23:30.570  &   +19:51:54.30  &  23.0  &  SAB(s)a pec & S1.5    &  156 & 68 & 42.8   & 42.4   & -1.2       &  21.0  \\  
   \noalign{\smallskip} 
\hline 
\hline
\end{tabular}}
\tablefoot{[a] Phase tracking centers of the band~7 ALMA observations used in this work, which cover the GATOS core sample of 10  galaxies, listed first, and three additional targets from 
the literature: NGC~1068 \citep{GB19}, NGC~1365 \citep{Com19}, and NGC~3227 \citep{Alo19}. [b] Distances are median values of redshift-independent estimates from the Nasa Extragalactic 
Database (NED) after excluding select measurements from unreliable or outdated references; Hubble  and AGN type taken from NED. In addition to the standard classification as a function of  AGN class (1, 2,  and intermediate types: 1.n, where n ranges from 5 to 9 and numerically larger subclasses have weaker broad-line components relative to the narrow lines, following the notation of \citet{Ost81}), some objects are classified as S1h or S1i if  broad polarized 
Balmer lines or  broad Paschen lines in the infrared are detected, respectively, according to the nomenclature of \citet{Ver06}. [c] Position angle ($PA_{\rm phot}$) and inclination ($i_{\rm phot}$) of the  optical disks 
based on photometric estimates taken from HyperLeda \footnote{http://leda.univ-lyon1.fr/}. [d] Luminosities of hard X-rays ($L_{\rm 14-150~keV}$ and $L_{\rm 2-10~keV}$) and gas column densities of obscuring 
material ($N(H)_{\rm Xabs}$) are taken from \citet{Ric17a}; $L_{\rm14-150~keV}$ and  $L_{\rm2-10~keV}$ are intrinsic luminosities (corrected for absorption)  and re-scaled to the adopted distances. [e] Eddington 
ratios ($\frac{L_{\rm AGN}}{L_{\rm Edd}}$) taken from \citet{Kos17}.} \label{Tab1} 
\end{table*}


\begin{table*}[ht!]
\caption{Parameters of the ALMA observations}
\centering
\resizebox{1.0\textwidth}{!}{ 
\begin{tabular}{lccccccc} 
\hline
\noalign{\smallskip} 
 Name        & $\sigma_{\rm cont [MSR-HSR]}$  & $\sigma_{\rm CO [MSR-HSR]}$  & $\sigma_{\rm HCO^+[MSR-HSR]}$   & 
  beam$_{\rm MSR}$   & beam$_{\rm HSR}$  & FOV & LAS \\
            & $\mu$Jy~beam$^{-1}$   &    mJy~beam$^{-1}$   &   mJy~beam$^{-1}$ &     $\arcsec$ $\times$ $\arcsec$$@^{\circ}$ ~ (pc~$\times$~pc) &     $\arcsec$ $\times$ $\arcsec$$@^{\circ}$~ (pc~$\times$~pc)& $\arcsec$  (kpc) &  $\arcsec$ (pc)  \\  
           \hline
 \noalign{\smallskip} 	   	   
  NGC6300       & 36-45 & 0.52-0.62 &  0.79-1.00 &  $0\farcs13\times0\farcs10@-44^{\circ}$ (9 $ \times $ 8) &  $0\farcs11\times0\farcs08@-21^{\circ}$ (8 $ \times $ 6) & 17 (1.2) & 4 (280) \\
  NGC5643     &  32-38 & 0.52-0.57 & 0.77-1.00  &   $0\farcs12\times0\farcs10@114^{\circ}$ (10 $ \times $ 8) & $0\farcs09\times0\farcs08@122^{\circ}$ (8 $ \times $ 7)   & 17 (1.4) & 4 (340) \\
  NGC7314      & 25-27 &  0.45-0.50 &  0.61-0.69 &  $0\farcs17\times0\farcs15@-4^{\circ}$ (15 $ \times $ 13) &$0\farcs14\times0\farcs10@11^{\circ}$ (12 $ \times $ 9)    & 17 (1.5) & 4 (350) \\
  NGC4388    &  36-45 & 0.62-0.63 &  0.84-0.99 &  $0\farcs14\times0\farcs12@89^{\circ}$  (13 $ \times $ 11) & $0\farcs12\times0\farcs09@102^{\circ}$ (11 $ \times $ 8) & 17 (1.5) &  4 (360) \\
  NGC4941    &  24-33 & 0.48-0.58 &  0.60-0.73 & $0\farcs15\times0\farcs11@103^{\circ}$ (15 $ \times $ 11) & $0\farcs13\times0\farcs07@103^{\circ}$  (13 $ \times $ 7) & 17 (1.7)  & 4 (410) \\ 
  NGC7213    &  54-60 & 0.45-0.60  &  0.56-NA & $0\farcs08\times0\farcs07@107^{\circ}$ (9 $ \times $ 8) & $0\farcs07\times0\farcs06@93^{\circ}$ (8 $ \times $ 7) & 17 (1.9)  & 4 (440) \\  
  NGC7582    &  43-46 &  0.55-0.65 &  0.75-0.91 & $0\farcs17\times0\farcs15@146^{\circ}$ (19 $ \times $ 17)  & $0\farcs13\times0\farcs11@5^{\circ}$ (15 $ \times $ 12) & 17 (1.9)  & 4 (450) \\ 
  NGC6814   &  24-29 &  0.42-0.54 &  0.53-NA & $0\farcs12\times0\farcs08@-61^{\circ}$ (14 $ \times $ 9) & $0\farcs09\times0\farcs07@99^{\circ}$ (10 $ \times $ 8) & 17 (1.9)  & 4 (460) \\ 
  NGC5506    &  31-36 & 0.51-0.59 &  0.65-0.79 &$0\farcs21\times0\farcs13@-61^{\circ}$ (28 $ \times $ 17) & $0\farcs18\times0\farcs09@-57^{\circ}$ (24 $ \times $ 12) & 17 (2.2) & 4 (530)  \\  
  NGC7465     & 23-30  &  0.42-0.56 &  0.49-0.67 &  $0\farcs12\times0\farcs08@-31^{\circ}$ (16 $ \times $ 11) &  $0\farcs08\times0\farcs07@143^{\circ}$ (11 $ \times $ 9) &17 (2.3) & 4 (540) \\  
  NGC1068     & 46-46 & 0.23-0.23 &  0.28-0.28	& $0\farcs13\times0\farcs07@100^{\circ}$ (9 $ \times $ 5 ) & $0\farcs04\times0\farcs03@74^{\circ}$ (3 $ \times $ 2)	 & 17 (1.2) & 1.8 (130) \\ 
  NGC1365    &  30  & 0.60 &  -- &  $0\farcs08\times0\farcs06@-59^{\circ}$ (12 $ \times $ 10) &  -- & 17 (1.6) & 3 (275) \\
  NGC3227    &  31  & 0.51 &  -- &  $0\farcs10\times0\farcs09@21^{\circ}$ (12 $ \times $ 10) &  -- & 17 (2.0) & 1 (120) \\
     \noalign{\smallskip} 
\hline 
\end{tabular}}
\tablefoot{Columns (2), (3), and (4) list the range of 1$\sigma$ sensitivities for the continuum as well as for the CO and HCO$^+$ line observations. Line sensitivities are derived for channels  of 10~km~s$^{-1}$-width (except for 
NGC\,1068, where \citet{GB19}  used 20~km~s$^{-1}$-wide channels). The range in columns
(2)--to--(4) accounts for the moderate spatial resolution data sets (MSR) and the high spatial resolution data sets (HSR) obtained by using a different robust weighting parameter
of the visibilities in the UV plane: $robust=1$ (0.1) for the MSR (HSR) configuration; these are available for all sources except for NGC\,3227. 
Columns (5) and (6) list the range of spatial resolutions (in $\arcsec$ and pc) reached for the MSR and HSR observations respectively. The field-of-view (FOV) and largest scale recovered (LAS) 
are listed in columns (7) and (8) in $\arcsec$ and kpc and pc, respectively.} \label{Tab2} 
\end{table*}

%

%
%
\section{Observations}\label{observations}
\subsection{ALMA data}\label{ALMA}
 We observed the emission of the CO(3--2) and HCO$^+$(4--3) lines and their underlying continuum emission in the CND of our targets with ALMA during Cycle~6 and 7 using  band~7 receivers (project-IDs: $\#$2017.1.00082.S and $\#$2018.1.00113.S; PI: S.~Garc\'{\i}a-Burillo). 
The phase tracking centers of the galaxies were taken from the All Sky--2MASS survey \citep{Skr06}. We note however that the positions of the AGN in each galaxy have been determined in this work through a fit of the continuum emission, as detailed in Sect.~\ref{continuum}. 
 We used a single pointing with a field-of-view (FOV) of 17$\arcsec$. Observations were designed to cover the emission of molecular gas and dust inside the CND ($r\leq 600-1200$~pc) of the ten selected Seyfert galaxies of the GATOS core sample using a common average angular resolution of $\sim0.1\arcsec$, which translates into a spatial  scale $\sim7-13$~pc for the range of distances to our targets. We combined two sets of configurations of the ALMA array: {\em extended} (C43-6) and {\em 
 compact} (C43-3). This mix of configurations assured that the largest angular scale  recovered in our maps is about $\sim4\arcsec=300-500$~pc, enough to 
 recover a sizeable fraction of the flux inside the FOV and virtually all the flux at scales that are mostly relevant to image both the dusty molecular tori and the connection to their hosts ($r\sim200$~pc). 
Observations typically required the execution of two tracks (one track per each configuration set) conducted between December 2017 and August 2019 and the use of 
$\sim43-49$ antennas of ALMA (see Table~\ref{Tab2} for details).

Four spectral windows  of 1.875~GHz-bandwidth were placed, two in the 
lower side band (LSB) and two in the upper sideband (USB). This setup allowed us to observe the  CO($J=3-2$) line (345.796~GHz at rest) and the continuum emission (343.901-344.101~GHz at rest) in the 
LSB bands, as well as  HCO$^+$($J=4-3$) (356.734~GHz at rest) and the continuum emission (355.845-358.000~GHz at rest) in the USB bands. We calibrated the data making use of the ALMA 
reduction package  {\tt CASA} \citep{McM07}\footnote{http//casa.nrao.edu/}. The calibrated uv-tables were exported to {\tt GILDAS\footnote{http://www.iram.fr/IRAMFR/GILDAS}}  to proceed with the 
continuum subtraction and imaging procedures as detailed below. We first subtracted the continuum from each of the spectral $(u,v)$ data sets using the {\tt GILDAS} task {\tt UV-BASELINE}. We fitted  
a baseline to the  $(u,v)$ data sets through a polynomial of degree zero masking the line emission around each transition with a range of velocity widths $\sim300-600$~km~s$^{-1}$ adapted for each target  
and subsequently obtained continuum-free spectral line images for  CO($J=3-2$) and HCO$^+$($J=4-3$). We derived images of the continuum emission by averaging in each of the two sub-bands 
centered around spectral lines those channels free of line emission using the  {\tt GILDAS} tasks {\tt UV-FILTER} and {\tt UV-CONT}, making use of the same velocity width masks employed by the {\tt UV-BASELINE} task. The line-free continuum emission images were combined with the genuine 
continuum images obtained in the remaining bands to produce a noise-weighted average image of the continuum emission at an average frequency range $=350.570-351.133$~GHz (at rest).

The flux accuracy is estimated to be  about 10--15$\%$, i.e., in line with the goal of standard ALMA observations at these frequencies.  We obtain for each galaxy two sets of angular resolutions by changing in the {\tt GILDAS} task {\tt UV-MAP} the robust parameter ($b$) from 1 (in the moderate spatial resolution data set, hereafter MSR) to 0.1 (in the high spatial resolution data set, hereafter HSR). 
The line data cubes were binned to a common frequency  resolution of 11.7~MHz (equivalent to $\sim$10~km~s$^{-1}$ in band~7). The point source sensitivities in the line data cubes were derived selecting areas free from emission in all channels. We summarize in Table~\ref{Tab2} the relevant parameters of the ALMA observations for the ten galaxies of the GATOS core sample. We also list the  parameters of the ALMA observations of the three galaxies used to complete the sample used in this work (NGC\,1068, NGC\,3227, and NGC\,1365). For these galaxies there are ALMA data published by \citet{GB19} (NGC\,1068), \citet{Alo19} (NGC\,3227), and \citet{Com19} (NGC\,1365), which have spatial resolutions and sensitivity requirements comparable to those of the GATOS core sample.

\subsection{Ancillary data}\label{ancillary}

\subsubsection{Archival {\it HST} images.}\label{HST-data}

We downloaded fully reduced optical images from the Hubble Legacy Archive (HLA)\footnote{https://hla.stsci.edu/} taken with the WFPC2/PC instrument on board
the Hubble Space Telescope ({\it HST}) using the broad-band filter F606W, except for NGC1068, for which we use the
narrower filter F547M. For the galaxy NGC7314 we also retrieved an image in the broad-band filter F450W. 
The HLA WFPC2 images are drizzled to a pixel size of 0.05\arcsec~using the MultiDrizzle
software. In order to construct $V-H$ color images of the targets, we downloaded either NICMOS or WFC3 near-infrared images in 
the broad-band filter F160W, except for the galaxy 
NGC7314, for which no near-infrared {\it HST} images are available and the two optical filters mentioned above were used instead. Details 
of the optical and near-infrared observations are shown in Table~\ref{tab_hst} of Appendix~\ref{App2}. The HLA NICMOS and WFC3 images are drizzled to 
pixel sizes of 0.05\arcsec~and 0.09\arcsec~respectively, whereas those downloaded from the Mikulski Archive for Space Telescopes 
(MAST) Portal\footnote{https://archive.stsci.edu/}, which were processed with either the CALNIC or CALWF3 pipelines, have pixel sizes of 0.075\arcsec~and 
0.128\arcsec~respectively. We corrected the astrometry of  the images using the position of reference stars in the field from Gaia data release 2 \citep[Gaia DR2;][]{Gai18}.
 Corrections typically ranged from $0\farcs1$ to $0\farcs5$.

\subsubsection{X-ray observations: {\tt NuSTAR}}\label{Xray-data}

The Nuclear Spectroscopic Telescope Array ({\tt NuSTAR}) is the first focusing hard X-ray telescope with high sensitivity\footnote{https://heasarc.gsfc.nasa.gov/docs/nustar/}. This gives the advantage to observe with a single mode from 3 to 79\,keV, perfectly suited to study the AGN reflection component.
Therefore, we used the hard band spectrum observed with {\tt NuSTAR} \citep{Har13}, including both $FPMA$ and $FPMB$ focal plane modules. We looked for {\tt NuSTAR} observations for our sample using the High Energy Astrophysics Archive Research Center (HEASARC) archive\footnote{https://heasarc.gsfc.nasa.gov}, finding observations for 11 AGN of our sample.  We chose the longest exposure if several observations were available.

{\tt NuSTAR} data reduction was done using the data analysis software {\tt NuSTARDAS} ($v.1.4.4$) distributed by HEASARC. The calibrated, cleaned, and screened event files were generated using the {\sc nupipeline} task. A circular region of 1 arcmin radius was taken to extract the source and background spectrum on the same detector and to compute the response files (RMF and ARF files) using the {\sc nuproducts} package within {\tt NuSTARDAS}. Finally, we used the {\sc grppha} task  to group the spectra with at least 60 counts per bin. We used the {\tt NuSTAR} data in the range 3--70~keV.

  \begin{figure*}[tbh!]
     \centering
    \includegraphics[width=0.75\textwidth]{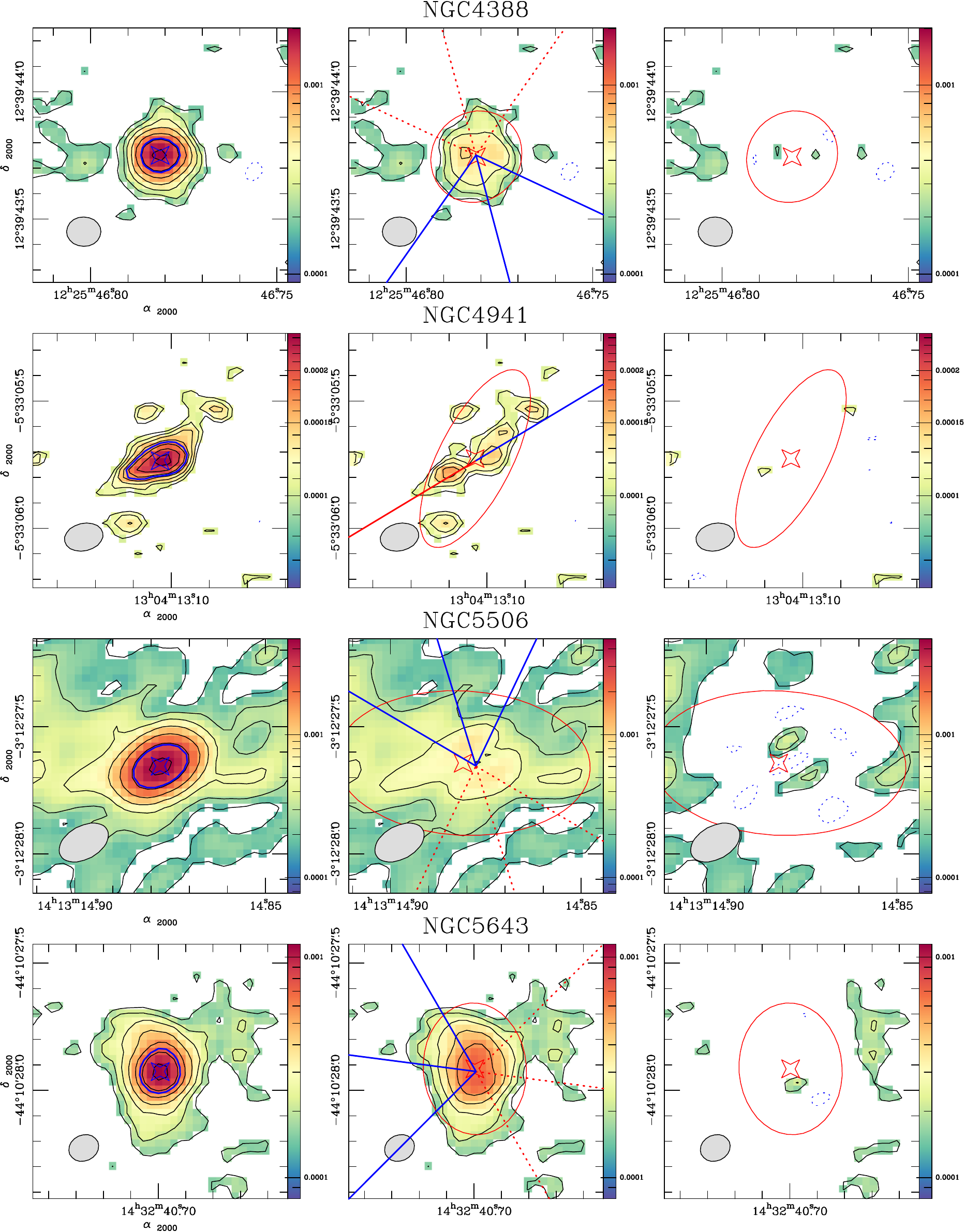}
   \caption{{\it Left panels:}~Continuum emission at the (rest) frequency range 350.6-351.1~GHz  ($I_{\rm cont}$) in the central $\Delta\alpha\times\Delta\delta
   =1\arcsec\times1\arcsec$ regions  of
   \object{NGC~4388}, \object{NGC~4941}, \object{NGC~5506}, and \object{NGC~5643}, obtained from the ALMA MSR datasets. The AGN positions, identified by the blue star markers, correspond to the location of the point sources determined 
   by the two-component fit to the continuum emission  described in Sect.~\ref{continuum}. The continuum maps  are shown in color scale and (black) contour levels with a logarithmic spacing from --2.5$\sigma_{\rm cont}$(dashed blue),  2.5$\sigma_{\rm cont}$ to 90$\%$ of the peak intensity ($I_{\rm cont}^{\rm max}$ in Jy beam$^{-1}$-units)  in steps of $\sim$0.18 dex on average. The (thick blue) contour corresponds to 1/2$\times I_{\rm cont}^{\rm max}$.  The values of  $\sigma_{\rm cont}$ and the parameters of the two-component fit for each galaxy are listed in Tables~\ref{Tab2} and \ref{Tab3}. {\it Middle panels:}~Same as {\it left panels} but after subtraction of the fitted point source. The (red) ellipses identify the Gaussian source fitted to the extended emission centered around the position of the red star markers. The lines highlight both the orientation (major axis) and the extent (opening angle) of the ionized winds identified in the literature for the targets (see Appendix~\ref{App1}), except for NGC~4941 and NGC~7213, where there is only an estimate of the orientation (major axis) of the wind. Lines are color-coded to reflect whether the measured velocities of the ionized wind lobes are either redshifted or blueshifted. Dashed lines indicate that the corresponding lobe is (mostly) obscured by the disk of the host. {\it Right panels:}~Residuals obtained after subtraction of the sum of the point and extended sources fitted to the data. The (grey) filled ellipses at the bottom left corners in all panels represent the beam sizes of the observations listed in Table~\ref{Tab2}.}  
   \label{cont-FITI}
    \end{figure*}
  \begin{figure*}
   \centering
    \includegraphics[width=0.75\textwidth]{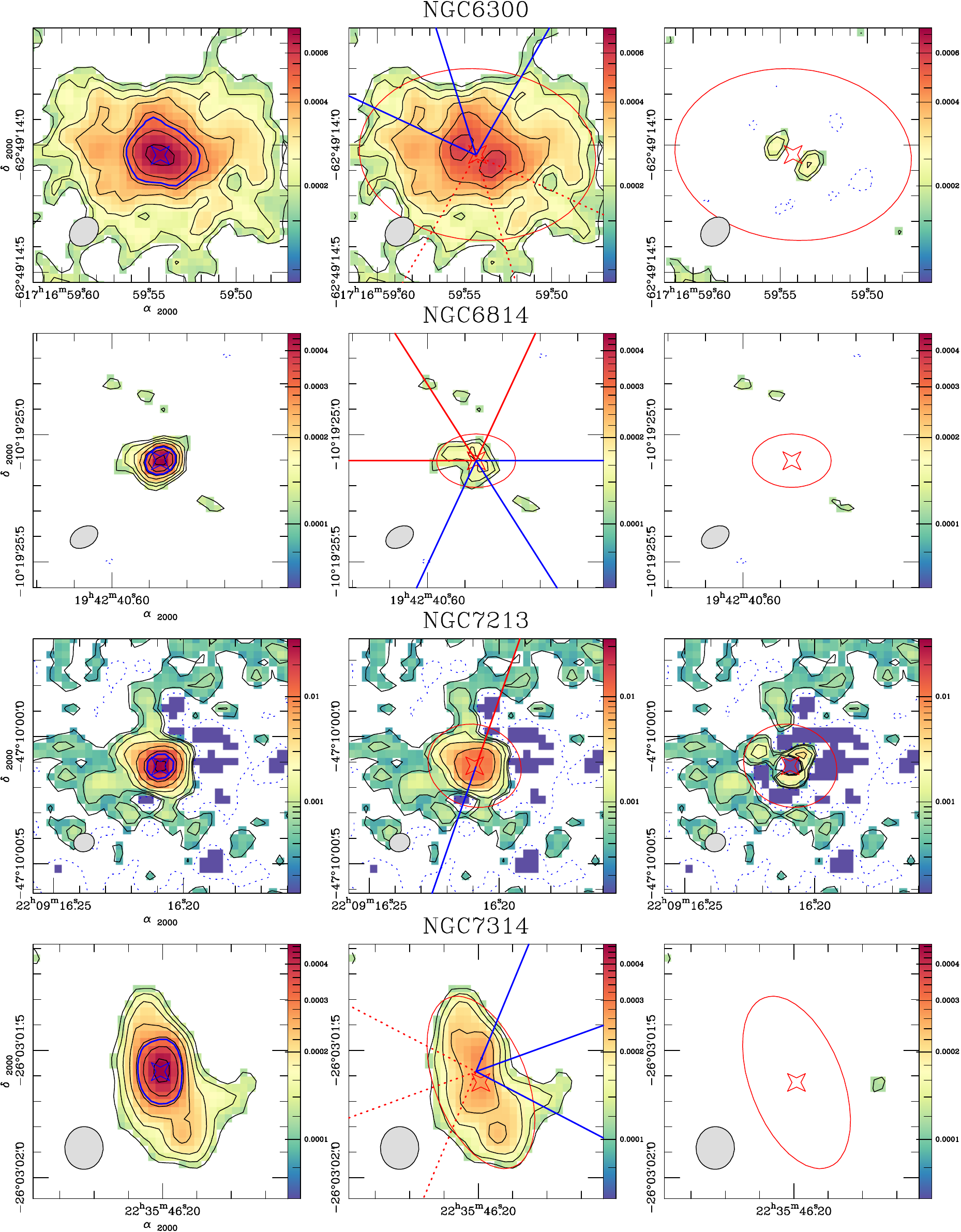}
   \caption{Same as Fig.~\ref{cont-FITI}, but for  \object{NGC~6300}, \object{NGC~6814}, \object{NGC~7213}, and \object{NGC~7314}.}  
   \label{cont-FITII}
    \end{figure*}

  \begin{figure*}
   \centering
    \includegraphics[width=0.75\textwidth]{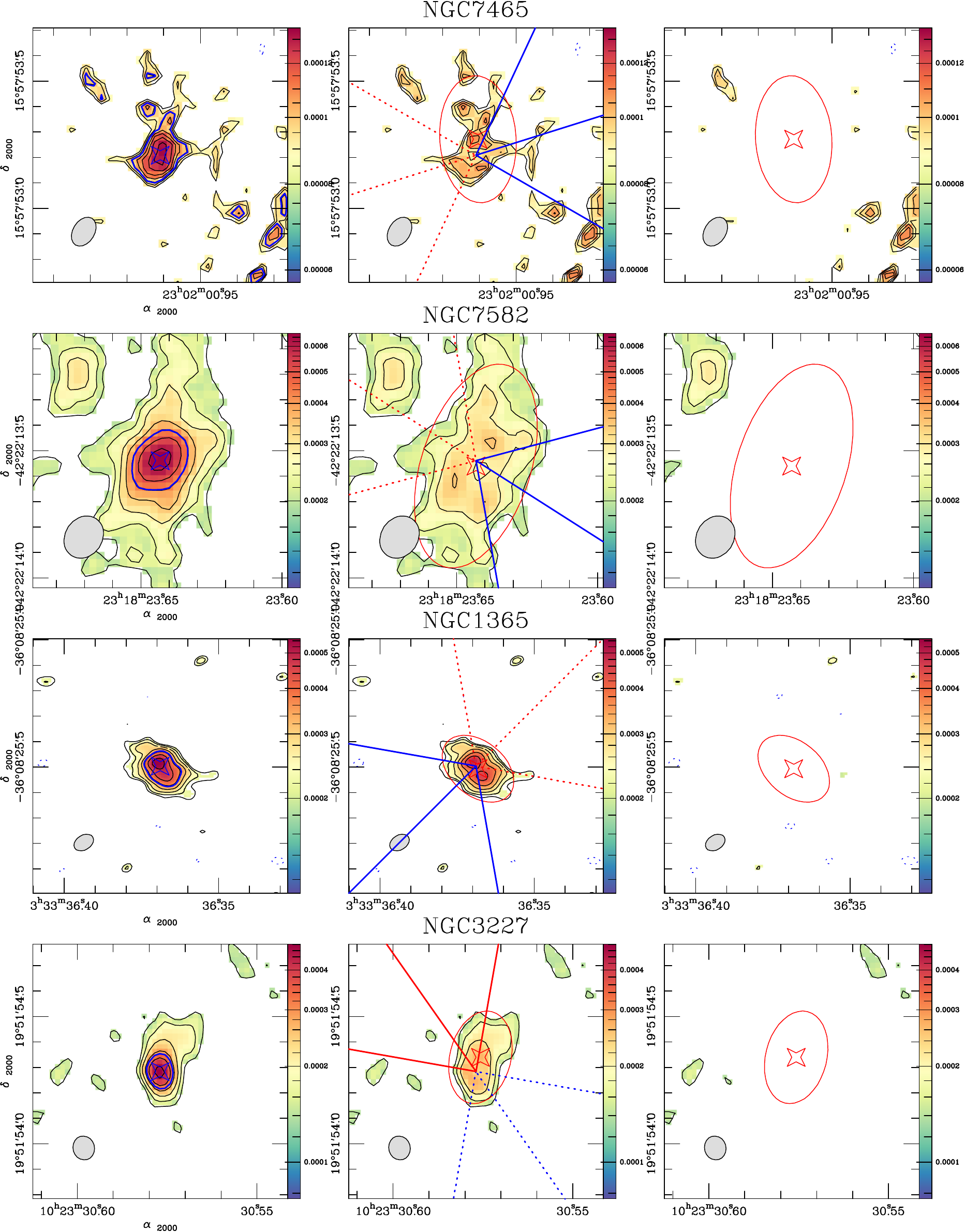}
   \caption{Same as Fig.~\ref{cont-FITI}, but for  \object{NGC~7465}, \object{NGC~7582}, \object{NGC~1365}, and \object{NGC~3227}.}  
   \label{cont-FITIII}
    \end{figure*}

\section{Continuum maps}\label{continuum}

\subsection{Morphology of the emission}\label{continuum-morphology}

Figures~\ref{cont-FITI}-to-\ref{cont-FITIII} show the continuum maps derived at  the (rest) frequency $\sim351$~GHz  (870~$\mu$m)   in the central $\Delta\alpha\times\Delta\delta
   =1\arcsec\times1\arcsec$  regions of the GATOS core sample galaxies. The maps were obtained from the ALMA MSR datasets. A feature common to all the images is the presence of disk-like morphologies where we identify by eye two main components: 1) a bright unresolved point source, and 2) a spatially-resolved  extended component characterized by lower surface brightness emission. A visual inspection of  Figs~\ref{cont-FITI}-to-\ref{cont-FITIII} indicates that  there is a large variance in the contributions of point sources and extended components to the total flux in the galaxies  of our sample\footnote{A similar conclusion is drawn from an inspection of the 870~$\mu$m continuum images obtained from the ALMA HSR dataset.}. The extended components adopt different elongated morphologies, which appear to be oriented mostly perpendicular relative to the AGN wind axes in the majority of our sources (see discussion in Sect.~\ref{orientation}).

We expect that  the contribution of thermal dust emission at (rest-frame) frequencies $\sim$ 351~GHz, which falls on the Rayleigh-Jeans portion of the submillimeter spectral energy distribution, dominates in normal star-forming and starburst galaxies. The submillimeter flux has been instrumental in deriving the total gas mass in different galaxy populations \citep[e.g.][]{Sie07, Dra07, Gal11, Sco14, Sco16}. Continuum emission in this frequency range  can be due the presence of dusty compact disks (`tori') located around the central engines. This emission has been seen to extend over scales of  tens  of parsecs in AGN \citep{Com19, GB19, Alo19}. Alternatively, dust emission may also arise from  components oriented in the polar direction. Polar dust emission can extend over scales of  tens or hundreds of parsecs.

  ALMA high-resolution images of the  submillimeter continuum  in the circumnuclear disks of Seyfert galaxies, similar to the GATOS sample galaxies shown in Figs~ \ref{cont-FITI}-to-\ref{cont-FITIII}, have nevertheless shown that this emission can also arise  from various  physical mechanisms not necessarily related to dust. A significant fraction of the submillimeter continuum  can also be attributed to non-thermal synchrotron  emission or to thermal free-free emission of the ionized plasma. The former originates either from small radio-jets or from compact subparsec-scale AGN cores. While these components are generally detected at cm wavelengths in radio-quiet AGN \citep{Nag99, Mun00,Ulv01,Ori10}, they can also contribute to the submillimeter flux at scales that would be unresolved by ALMA \citep{GB14,GB19, Alo19, Com19, Pas19, Ros19}.

NGC~1068 is a well-studied case that illustrates the need for high-resolution observations to disentangle this complexity.  \citet{GB19} used the information available from cm and mm-wavelengths \citep[e.g.;][and references therein] {Hoe08, Kri11} and estimated that the fraction of the 870~$\mu$m-emission  that can be attributed to dust  at the location of the AGN source of NGC~1068 is $\leq10\%$  \citep[knot S1, according to the nomenclature used by][]{Gal96}. Most of the submillimeter flux in the S1 knot arises from free-free emission of the ionized plasma at scales $\leq1$~pc. However, after excluding  the circumnuclear disk regions close to the radio-jet trail and the S1 knot \citep[imaged by ][]{Gal96, Gal04}, \citet{GB19} concluded that the bulk of the submillimeter continuum can be attributed to dust emission.

   \begin{figure}
   \centering
    \includegraphics[width=0.85\textwidth]{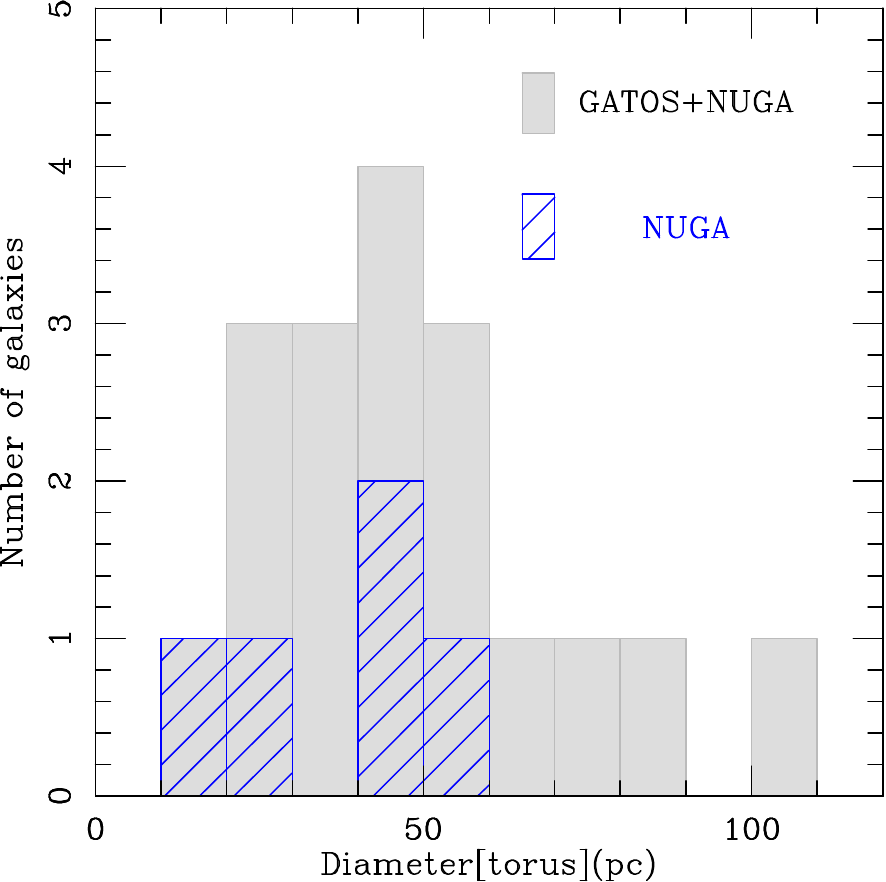} 
    \caption{Histograms showing the distribution of the deconvolved diameters of the dusty molecular tori (or polar components) derived from the combined sample of NUGA and  GATOS (gray-filled histograms). We adopt for GATOS the average sizes  estimated from the MSR and HSR datasets. The blue-filled histogram identifies the distribution of sizes of the tori derived for NUGA galaxies \citep{Com19}.}       
 \label{histo-sizes}
   \end{figure}


\begin{table*}[t!]
\caption{Parameters of the point and extended sources fitted to the continuum emission of the MSR and HSR datasets}
\centering
\resizebox{1\textwidth}{!}{ 
\begin{tabular}{lccccccccc} 
\hline
\noalign{\smallskip} 
  Name        & $\alpha_{\rm 2000}^{[a]}$ & $\delta_{\rm 2000}^{[a]}$ & $I_{\rm point}$$^{[b]}$  &  $I_{\rm Gauss}$$^{[c]}$ & $I_{\rm Gauss}/I_{\rm point}$$^{[d]}$ & FWHM$_{\rm x}$$^{[e]}$  & FWHM$_{\rm y}$$^{[e]}$  &  $PA_{\rm Gauss}$$^{[e]}$   &   $i_{\rm Gauss}$$^{[e]}$  \\
              	   &       $^h$ $^m$ $^s$                       &          $^{\circ}$ $\arcmin$ $\arcsec$                   & mJy   &         mJy        &          & $\arcsec$            &    $\arcsec$        &                $^\circ$                             &    $^\circ$       \\  
	   \hline
	   \hline
	            &							  &									     &      (MSR,~HSR)         & 	 (MSR,~HSR)	& (MSR,~HSR)	   & 		 (MSR,~HSR)		&  (MSR,~HSR)	 &   (MSR,~HSR)  &   (MSR,~HSR) \\
	   \hline
	   \hline
 \noalign{\smallskip} 	   	   
  NGC6300     &  17:16:59.543    &   -62:49:14.04 & (0.37$\pm$0.04, 0.22$\pm$0.05) & (10.52$\pm$0.17, 11.95$\pm$0.26) &  	(30, 55)$\pm7$	&	(0.65, 0.64)$\pm$0.02    &  (0.47, 0.46)$\pm$0.02       &  (85, 85)$\pm$2    &  ($>$44, $>$44) \\
  NGC5643     & 14:32:40.699   &  -44:10:27.93  & (0.69$\pm$0.03, 0.67$\pm$0.04) & (3.95$\pm$0.08, 4.27$\pm$0.12)  & 	(6, 10)$\pm1$	&	(0.32, 0.32)$\pm$0.02    &  (0.25, 0.23)$\pm$0.02    &  (4, 5)$\pm$2    &  ($>$39,  $>$44)  \\
  NGC7314     &  22:35:46.201 &   -26:03:01.58  &  (0.29$\pm$0.03, 0.29$\pm$0.03)  & (1.43$\pm$0.06, 1.53$\pm$0.09)  &  	(5 ,5)$\pm1$	&	(0.57, 0.59)$\pm$0.01    &  (0.29, 0.31)$\pm$0.01       &  (21, 30)$\pm$3     &  ($>$59, $>$58)   \\  
  NGC4388     &  12:25:46.781  &   +12:39:43.75   &  (1.86$\pm$0.04, 1.79$\pm$0.05)  & (1.75$\pm$0.07, 1.92$\pm$0.11)   &  	(1 ,1)	$\pm0.1$	& 	(0.27, 0.26)$\pm$0.01    &  (0.25, 0.23)$\pm$0.01       &  (-43, -7)$\pm$5     &  ($>$22, $>$28)  \\
  NGC4941     &  13:04:13.103  &    -05:33:05.73   &  (0.19$\pm$0.03, 0.16$\pm$0.03)  & (1.13$\pm$0.10, 1.34$\pm$0.16)  &  	(6, 9)	$\pm1$	&	(0.77, 0.73)$\pm$0.02   &  (0.29, 0.28)$\pm$0.02       &  (-27, -29)$\pm$5     &  ($>$68, $>$67)   \\
  NGC7213     &   22:09:16.209 &    -47:10:00.12 &  (36.5$\pm$0.05, 36.10$\pm$0.06)  & (39.59$\pm$0.10, 19.10$\pm$0.10)  &  (1, 0.5)$\pm0.1$  	&	(0.15, 0.14)$\pm$0.01    &  (0.13, 0.09)$\pm$0.01       &  (71, 44)$\pm$8     &  ($>$30, $>$50) \\
  NGC7582     &    23:18:23.643 &   -42:22:13.54 &  (0.55$\pm$0.04, 0.52$\pm$0.05)  & (3.32$\pm$0.16, 3.55$\pm$0.21)  &  	(6, 7)	$\pm1$	&	(0.83, 0.74)$\pm$0.02    &  (0.43, 0.40)$\pm$0.01       &  (-18, -20) $\pm$2    &  ($>$59, $>$57) \\
  NGC6814     &    19:42:40.587 &   -10:19:25.10 &  (0.42$\pm$0.02, 0.43$\pm$0.03)  & (0.63$\pm$0.06, 0.49$\pm$0.07)  &  	(1.5, 1)$\pm0.1$	&	(0.31, 0.35)$\pm$0.02    &  (0.21, 0.12)$\pm$0.02       &  (90, 61)$\pm$10     &  ($>$47, $>$70)  \\
  NGC5506     &   14:13:14.878  &   -03:12:27.66 &  (5.07$\pm$0.03, 4.73$\pm$0.04)  & (4.06$\pm$0.09, 3.96$\pm$0.11)  &  (1, 1)$\pm0.1$		&	(0.67, 0.52)$\pm$0.01    &  (0.38, 0.31) $\pm$0.01     &  (87, 88)$\pm$2  &  ($>$55, $>$53) \\
  NGC7465     &   23:02:00.961  &  +15:57:53.21  &   (0.08$\pm$0.02, $<$0.09)  & (1.19$\pm$0.09, 1.13$\pm$0.13)  &  	(14, 22)$\pm4$		&	(0.50, 0.42) $\pm$0.02   &  (0.30, 0.24)$\pm$0.01       &  (4, 4)$\pm$4   & ($>$53, $>$55)  \\

 \noalign{\smallskip}
  \hline
  \hline
   \noalign{\smallskip}
  NGC1365     &    03:33:36.369 &   -36:08:25.50   &   0.15$\pm$0.04  & 2.76$\pm0.11$  & 	19$\pm5$							& 	0.21$\pm$0.03    &  0.14$\pm$0.03       &  50$\pm$10    &  $>$48  \\
  NGC3227     &   10:23:30.577  &   +19:51:54.28  &   0.36$\pm$0.03  & 2.17$\pm0.11$  &  	6$\pm1$							&	0.37$\pm$0.01    &  0.24$\pm$0.01       &  166$\pm$2    &  $>$50  \\  
  
   \noalign{\smallskip} 
\hline 
\hline
\end{tabular}}
\tablefoot{[a] Positions of  the unresolved (point-like) components fitted to the continuum emission derived from the MSR and HSR datasets of the GATOS core sample (listed first); we also include new fits to the data of NGC~1365 \citep{Com19}, and NGC~3227 \citep{Alo19}. [b] Intensity of the fitted point sources. [c] Spatially-integrated intensity of the elliptical Gaussians that fit the extended emission. [d] Flux ratios between the extended components and the points sources. [e] Extent (major and minor FWHM: FWHM$_{\rm x,y}$) and orientation ($PA_{\rm Gauss}$) of the Gaussians. We also list the lower limit to the inclinations of the extended components derived assuming axisymmetry and a negligible thickness-to-radius ($H/R$) ratio for these features: $i_{\rm Gauss}$=acos(FWHM$_{\rm y}$/FWHM$_{\rm x}$).} \label{Tab3} 
\end{table*}



\begin{table*}[bth]
\caption{Sizes, masses, and orientations of the extended components of continuum emission.}
\centering
\resizebox{.8\textwidth}{!}{ 
\begin{tabular}{lccccccccc} 
\hline
\noalign{\smallskip} 
  Name        & Size$^{[a]}$  & log$_{\rm 10}M_{\rm gas}^{\rm dust}$ $^{[b]}$ &  log$_{\rm 10}M_{\rm gas}^{\rm CO}$ $^{[c]}$ &  log$_{\rm 10}N_{\rm H_2}^{\rm AGN}$ $^{[d]}$ & $PA_{\rm Gauss}$$^{[e]}$   &   $PA_{\rm out}$$^{[f]}$  &   $\Omega_{\rm out}$$^{[f]}$  & $\Delta$$^{[g]}$ & Geometry$^{[h]}$ \\
              	   &    pc    	  &   M$\sun$      & M$\sun$ &   mol.~cm$^{-2}$ & $^\circ$   				&    $^\circ$      			  &    $^\circ$  &    $^\circ$ 		    &  			 ---      \\  
	   \hline
	   \hline
	            &	(MSR,~HSR)	  &  (MSR,~HSR) &      (MSR,~HSR)    &      (MSR,~HSR)     & 	(MSR, ~HSR)	   & 		--	&  -- & (MSR,~HSR) & -- \\
	   \hline
	   \hline
 \noalign{\smallskip} 	   	   
  NGC6300     &  (64, 53) & (6.5, 6.6) & (6.8, 6.7) & (23.4, 23.4)  & (85, 85)    &  18 & 95 & (67, 67) & torus \\
  NGC5643     &  (43, 41) & (6.3, 6.5) & (6.7, 6.7) & (23.6, 23.6) &  (4, 5)    	  &  83 & 105 & (79, 78)  & torus  \\
  NGC7314     &  (60, 50) & (5.9, 5.8) & (5.8, 5.7) &  (22.1, 22.1) & (21, 30)    &  -70 & 95 &(89, 80) & torus \\  
  NGC4388     &  (32, 29) & (6.0, 6.0) & (5.5, 5.4) &  (22.3, 22.4) & (-43, -7)    &  15 & 100  & (58, 22) & {\it mixed/ polar}  \\
  NGC4941     &  (78, 74) & (5.9, 6.0) & (5.9, 5.7) &  (21.9, 21.9) & (-27, -29)    &  -59 & -- & (32, 30) & mixed / polar  \\
  NGC7213     &  (40, 34) & (7.5, 7.4) & (4.8, 4.5) &   ($<$22.0, $<$22.3) & (71, 44)    &   {\it -19} &  -- & (90, 63) & {\it torus}  \\
  NGC7582     &  (91, 82) & (6.4, 6.6) & (6.6, 6.5) &  (22.6, 22.5) & (-18, -20)    &  58 &  95 & (76, 78) & torus   \\
  NGC6814     &  (33, 39) & (5.7, 5.8) & (4.9, 4.9) &  ($<$21.8, $<$22.3)  &  (90, 61)    &   33 & 115 & (57, 28) & {\it mixed / polar}  \\
  NGC5506     &  (129, 91) & (6.7, 6.9) & (7.0, 6.7) & (22.6, 22.7) & (87, 88)    & 17  & 85 & (70, 71) & torus  \\
  NGC7465     &  (67, 56) & (6.2, 6.3) & (6.3, 6.2) &  (22.6, 22.7)  & (4, 4)    &  -72 &  95  & (76, 76) & torus   \\

 \noalign{\smallskip}
  \hline
  \hline
   \noalign{\smallskip}
  NGC1068    &  26  &  5.0 & 5.5 & 23.6 & -55    &  30  &  80  &  85 & torus   \\  
  NGC1365    &  28  & 6.2 & 5.4 & 22.3 &  50    &  -45 &  110 & 85 & torus  \\
  NGC3227    &  41  & 6.3  & 6.2 & 22.7 & -14    &  35 &  90 & 49 & mixed  \\  
   \noalign{\smallskip} 
\hline 
\hline
\end{tabular}}
\tablefoot{[a] Deconvolved diameters of the extended components derived from the MSR and HSR continuum datasets of the GATOS core sample (listed first); we also include fits to the data of NGC~1068 \citep{GB16}, NGC~1365 \citep{Com19} and NGC~3227 \citep{Alo19}. [b] Molecular gas masses  derived from the continuum emission of the extended components (see details in Sect.~\ref{continuum}). [c] Same as [b] but derived from the CO(3--2) emission. [d] Line-of-sight H$_{\rm2}$ column densities measured towards the AGN. [e] Position angle of the elliptical Gaussian fits to the extended components (see details in Sect.~\ref{continuum}). [f] Position angle  ($PA_{\rm out}$) and opening angle ($\Omega_{\rm out}$) of the ionized outflows (see references in Appendix~\ref{App1} and Sect.~\ref{GATOS}). [g] Projected relative orientation angle between the extended component and the outflow axis, defined as $\Delta\equiv$ min[$\vert PA_{\rm Gauss}-PA_{\rm out} \vert,~180^{\circ}-\vert PA_{\rm Gauss}-PA_{\rm out} \vert]$. [h] Nature of the extended components classified according to the geometry derived from the value of $\Delta$: polar ($\Delta\leq30^{\circ}$), equatorial or torus-like ($\Delta\geq60^{\circ}$), and mixed classification ($30^{\circ}<\Delta<60^{\circ}$). Doubtful cases (highlighted in italics) correspond to galaxies where the estimates derived from the MSR and HSR dataset are not in agreement or where the orientation of the outflow axis is uncertain (NGC~4388, NGC~6814, and NGC~7213).} \label{Tab4} 
\end{table*}

\subsection{Decomposition of continuum emission: point sources and extended components}\label{continuum-fits}

In this section we make a morphology-wise decomposition of the submillimeter (870$\mu$m) continuum images of the GATOS sample galaxies, as a first step to gauge the geometry (extent, orientation) and mass content of dusty disks (`tori') and their connections. This type of decomposition, made possible thanks to the high spatial resolution and high-sensitivity capabilities of ALMA, is instrumental in minimizing the contribution of sources unrelated to dust emission \citep[see also discussion in ][]{Pas19}. With this aim we fitted the continuum images of the MSR and HSR datasets  using a combination of point-like and 2D elliptical Gaussian sources in order to quantify the contribution of each family of components. The precise knowledge of the ALMA beams allows for an accurate estimate of the contribution of the unresolved components. In particular, the uncertainties on the sizes and orientations of the ALMA beams, which are derived from Gaussian fits to the main lobes are typically $\simeq0\farcs001-0\farcs002$ and $\simeq1^{\circ}-3^{\circ}$, respectively. The fit is performed in the plane of the sky to take advantage of the high-fidelity of the images obtained by ALMA. We also applied the same procedure to determine new fits to the data of NGC~1365 \citep{Com19}, and NGC~3227 \citep{Alo19}. In the case of NGC\,1068 we used the torus parameters derived by  \citet{GB16} from the ALMA continuum images at 440~$\mu$m.  We added these three sources  to the GATOS core sample for the ensuing analysis.

We determined the best-fit solution for each galaxy using a three-step scheme designed, first, to minimize any potential contribution of synchrotron or free-free emission to the flux of the extended component while at the same time avoiding an unphysical over-subtraction of the unresolved component. In a first step ({\it step-1}) we estimated the intensity  of the unresolved component ($I_{\rm point}$) using a pair of 1D-Gaussians accounting for the fluxes of the unresolved and the extended components across two orthogonal strips chosen to intersect the emission peak along $PA=0^{\circ}$ and $90^{\circ}$. We fixed the sizes of the point sources for each target by projecting the ALMA beams along the corresponding strips. In a second step ({\it step-2}) we freely fitted both the position and the intensity of the point source in 2D space using the average obtained for $I_{\rm point}$ along both strips in {\it step-1} 
 as the initial value. In a third step we freely fitted a 2D elliptical Gaussian (of intensity $I_{\rm Gauss}$) to the residual obtained after subtraction of the point source estimated in {\it step-2}. We evaluated the goodness of the final fits  by an inspection of the residuals, shown in Figs~\ref{cont-FITI}-to-\ref{cont-FITIII}. The two-component model is a good representation of the emission in all the galaxies except in NGC\,7213, where the residuals are seen to be exceedingly large ($\geq10\sigma$).  
 This scheme is similar to the Point Spread Function (PSF) scaling technique used  to separate the unresolved and extended emissions in  ground-based mid-IR images of Seyfert galaxies, also adopted in the analysis of AH21.  
 
Table~\ref{Tab3} lists the main parameters obtained in the fitting procedure described above. The coordinates derived from the MSR and HSR datasets for the point sources agree within $0\farcs01$. 
We hereafter assume that these positions likely correspond to the AGN loci in our targets.
Furthermore, the centroids derived for the extended components also coincide  within $0\farcs05$ with the assumed AGN coordinates.  This coincidence suggests that the extended components have a strong physical link to the central engines. 
The equivalent full-sizes of the extended components, defined as the sizes of the disks measured at a  $\sim3\sigma$ intensity level of the Gaussians used in the fits, range from $\sim25$~pc to $\sim130$~pc with a median value of $\sim42$~pc (or an equivalent radius of $\sim21$~pc). The latter is within the range of sizes estimated for the dusty molecular tori that have been imaged by ALMA in other Seyfert galaxies \citep{GB16, GB19, Gal16, Ima18, Ima20, Alo18, Alo19, Com19, Imp19, Aud19}. Furthermore, these sizes  are also similar to those derived for the compact gas disks imaged in the 1--0 S(1) line of H$_2$ in a sample of nearby Seyferts \citep[$D\sim60$~pc;][]{Hic09}.

Figure~\ref{histo-sizes} compares the distribution of deconvolved sizes (diameters) for the dusty molecular  tori (or polar components) derived from the combined sample of NUGA and GATOS. Although the  GATOS sources seem  to populate the tail showing the largest sizes, both samples are characterized by an identical median value of $\sim42$~pc.

  \subsection{Spectral index maps}\label{spectral}

We describe in this section the use of spectral index maps  to characterize the nature of submillimeter continuum emission  in our Seyfert galaxies.

\begin{figure*}[bt!]
   \centering
    \includegraphics[width=0.85\textwidth]{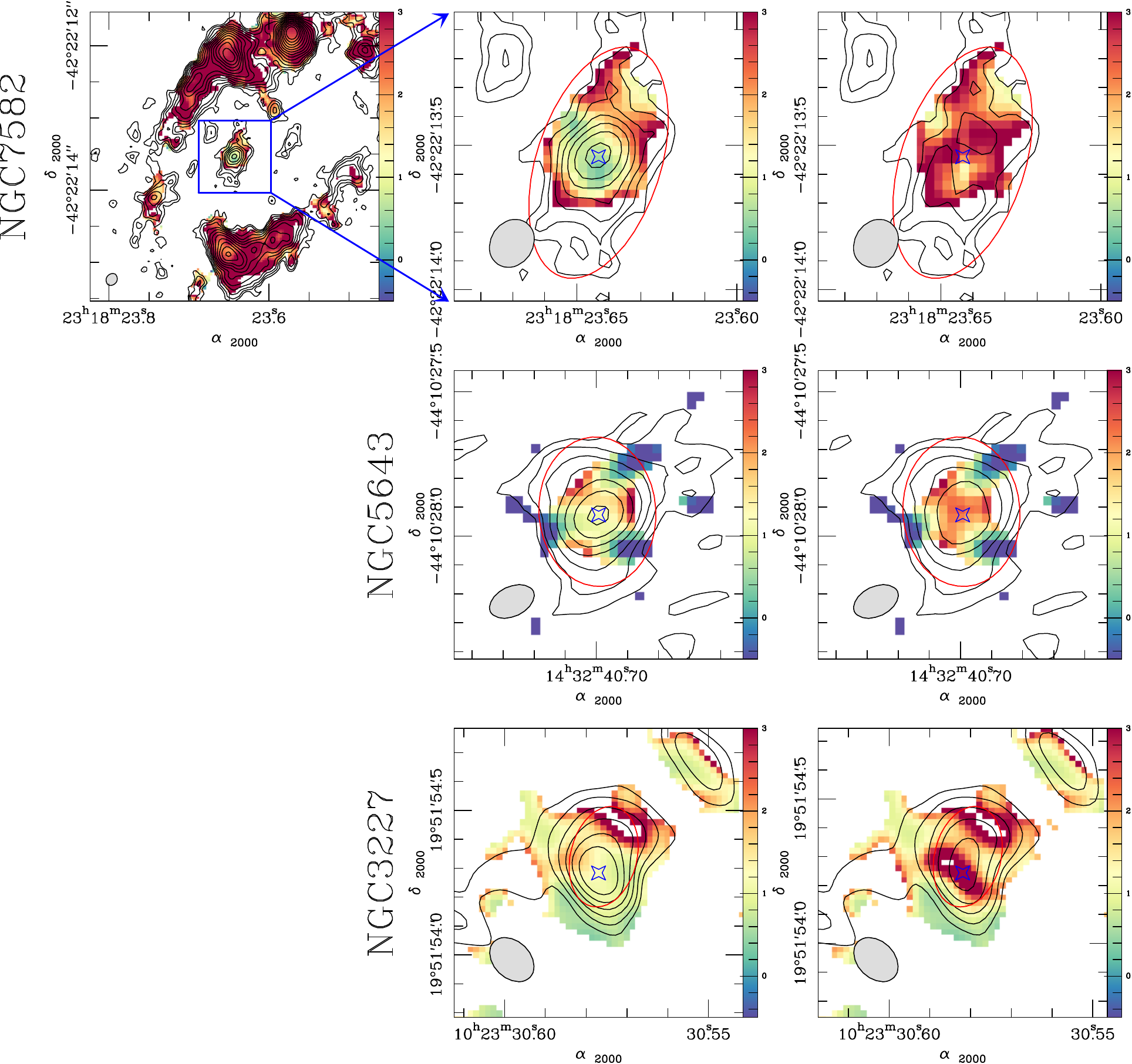}
   \caption{{\it Top row:}~Overlay of the ALMA 351~GHz continuum map of the  central $\Delta\alpha\times\Delta\delta$=
  $4\arcsec\times4\arcsec$ region of \object{NGC~7582} (contours) on the spectral index map of the galaxy ($\alpha$, with $S_{\nu}\propto\nu^{\alpha}$; in color) obtained from the images taken at $\nu$=230~GHz \citep{Alo20}  and 351~GHz (this work)  ({\it left panel}).  We zoom in on the  central $\Delta\alpha\times\Delta\delta
   =1\arcsec\times1\arcsec$ region of \object{NGC~7582} in {\it middle panel} and display the overlay after subtraction of the point-source in {\it right panel}. {\it Middle row:}~Same as {\it top row} but  zooming in on the 
   central $\Delta\alpha\times\Delta\delta
   =1\arcsec\times1\arcsec$ region of \object{NGC~5643}; spectral index map obtained from the images taken at $\nu$=230~GHz \citep{Alo18}  and 351~GHz (this work).  
   {\it Bottom row:}~Same as {\it middle row} but for \object{NGC~3227}; spectral index map obtained from the images taken at $\nu$=230~GHz \citep{Alo19}  and 351~GHz (this work). 
   The common apertures (grey ellipses) adopted to derive the spectral index maps are: $0\farcs17\times0\farcs15$ at $PA=146^{\circ}$ (NGC~7582), $0\farcs16\times0\farcs10$ at $PA=116^{\circ}$ (NGC~5643), and
   $0\farcs18\times0\farcs13$ at $PA=46^{\circ}$ (NGC~3227).   In all panels contour levels span the range [3$\sigma$,  90$\%$ of the peak intensity] in logarithmic steps of 0.13~dex (NGC~7582), 0.19~dex (NGC~5643), and  0.12~dex (NGC~3227). The red ellipses identify the extended components associated with the dust emission of the tori. The star markers identify the AGN positions.}  
   \label{alpha-cont}
    \end{figure*}

Following a methodology similar to the one applied by  \citet{GB19} in their analysis of the continuum emission of NGC~1068, we used the  230~GHz-continuum images of  NGC\,7582, NGC\,5643 and NGC\,3227, published by \citet{Alo18, Alo19, Alo20} in combination with our 351~GHz continuum images,  to derive  the maps of the spectral index $\alpha$ (where $\alpha$ is defined such as $S_{\nu}\propto\nu^{\alpha}$). For these sources we have comparable high-resolution ALMA images at both frequencies.  We convolved all the images employed in this analysis to obtain a common range of  spatial resolutions $0\farcs13$--$0\farcs16$ (11-18~pc). The result is shown in Fig~\ref{alpha-cont}.

NGC\,7582 is the only target of the GATOS core sample where we detected strong continuum emission  from a $r\sim1\farcs5$ (170~pc)--starburst ring located well outside the central $r\sim50$~pc of the galaxy. The value derived for  $\alpha$ throughout the starburst ring, $\geq2-3$, can be accounted for by dust thermal emission. A zoom onto the inner $r\sim0\farcs5$ (55~pc) of the galaxy shows that continuum emission detected in the compact disk feature exhibits significantly lower $\alpha$  values at the AGN locus: $\sim0-1$. However, we detect a gradient showing higher values of $\alpha$ ($\sim$1.5--3) at larger radii, which suggests that dust emission starts to dominate the submillimeter flux budget  farther out in the extended component. We  identify a similar trend in NGC\,5643 and NGC\,3227: low values of the spectral index at the AGN position, $\alpha\sim$0-1, in contrast to significantly higher values of $\alpha$ ($\sim$1.5--3) farther out over a sizeable fraction of the extended component. \citet{Alo19} used a slightly lower resolution version of the 230~GHz-continuum maps of NGC\,3227 and found an equivalent trend for the spectral index of the submillimeter continuum emission.

 The trends identified in the spectral index maps of Fig.~\ref{alpha-cont} confirm that in AGN that are not radio-silent, such as the GATOS sample galaxies, other mechanisms different from thermal dust emission can contribute to a large extent to the $\sim850-870$~$\mu$m continuum flux. In particular, this contribution can be significant mostly close to the central engine on spatial scales ($\sim1$~pc) that remain unresolved by ALMA.  
 Apart from NGC~1068 and the three galaxies of Fig.~\ref{alpha-cont}, 230~GHz continuum images with a spatial resolution $\sim0\farcs1-0\farcs2$ comparable to that of our 351~GHz maps are not available for any of the remaining sources in our sample. Therefore, meaningful spectral index maps cannot be obtained for the rest of the GATOS galaxies.

  Fig.~\ref{alpha-cont} shows also a new version of the spectral index maps of the submillimeter continuum emission in NGC~7582, NGC~5643, and NGC3227 obtained after subtraction of the point sources found in the fitting scheme described in Sect.~\ref{continuum-fits} and applied to the  351~GHz and 230~GHz continuum images. High values of the spectral index $\sim1.5-3$ are now found more systematically throughout the extended components in our sources, an indication that the continuum emission comes primarily from dust in these regions. In particular, the distribution of $\alpha$ values  in the `corrected'  maps does not show any marked depression close to the AGN. On the other hand, the spectral indices derived from the subtracted point sources are  either flat ($\sim0.1$ [NGC~7582];  $\sim0.3$ [NGC~5643]) or steep  ($\sim-1.4$ [NGC~3227]), a result that seems to confirm that the unresolved components are likely associated with free-free or a combination of dust and synchrotron emission.

\subsection{Masses of the extended components}\label{tori-masses}

Continuum emission in the extended components is associated with CO(3--2) emission in the majority of our sources, and in some galaxies with dense molecular gas traced by HCO$^+$(4--3) emission, as shall be discussed in detail in Sects.~\ref{CND} and \ref{faces}.  In this section we further test our hypothesis that the bulk of the continuum emission of the extended components can be attributed to dust by comparing the H$_{\rm 2}$ mass derived from the 351~GHz continuum and the CO(3--2) line emission, as described below.

We assumed a dust temperature $T_{\rm dust} \sim T_{\rm gas}=100$~K and a dust emissivity $\kappa_{{\rm 351~GHz}}=0.0865$~m$^2$~kg$^{-1}$ \citep{Kla01} to derive dust masses using  the values of $I_{\rm Gauss}$ listed in Table~\ref{Tab3} and a modified black-body function. In the absence of estimates for each galaxy, we were constrained to adopt a common $T_{\rm dust}$ for all sources similar to the one used by \citet{GB14} and \citet{Vit14} to fit the CO line 
excitation and the continuum SED of the dusty molecular torus of NGC\,1068. Furthermore, we assumed a canonical (neutral) gas--to--dust ratio  $\sim100$, which is typical of the solar metallicity environment of the central regions of spiral galaxies \citep[e.g;][]{Dev19}. The molecular gas mass derived for the extended components are listed as $M_{\rm gas}^{\rm dust}$ in Tab.~\ref{Tab4}.

Moreover, we obtained an independent estimate of the molecular gas mass associated with the extended components by integrating the CO(3--2) line emission inside the areas defined by the full-sizes of the best-fit Gaussian disks derived in Sect.~\ref{continuum-fits}. In our estimate we adopted a 3--2/1--0 brightness temperature ratio $\sim2.9$, which is similar to the overall ratio measured by \citet{GB14} and \citet{Vit14} in the NGC\,1068 torus. We purposely used NGC\,1068 as template for the sake of consistency with the assumptions made in the dust-based estimate of  $M_{\rm gas}$. Second, we assumed a galactic CO--to--H$_2$ conversion factor  ($X_{\rm CO}=2\times10^{20}$mol~cm$^{-2}$~(K~km~s$^{-1}$)$^{-1}$). The CO-based molecular gas masses derived from Equation~(3) of \citet{Bol13}, after including the mass of Helium, are listed as  $M_{\rm gas}^{\rm CO}=1.36\times M({\rm H_2})$ in Tab.~\ref{Tab4}. 
The molecular gas masses derived for the extended components in GATOS sources encompass a wide range $\simeq5\times10^{4}-1\times10^{7}$M$_{\sun}$, with a median value of $\sim6\times10^{5}$M$_{\sun}$.  

\begin{figure}
   \centering
   \includegraphics[width=0.85\textwidth]{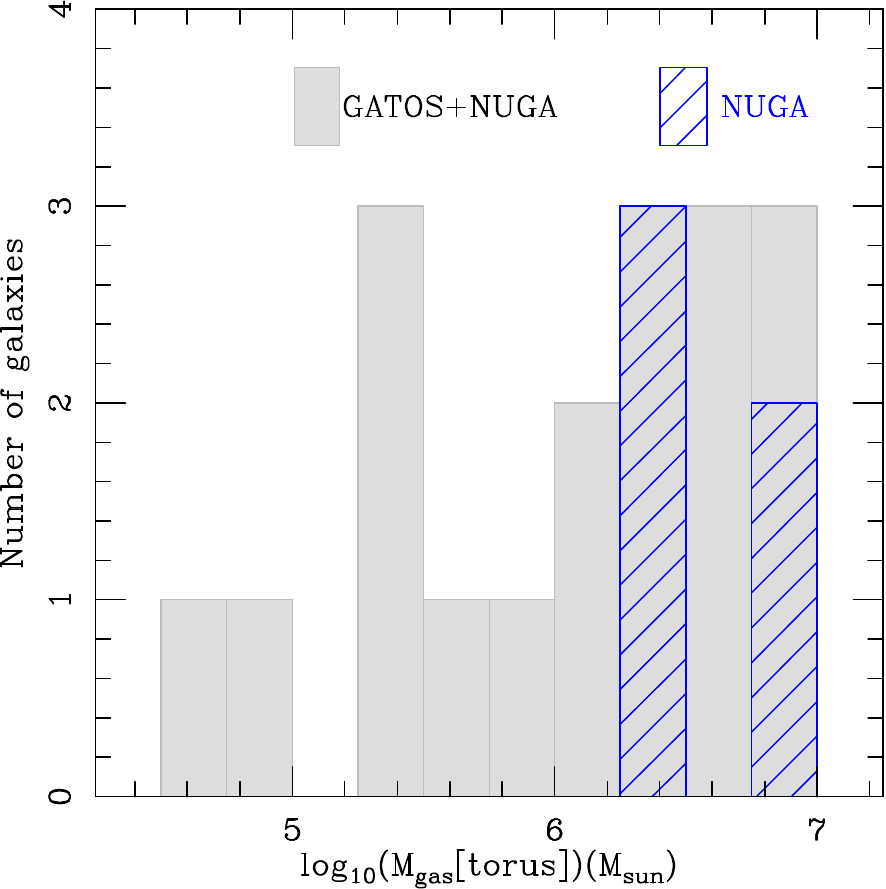}  
   \includegraphics[width=7.8cm]{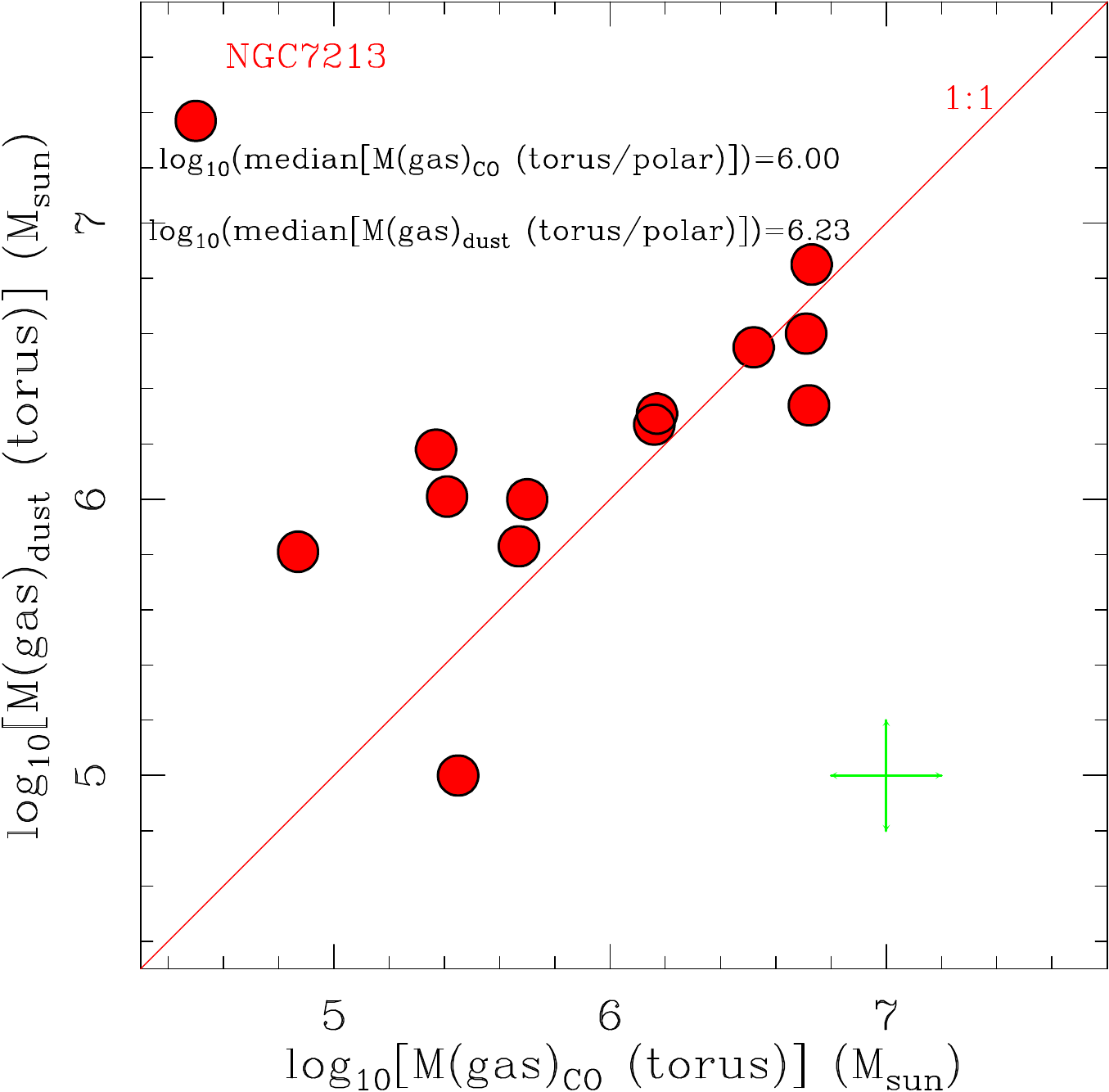}  
    \caption{{\it Upper panel:}~Histograms showing the distribution of the masses of the dusty molecular tori (or polar components) derived from the combined sample of NUGA and  GATOS (gray-filled histograms). We adopt for GATOS the average masses estimated from the MSR and HSR datasets. The blue-filled histogram identifies the distribution of masses of the tori derived for NUGA galaxies \citep{Com19}. {\it Lower panel:}~Comparison of the molecular gas mass  derived from the CO(3--2) integrated emission and from the dust emission inside the torus (or polar) features detected by ALMA in the galaxies listed in Table~\ref{Tab3}; for this comparison the HSR dataset is used. The torus (or polar) regions are  identified by the elliptical Gaussians fit to the extended components of the continuum emission shown in Figs.~\ref{cont-FITI} to ~\ref{cont-FITIII}, as described in Sect.~\ref{continuum-fits}. The median values of the distributions exclude NGC~7213, highlighted as an outlier. A comparison with the MSR dataset yields virtually identical results. Errorbars at the lower right corner account for an uncertainty of $\pm$0.2~dex due to the assumed  conversion factors.}       
 \label{histo-masses-CO-dust}
   \end{figure}

Figure~\ref{histo-masses-CO-dust} (upper panel) shows also the gas masses of the molecular tori in NUGA sources, derived using the same conversion factors. The masses of NUGA tori are among the highest in the combined sample: $\simeq2\times10^{6}-1\times10^{7}$M$_{\sun}$. Based on the similar median values of the molecular tori size distributions of NUGA and GATOS sources derived in Sect.~\ref{continuum-fits}, the comparatively higher gas masses of NUGA tori point to correspondingly higher gas surface densities (see discussion in Sect.~\ref{gas-tori}).

Figure~\ref{histo-masses-CO-dust} (lower panel) compares
the values obtained for $\log M_{\rm gas}^{\rm dust}$  and $\log M_{\rm gas}^{\rm CO}$ in our sample derived from the HSR dataset. Leaving aside NGC~7213 the two distributions are in reasonable agreement: their median values differ only by $\sim0.23$~dex ($\sim0.16$~dex if the MSR dataset is used instead), an indication that 
the bulk of the continuum emission of the extended components stems from dust. In particular, after excluding NGC~7213, $M_{\rm gas}^{\rm dust}$  and $M_{\rm gas}^{\rm CO}$ show Pearson correlation parameters
$\sim+0.74$ and two-sided   $p$ values $\sim6\times10^{-3}$ for both datasets. This result is to some extent expected since the extended components contain most of the continuum emission in eight of the twelve sample galaxies. Although the flux ratios between the extended components and the point sources show a wide range in our sample ($I_{\rm Gauss}$/$I_{\rm point}\sim0.5-55$), their median value, $\sim$6, is high.  
The point source contribution is relevant only in NGC~4388, NGC~5506, NGC~6814, and NGC~7213, where $I_{\rm Gauss}$/$I_{\rm point}$ ratios $\leq 1$. For these sources the underlying assumptions behind the subtraction scheme described in Sect~\ref{continuum-fits} may compromise the H$_2$ mass estimates for the extended components on several counts. NGC\,7213 actually appears as an extreme outlier in Fig.~\ref{histo-masses-CO-dust} (lower panel). In NGC\,7213, $M_{\rm gas}^{\rm dust}\sim500-900\times M_{\rm gas}^{\rm CO}$.  A less dramatic mismatch is found in  NGC\,6814 where $M_{\rm gas}^{\rm dust} \sim8 \times M_{\rm gas}^{\rm CO}$.  The extended component seems to include a significant contribution of either free-free or synchrotron emission of ionized gas in NGC\,7213, and possibly also in NGC\,6814. Alternatively, the assumption of a 3--2/1--0 ratio similar to that of NGC~1068 may underestimate the H$_{\rm 2}$ mass derived from CO(3--2) if diffuse molecular gas
makes a significant contribution in sources like NGC\,7213 and NGC\,6814.
  
In summary, we can conclude that the continuum emission of the extended components can be attributed in all likelihood to dust in all our sources with the exception of NGC\,7213, and, to a lesser extent, NGC\,6814, which represent more doubtful cases. 


\subsection{Orientation of the extended components:  equatorial (torus-like) or polar?}\label{orientation}

The elevation angle of an outflow from the torus midplane ($\theta$) is a key geometrical parameter in the characterization of launching mechanisms. Its closest observational analog is the relative angle between the torus major axis and the outflow axis measured on the sky plane ($\Delta$). However, there is no one-to-one relation between these angles. Projection effects imply that besides $\theta$, $\Delta$ also depends on two additional  parameters: the inclination of the torus ($i_\mathrm{torus}$) and the azimuth of the outflow measured in the torus midplane ($\Phi$). Although we lack sufficient constraints on the three-dimensional geometry of our targets, the distributions of measured $\Delta$ values still encode some important information on the alignment between AGN tori and outflows detected either in ionized or molecular gas. 

\begin{figure}[t!]
   \centering
    \includegraphics[width=0.85\textwidth]{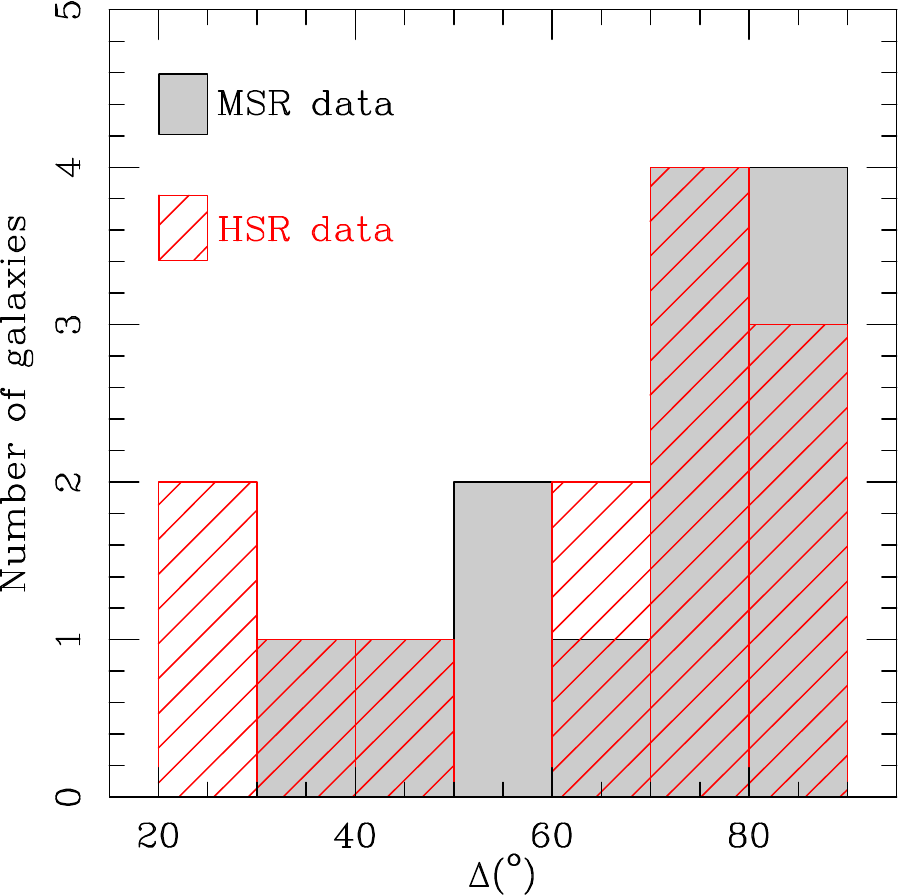}   
    \caption{Histograms showing the distribution of the projected relative orientation angles between the extended components of continuum emission and the ionized outflow axes, defined as $\Delta\equiv$ min[$\vert PA_{\rm Gauss}-PA_{\rm out} \vert,~180^{\circ}-\vert PA_{\rm Gauss}-PA_{\rm out} \vert]$ derived from the MSR and HSR ALMA datasets of GATOS.}       
 \label{histo-PA}
   \end{figure}

We analyzed the orientations of the extended components projected on the plane of the sky,  defined by $PA_{\rm Gauss}$, relative to those of the ionized outflows, defined by $PA_{\rm out}$. Figure~\ref{histo-PA} shows the distribution of $\Delta$ values, defined as $\Delta\equiv$ min[$\vert PA_{\rm Gauss}-PA_{\rm out} \vert,~180^{\circ}-\vert PA_{\rm Gauss}-PA_{\rm out} \vert]$, derived from the MSR and HSR datasets (listed in Tab.~\ref{Tab4}). Within the limits imposed by the sparse statistics, we see that $\Delta$ values do not appear to be uniformly distributed within the the range $0^{\circ}-90^{\circ}$, especially when the HSR datasets are considered.

A group of 9 galaxies show values of  $\Delta\sim60-90^{\circ}$, suggestive of an orientation perpendicular to the outflow axes for the extended component, as expected for a torus geometry for the dust. In contrast, a smaller subset of 4 galaxies display lower values  of  $\Delta\sim20-60^{\circ}$; these lower values would be indicative of polar-like ($\Delta\leq30^{\circ}$) or mixed geometry ($30^{\circ}<\Delta<60^{\circ}$) for the dust in these sources. In particular, in galaxies like NGC\,3227 we cannot exclude the existence of cases with mixed geometry where dust emission is detected in the torus and in the polar components.     


   \begin{figure}
   \centering
    \includegraphics[width=0.99\textwidth]{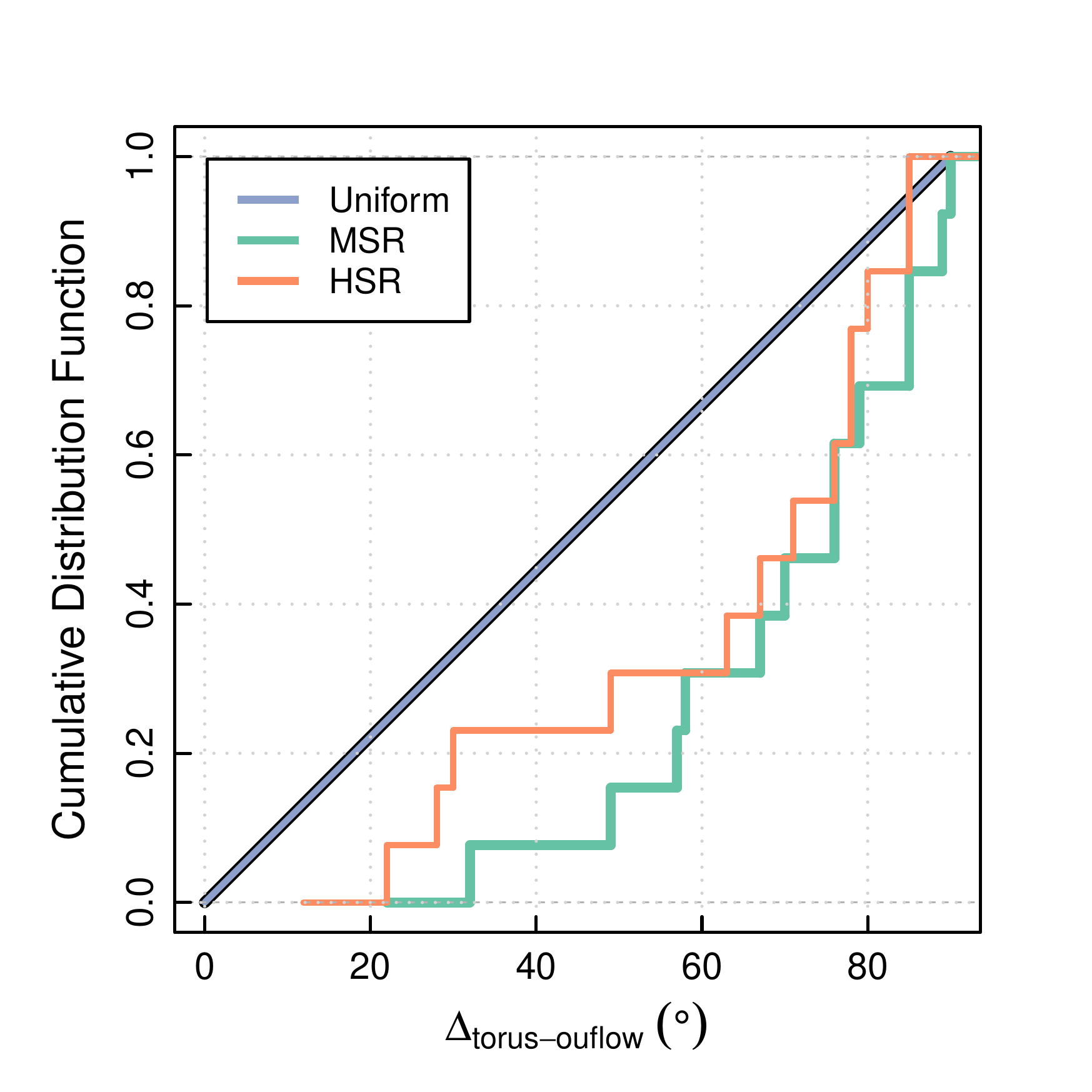}   
    \caption{Comparison of the uniform cumulative distribution function (CDF)  of $\Delta$ values in the 0$^\circ$-90$^\circ$ range (blue)  with those derived from our MSR (green)  and HSR  (reddish) datasets.}       
 \label{KS-tests}
   \end{figure}


  \begin{table*}[tb!]
    \centering
    \resizebox{.8\textwidth}{!}{     
    \begin{tabular}{llcccc}
    \multicolumn{6}{c}{\sc statistic tests on $\Delta$
    measurements}\\
    \noalign{\smallskip}
    \hline
    \noalign{\smallskip}
    &  & \multicolumn{2}{c}{MSR Sample}& \multicolumn{2}{c}{HSR Sample}\\
     Test & Null Hypothesis & Statistic & p-value &  Statistic & p-value \\
    \noalign{\smallskip}
    \hline
        \noalign{\smallskip}
    AD (two-sided) & sample comes from  uniform distribution &
     $\infty$ & $4.6\times10^{-5}$ &  $3.3\times10^{-1}$ & $1.9\times10^{-2}$\\
    KS (two-sided) & sample comes from uniform distribution & 
     $4.8\times10^{-1\phantom{0}}$ & $2.9\times10^{-3}$ & $3.9\times10^{-1}$ & $2.6\times10^{-2}$\\
    KS ('less') & sample CDF not below uniform CDF &
     $4.8\times10^{-1\phantom{0}}$ & $1.4\times10^{-3}$ &  $3.9\times10^{-1}$  & $1.3\times10^{-2}$\\
    KS ('greater') & sample CDF not above uniform CDF &
     $5.6\times10^{-17}$ &  1 & $5.6\times10^{-2}$ & $8.9\times10^{-1}$ \\
    \noalign{\smallskip}
        \hline

         &  \\
         & 
    \end{tabular}}
    \caption{We list the statistic parameters and $p$ values obtained from the  KS  and AD tests performed on the set of $\Delta$ values derived for the galaxies listed in Table~\ref{Tab4}, as obtained  from the MSR and HSR data.}
    \label{Tab5}
\end{table*}

   \begin{figure*}
   \centering
    \includegraphics[width=1\textwidth]{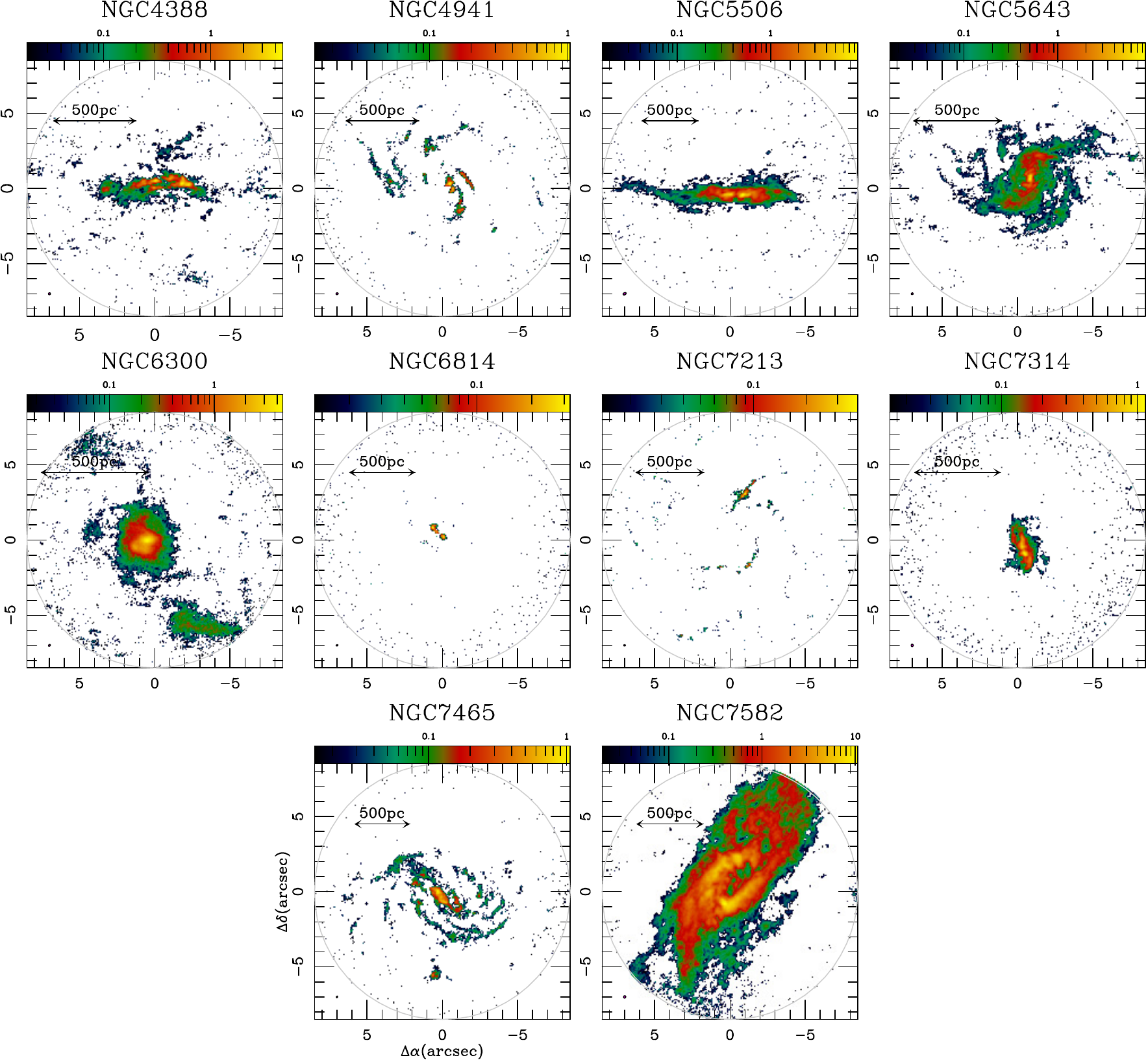}
    \caption{Primary-beam corrected velocity-integrated intensity maps of the 3--2 line of CO ($I_{\rm CO}$) inside the 17$\arcsec$ ALMA field-of-view (FOV; represented by the gray circle) for the galaxies of the core sample of GATOS. Maps were derived using a 3$\sigma$ 
    clipping. Intensities are expressed in Jy~km~s$^{-1}$-units using a logarithmic (color) scale. ($\Delta\alpha$,~$\Delta\delta$)-offsets are relative to the phase tracking centers used during the observations, listed in Table~\ref{Tab1}. Filled-ellipses at the bottom left corners identify the sizes of the ALMA beams.}      
 \label{COmaps-full}
   \end{figure*}

It can be shown that, if outflows were randomly oriented in space with respect to tori, the probability distribution of $\Delta$ would be uniform from 0$^\circ$ to 90$^\circ$. This holds true even for arbitrary values of $i_\mathrm{torus}$, and so it is robust against potential selection biases in the orientation of putative tori relative to the line of sight. Using standard statistical tools, like Kolmogorov-Smirnov (KS) and Anderson-Darling (AD) tests, we quantify below the likeliness that  the $\Delta$ values measured for the extended components are drawn from a uniform distribution. We can therefore assess the null hypothesis of random three-dimensional alignment between outflows and the extended components. 

In  Fig.~\ref{KS-tests} we compare  the uniform cumulative distribution function (CDF) of $\Delta$ values in the 0$^\circ$-90$^\circ$ range  with those of our MSR  and HSR  datasets. A visual inspection suggests that the measured values are relatively skewed towards the upper end. In particular, there is a clear dearth of measurements below $\sim$25$^\circ$. Table~\ref{Tab5} summarizes the results of KS and AD tests applied to the MSR and HSR samples. The low $p$ values of the two two-sided tests, in particular those for the MSR sample, imply that the agreement with a uniform distribution from 0$^{\circ}$ to 90$^\circ$ can be rejected at a very high significance. The one-side KS tests allow us to identify the sign of this disagreement. For the ``less'' KS tests, $p$ values are only $\sim0.1-1\%$, thus rejecting their null hypothesis at a higher significance than before. The implication is that the sample CDFs very likely lie below that of the uniform distribution. In this regard, the tests just substantiate the behaviour apparent in Fig.~\ref{KS-tests}. As expected, for the ``greater'' KS tests, however, $p$ values are close to unity, so the plausibility of their null hypothesis remains undecided. 

In short, we can draw two main conclusions from our statistical tests. In the first place, our data support the idea that the extended components detected in dust continuum emission in our sources have a preferential orientation relative to the AGN outflows. In the second place, the sense of this preference is that outflows and extended  components are more likely to appear (close to) perpendicular than parallel on the plane of the sky relative to the AGN outflows, as expected for a torus-like equatorial geometry. Hereafter we will therefore use the term `dusty molecular tori' to refer to the compact disks detected in dust continuum and molecular line emission around the central engines in our targets. For the reasons mentioned above, this does not necessarily imply that tori and outflows are strictly perpendicular in the three-dimensional space.

   \begin{figure}
   \centering
    \includegraphics[width=.9\textwidth]{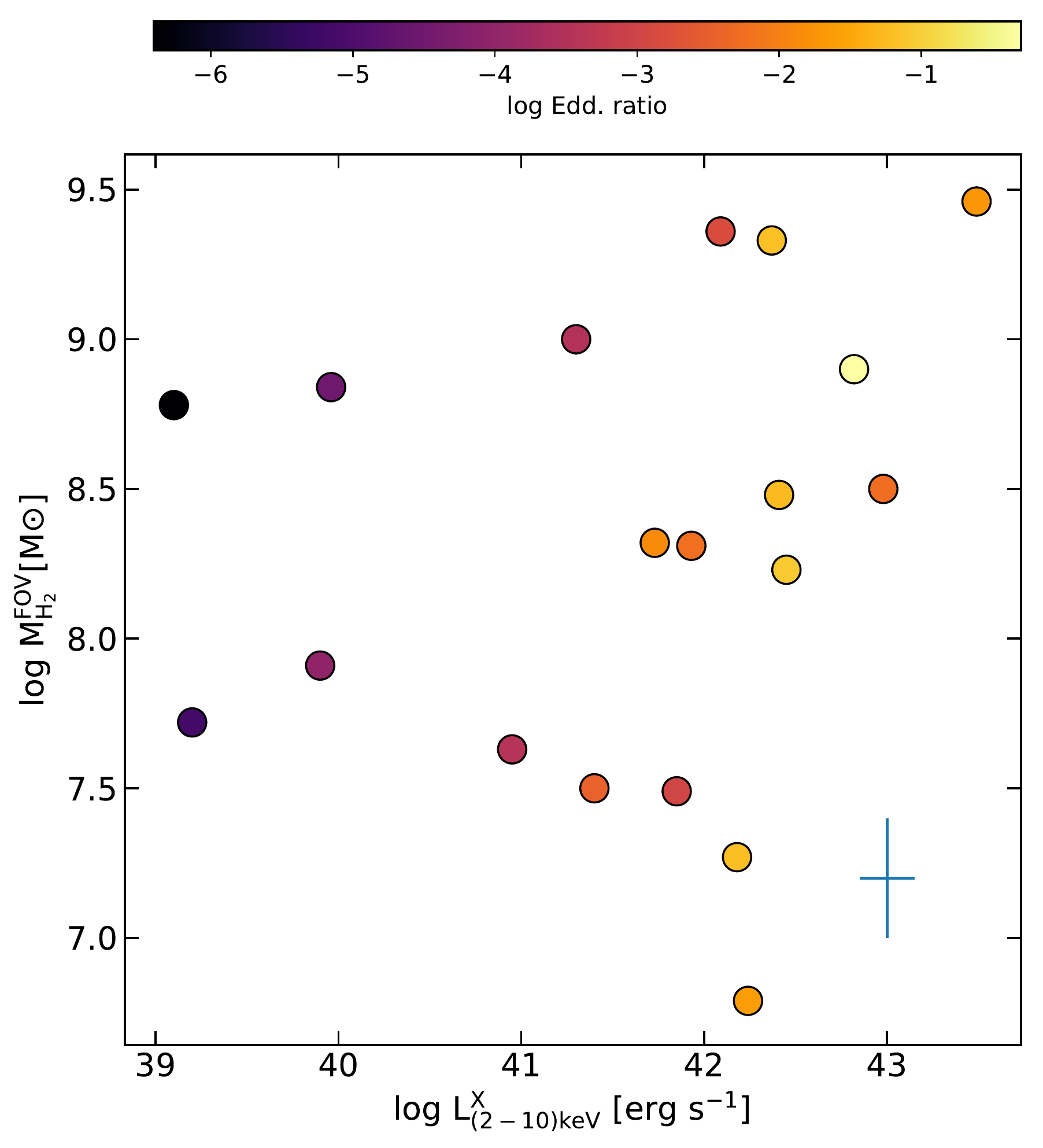}
    \caption{Molecular gas mass detected inside the $17\arcsec$-FOV of the ALMA CO(3--2) images of the GATOS and NUGA  galaxies  as a function of the intrinsic AGN luminosities measured in the 2--10~keV band for the galaxies of our combined sample. The regions mapped by ALMA span the central $r\leq0.4-1.2$~kpc. Symbols are color-coded as a function of the Eddington ratio of the sample galaxies.  Errorbars account for the range of uncertainties on the molecular gas mass estimates, due to the assumed CO-to-H$_2$ and 3--2/1--0 conversion factors ($\sim\pm$0.2~dex), and  on the AGN luminosities ($\sim\pm$0.15~dex).}     
 \label{global-AGN}
   \end{figure}

 \section{Molecular line emission}\label{co}

 \subsection{The $\sim$kiloparsec scale region}\label{CO-global}

Figure~\ref{COmaps-full} shows the  velocity-integrated intensity maps of the 3--2 line of CO ($I_{\rm CO}$) obtained in the GATOS core sample galaxies.   The maps were derived using a 3$\sigma$  clipping on the emission extracted from the MSR dataset and corrected for primary beam attenuation. The ALMA FOV images the inner regions of the galaxy disks up to  $r\leq$0.6--1.2~kpc for the range of distances to our sources (the median value of the sample is $\sim$0.9~kpc).   

 We integrated $I_{\rm CO}$ inside the region covered by the ALMA FOV to derive the corresponding H$_{\rm 2}$ masses for the galaxies shown in Fig.~\ref{COmaps-full}. To this end we assumed a 3--2/1--0 brightness temperature ratio $\sim0.7$. This value lies in the range of the mean ratios measured in the central kpc-region of NGC\,1068  and in other nearby normal and starburst galaxies on comparable spatial scales: $\sim$0.6--1.0 \citep{Dev94, Mau99, Mao10, GB14, Vit14}. As in Sect.~\ref{continuum-fits} we also adopted a Milky-Way CO--to--H$_2$ conversion factor ($X_{\rm CO}=2\times10^{20}$mol~cm$^{-2}$~(K~km~s$^{-1}$)$^{-1}$). The derived molecular gas masses, after including the mass of Helium  span nearly  three orders of magnitude, from $6.2\times10^{6}M_{\sun}$ to $2.9\times10^{9}M_{\sun}$.

 Fig.~\ref{COmaps-full} shows a wide range of morphologies in the distribution of molecular gas  in the  central $r\leq$0.6--1.2~kpc-regions of the core sample of GATOS.   On the global $\sim$kiloparsec scales displayed in Fig.~\ref{COmaps-full} we identify two-arm spiral structures connected to rings or compact disks. These morphologies can be explained in the framework of the expected gas response to the stellar bar potentials, which are present in some galaxies as described in  Appendix~\ref{App1} (NGC\,5643, NGC\,6300, NGC\,7314, NGC\,7465, and NGC\,7582). In other sources molecular gas displays a twin peak morphology at the center, which is connected to a weak two-arm spiral pattern in NGC\,6814 or to a filamentary multiple-arm structure in NGC\,4941. Molecular gas in NGC\,7213 shows an asymmetric two-arm spiral feature with a conspicuous hole at the center.  Although the in-plane distribution of molecular gas in NGC\,4388 and NGC\,5506 is more difficult to discern due to their high inclinations, we nevertheless identify in NGC\,4388  a ring-like morphology in molecular gas. Moreover, the vertical distribution of molecular gas  shows  a network of gas filaments  coming out of the plane in NGC\,4388. In particular, there is a conspicuous  3$\arcsec$ (270~pc)-long gas lane located northwest above the galaxy plane. The radial distribution of molecular gas in NGC\,5506 also shows hints of a ring-like morphology and evidence of an asymmetric warped extension of molecular gas at the eastern edge of the disk at $r>5\arcsec$ (800~pc).   
 
 We refer to Sect.~\ref{CND} for a detailed description of the distribution of molecular gas and its relation to continuum emission on two spatial scales: $r\leq200$~pc  and  $r\leq50$~pc, hereafter identified as the CND and 
 the torus regions, respectively. 
 

 Figure~\ref{global-AGN} explores the dependence of  $M_{\rm gas}$[FOV] measured  in the GATOS and NUGA sample galaxies on their intrinsic AGN luminosities, measured in the 2--10~keV band ($L_{\rm AGN}$(2-10~keV)) or on the Eddington ratios. We derived the  molecular gas masses corresponding to the central $r\leq$0.4-0.8~kpc regions of the NUGA galaxies published by \citet{Com19} using the conversion factors applied to the GATOS Seyferts, described in Sect.~\ref{CO-global}. As shown by Fig.~\ref{global-AGN} there is no significant correlation between  $M_{\rm gas}$[FOV] and $L_{\rm AGN}$(2-10~keV) or with the Eddington ratios.
This result seems to contradict the predictions of the theoretical models and numerical simulations of \citet{Kaw05}, who anticipated that X-ray and CO luminosities should be correlated in the central kpc-scale regions of active galaxies. 

The lack of correlation illustrated by Fig.~\ref{global-AGN} can be explained taking into account the very different spatial scales and timescale involved in the last steps of the AGN fueling process on the one hand ($\sim$1--to--a few 10pc), and in the build-up of kpc-scale reservoirs of molecular gas on the other hand. AGN variability, which reflects  the episodic and chaotic nature of the BH feeding process, would naturally tend to wash out any correlation between the kpc-scale molecular gas content  and $L_{\rm AGN}$ for small galaxy samples like ours. A similar argument can be used to anticipate the lack of correlation of 
 $M_{\rm gas}$[FOV]  with the Eddington ratio.  In agreement with our findings, \citet{Ros18} found no correlation between the kpc-scale molecular gas content and the power of the AGN measured by X-ray luminosities using a sample of 26 nearby AGN. Similarly, \citet{Izu16b} found a weak correlation between the dense molecular gas content probed by the HCN(1--0) luminosities measured on $\geq$~kpc scales and the black hole accretion rates in a sample of 32 AGN.

 After the pioneering work of \citet{Yam94}, the use of larger samples have nevertheless led other groups to find evidence of different degrees of correlation  between the (soft and hard) X-ray and the total CO luminosities \citep{Mon11, Kos21}.  Of particular note, the spatial scales probed in these surveys, $\sim$5--25~kpc, are significantly larger than the 1~kpc-scale regions probed in our sample. Moreover, while distances span a factor of 2 in the sample used in our work, they span a factor 5-30 in the surveys of \citet{Mon11} and \citet{Kos21}. Therefore, we cannot exclude that the correlation found by  \citet{Mon11} and \citet{Kos21} is partly driven by a luminosity distance bias ($\times D^2$), which affects the two axes of their explored parameter space.

  \begin{figure*}
   \centering
    \includegraphics[width=0.92\textwidth]{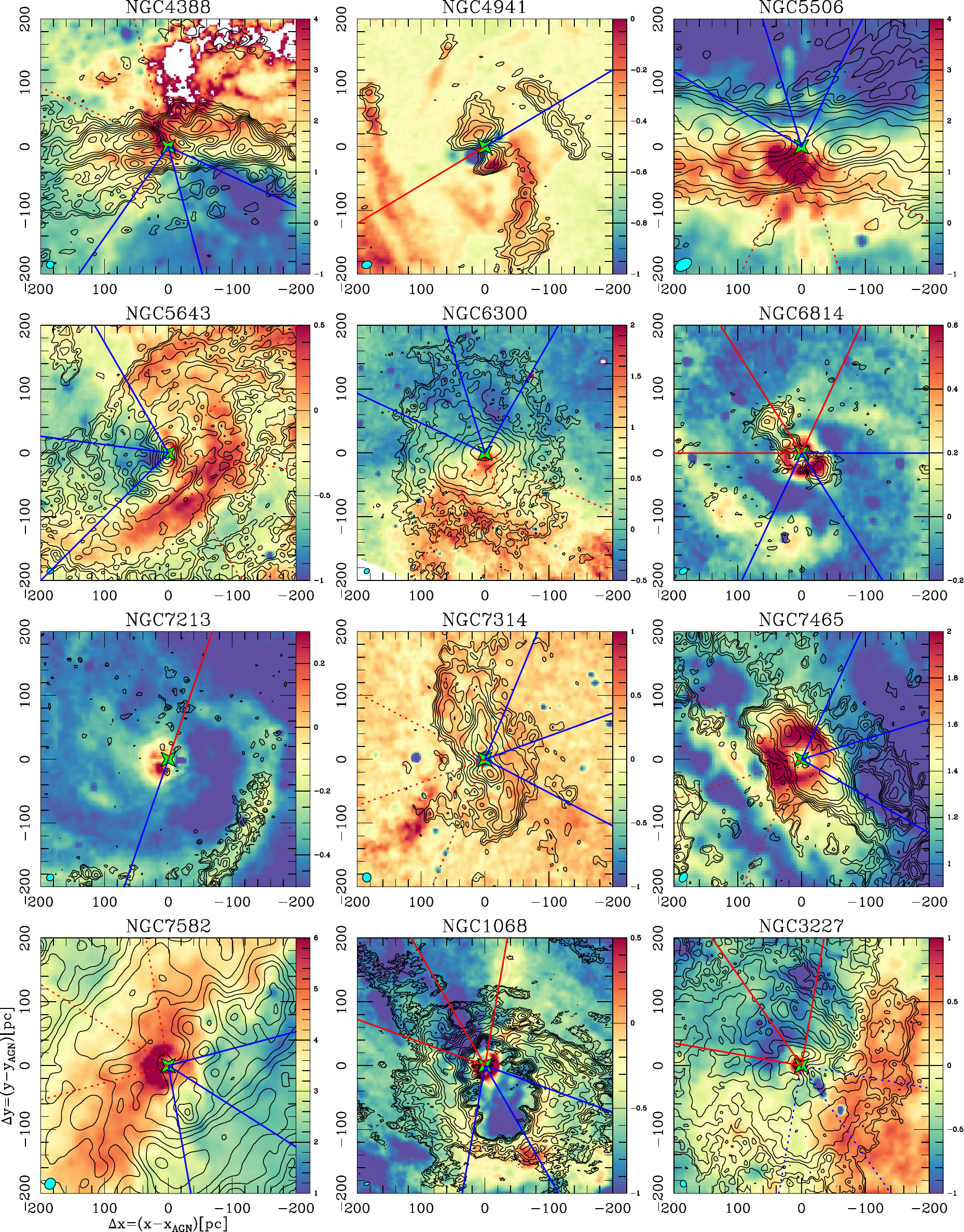}
   \caption{Velocity-integrated CO(3--2) emission images ($I_{\rm CO}$; contours), obtained from the ALMA MSR datasets, are overlaid on the $V-H$ color maps (shown in color scale), derived from the images obtained by {\it HST} with the F606W (V) and F160W (H) filters, in the central  $\Delta x\times\Delta y$=400~pc~$\times$~400~pc regions of the galaxies of the core sample of GATOS. For \object{NGC~7314} we used data of the F450W filter to replace the unavailable F160W. We also include the CO and $V-H$ images of NGC~1068 \citep{GB19} and NGC~3227 \citep{Alo19}. Contour levels have a logarithmic spacing from 3$\sigma_{\rm co}$ to 90$\%$ of the peak CO intensity inside the displayed field-of-view ($I_{\rm CO}^{\rm max}$)  in steps of $\sim$0.18~dex on average. The (cyan) filled ellipses at the bottom left corners in all panels represent the beam sizes of the CO observations. The AGN positions are highlighted by the star markers. The orientation and extent of the ionized winds are illustrated as in Figs.~\ref{cont-FITI}, ~\ref{cont-FITII} and ~\ref{cont-FITIII}. Lines are color-coded to reflect whether the measured velocities of the ionized wind lobes are either redshifted or blueshifted. Dashed lines indicate that the corresponding lobe is (mostly) obscured by the disk of the host.}    
   \label{CO-colors}
    \end{figure*}

  \begin{figure*}
   \centering
    \includegraphics[width=0.92\textwidth]{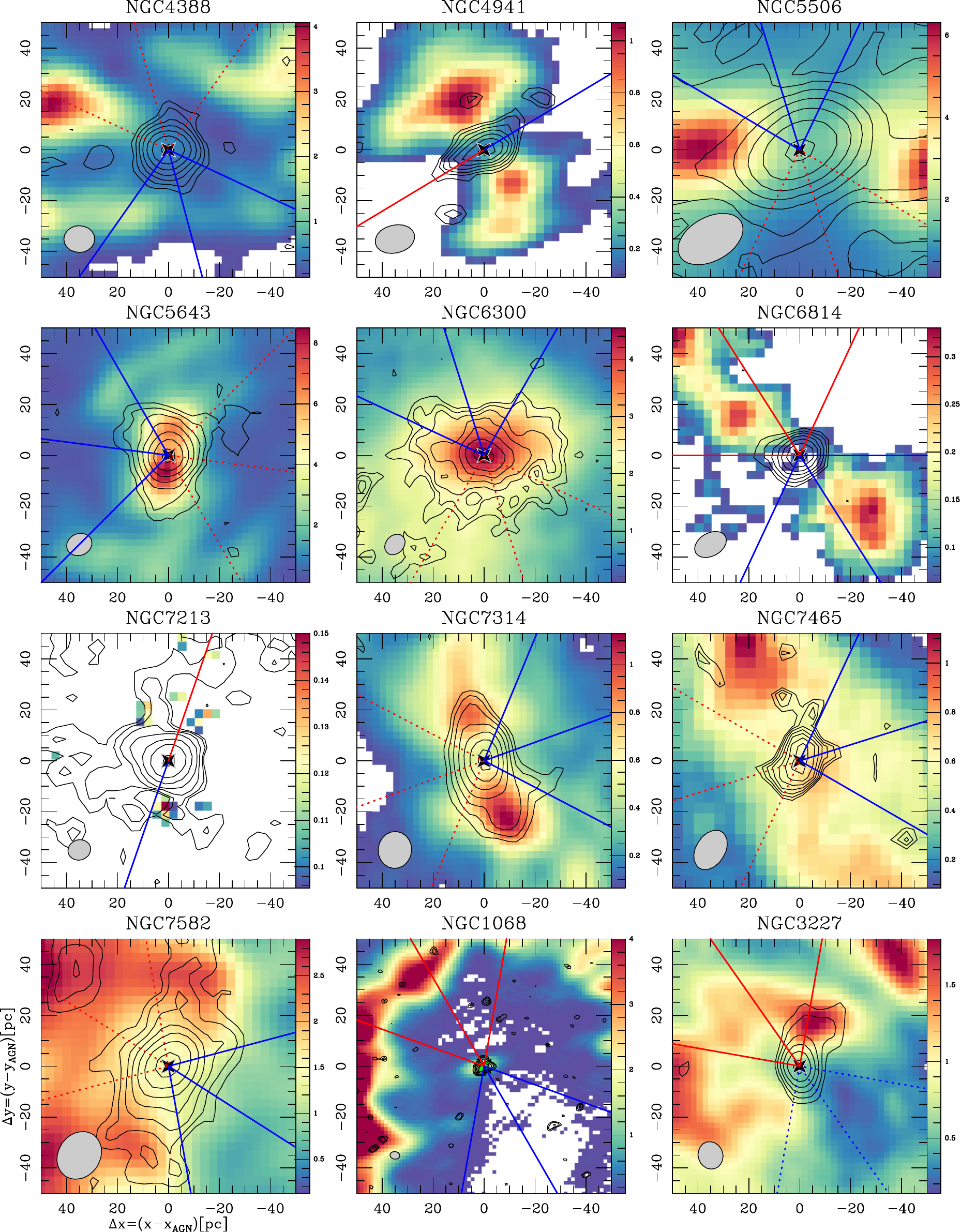}
   \caption{Overlay of the continuum emission, at the (rest)  frequency range 350.6-351.1GHz ($I_{\rm cont}$; contours), on the velocity-integrated CO(3--2) maps ($I_{\rm CO}$; color scale) in the central 
  $\Delta x\times\Delta y$=100~pc~$\times$~100~pc  regions of the core sample of GATOS galaxies. We also include the CO and continuum images of NGC~1068 \citep{GB19} and NGC~3227 \citep{Alo19}. In NGC~1068 we use the 694~GHz continuum map of \citet{GB16} to better identify the dusty torus. The continuum maps  are shown using  (black) contour levels  with a logarithmic spacing from 3$\sigma_{\rm cont}$ to 90$\%$ of the peak intensity in steps of $\sim$0.16 dex on average. The CO(3--2) maps are displayed in (linear) color scale spanning the range [$3\sigma_{\rm CO}$, $I_{\rm CO}^{\rm max}$] in units of Jy~km~s$^{-1}$.   The values of  $\sigma_{\rm cont}$ and  $\sigma_{\rm CO}$ for each galaxy are listed in Table~\ref{Tab2}. The (grey) filled ellipses at the bottom left corners of the panels represent the beam sizes of the observations. The AGN positions are highlighted by the star markers. The orientation, extent and geometry of the ionized winds are illustrated as in Figs.~\ref{cont-FITI}, ~\ref{cont-FITII} and ~\ref{cont-FITIII}. All images have been obtained from the ALMA MSR datasets.}  
   \label{cont-CO}
    \end{figure*}


\subsection{The circumnuclear disk (CND) and tori scales: $r\leq200$~pc and $r\leq50$~pc}\label{CND}

 In this section we compare the CO(3--2) and continuum emissions obtained by ALMA  in the ten galaxies of the GATOS core sample with the morphology of the {\it HST} $V-H$ images on two spatial scales, which are characteristic of the CND  ($r\leq200$~pc) and the dusty molecular tori ($r\leq50$~pc) regions in our sources.

Figure~\ref{CO-colors} compares the $I_{\rm CO}$ emission derived by ALMA  in the central  
$\Delta x\times\Delta y$=400~pc~$\times$~400~pc regions  of the GATOS core sample galaxies with their {\it HST} $V-H$ color images. $V$ and $H$ band images were  obtained with the F606W (V) and F160W (H) filters of {\it HST}\footnote{For \object{NGC~7314} we used data of the F450W filter to replace the F160W filter, which is not available.}. ALMA images were derived from the MSR datasets. We also include the CO and $V-H$ images of NGC~1068 \citep{GB19} and NGC~3227 \citep{Alo19}.
Figure~\ref{cont-CO} zooms in on the inner $\Delta x\times\Delta y$=100~pc~$\times$~100~pc  regions of the core sample of GATOS galaxies to show the overlay of the 351~GHz continuum emission on the $I_{\rm CO}$ maps. We also include the images of NGC~1365 \citep{Com19} and NGC~3227 \citep{Alo19}.  In both figures we used as reference the AGN positions determined from the fit to the continuum point sources derived in Sect.~\ref{continuum-fits}. 

We make below a detailed case-by-case description of this comparison.

\subsubsection{NGC~4388}
Despite its high inclination ($i=79^{\rm o}$, as determined in Appendix~\ref{kinemetry}),  an elongated ringed disk of $\sim100$~pc deprojected radius can be discerned in the inner distribution of molecular gas   traced by CO(3--2) in this Seyfert 1.9, as can be better visualized in the zoomed view of Fig~\ref{cont-CO}. The ring is centered 
around the AGN continuum source and it is likely associated with the gas response at the Inner Lindblad Resonance (ILR) of the stellar bar identified by  \citet{Vei99a}. The ring shows a clumpy morphology, with conspicuous minima located northeast and 
southwest of the AGN along $PA\sim15-20^{\circ}$, i.e. close to the projected orientation of the ionized wind and radio jet \citep{Fal98, Sch03}. Although a distinct torus feature cannot be identified, the region inside the CO ring is nevertheless not entirely emptied of molecular gas:  yet at lower level, CO emission is detected both at and around the AGN up to  $r\sim100$~pc. The CO flux measured at 
the AGN position translates into a molecular gas mass $M_{\rm gas}[r\leq5-6$~pc]~$\sim(5.0-6.8)\times10^4 M_{\sun}$ and a corresponding column density\footnote{Hereafter the ranges in 
$M_{\rm gas}$ and  $N$(H$_{\rm 2})$ span the values derived from the MSR and HSR datasets.} $N$(H$_{\rm 2})\sim(2.1-2.3)\times10^{22}$cm$^{-2}$.

The extended component of the 351~GHz continuum shows a mixed geometry (i.e., intermediate between equatorial and polar) with an undefined 
angle ($PA_{\rm Gauss}$=[-7$^{\circ}$, -43$^{\circ}$]; Table~\ref{Tab4}) relative to the orientation of the outflow ($PA_{\rm out}\sim10^{\circ}-20^{\circ}$). As mentioned in Sect.~\ref{CO-global},  
the vertical distribution of molecular gas  shows  a network of lanes and filaments detected above and below the plane of the galaxy.  In particular, emission from the inner section of a 3$\arcsec$ (270~pc)-long gas lane located 
northwest is detected $\geq150$~pc above the galaxy midplane. The $V-H$ map shows red (blue) colors on the northern (southern) side of the galaxy, which indicates that the northern (southern) side corresponds to the near (far) 
side. This also explains why the northeast  side of the ionization cone detected in the optical is mostly obscured by the host galaxy \citep{Sch03}.  The $V-H$ map shows red colors at the AGN position, in agreement with the Seyfert 1.9 classification of the galaxy. Sub-arcsecond resolution  millimeter NOEMA observations of the CO(2--1) transition were recently presented by  \citet{Dom20}. Although at a lower resolution than the ALMA data in this work,  the CO(2-1) integrated emission also shows a marked ring-like
morphology. 

\subsubsection{NGC~4941}
Molecular gas in this Seyfert~2 shows a twin-peak morphology inside $r<$50~pc, with a deficit of molecular gas at the AGN. CO(3--2) emission is nevertheless detected at the AGN, which implies a molecular gas mass $M_{\rm gas}[r\leq5-7$~pc]~$\sim(1.9-3.3)\times10^4 M_{\sun}$ and a corresponding column density $N$(H$_{\rm 2})\sim(7.8-8.1)\times10^{21}$cm$^{-2}$. The two CO peaks show a conspicuous butterfly shape and appear located symmetrically around the AGN along 
$PA\sim40^{\circ}$. Despite its Seyfert~2 classification, the CO peaks surround a region characterized by blue $V-H$ colors. This region extends $\pm30$~pc on both sides of the AGN in a direction roughly perpendicular to the CO peaks: $PA\sim-60^{\circ}$. The orientation of this blue color feature lies in the range of values derived for the ionized outflow \citep[$PA_{\rm out}\sim-40^{\circ}$ to $-60^{\circ}$;][]{Bar09} and the radio jet \citep[$PA\sim-25^{\circ}$;][]{The00, Sch01}. Moreover, the orientation of the radio jet is similar to the orientation of the extended component of the 351~GHz continuum emission detected by ALMA ($PA_{\rm Gauss}\sim-28^{\circ}$; Table~\ref{Tab4}). This is a borderline case between a mixed and a polar geometry, an indication that synchrotron emission could be contributing to the 351~GHz continuum. Furthermore, the CO  peaks  are connected at larger radii to an asymmetric two-arm molecular spiral structure detected in the $V-H$ color map.

\subsubsection{NGC~5506}
The distribution of molecular gas in this highly-inclined Seyfert~1.9 galaxy ($i=80^{\rm o}$; Appendix~\ref{kinemetry}) shows a ring-like morphology with CO(3--2) maxima located at a radial distance $r\sim50$~pc.  Like in other GATOS sources, CO does not peak at the AGN source, which is nevertheless characterized by spatially-resolved 351~GHz continuum emission with an equatorial (torus-like) geometry. There is however CO emission detected at 
the AGN, which implies a molecular gas mass $M_{\rm gas}[r\leq9-11$~pc]~$\sim(3.2-3.7)\times10^5 M_{\sun}$ and a column density $N$(H$_{\rm 2})\sim(4.0-4.6)\times10^{22}$cm$^{-2}$. Furthermore, it is one of three GATOS sources where we detected significant HCO$^{+}$(4--3) emission stemming from very dense molecular gas in the torus (see Sect.~\ref{faces}). The $V-H$ map shows red (blue) colors on the southern (northern) side of the galaxy, which allows us to identify the northern (southern) side as the far (near)  side. This also accounts for the obscuration of the southern counterpart of the ionization cone detected in the optical northeast along $PA_{\rm out}\sim15-20^{\circ}$ \citep{Fis13}.  The $V-H$ map shows reddish colors at the AGN position, in agreement with the Seyfert 2 classification of the galaxy. The extended component of the ALMA 351~GHz continuum shows an equatorial geometry, i.e., the dusty disk is oriented at a large angle relative to the outflow axis ($PA_{\rm Gauss}\sim87-88^{\circ}$; Table~\ref{Tab4}).

\subsubsection{NGC~5643}
At CND scales ($r\sim$50-200~pc), molecular gas traced by CO(3--2) in this Seyfert 2 nucleus is concentrated in a two-arm nuclear spiral structure. 
This morphology reflects the gas response to the 5.6~kpc-size  stellar bar detected in the 
near-IR  along $PA\sim85^{\circ}$ \citep{Mul97}. Closer to the nucleus, the nuclear spiral is connected to a distinct bright CO disk at $r\leq20$~pc oriented along $PA\sim4-5^{\circ}$, which
is likely responsible for the red $V-H$ color towards the nucleus.
The compact CO disk  shows an excellent correspondence in size and orientation with the dusty torus detected by ALMA at 351~GHz ($PA_{\rm Gauss}\sim4-5^{\circ}$; Table~\ref{Tab4}).  Moreover,  there is a counterpart of the molecular torus detected in 
very dense gas as probed by the HCO$^{+}$(4--3) line (see Sect.~\ref{faces}).  The CO emission flux measured at 
the AGN translates into a molecular gas mass $M_{\rm gas}[r\leq4-5$~pc]~$\sim(5.5-8.7)\times10^5 M_{\sun}$ and a column density $N$(H$_{\rm 2})\sim(4.0-4.4)\times10^{23}$cm$^{-2}$. 
A comparison with the close to east-west orientation of the ionized outflow ($PA_{\rm out}\sim80^{\circ}-85^{\circ}$) and the radio jet \citep{Mor85, Sim97, Lei06, Cre15, Min19, Gar21} indicates that the dusty molecular torus geometry is roughly equatorial.
Although the outflow is nearly edge-on, the $V-H$ color map indicates that the western redshifted lobe is obscured by the galaxy disk. 
The ALMA sub-arcsecond CO(2--1) observations presented by \citet{Alo18}  showed a similar morphology to that seen in the CO(3--2) transition, on both the nuclear and circumnuclear scales. 

 \subsubsection{NGC~6300}
 Molecular gas in this  Seyfert~2 at moderate inclination ($i=57^{\rm o}$; Appendix~\ref{kinemetry}) shows a structured $r\leq150$~pc inner CO(3--2) disk. This disk seems morphologically 
 detached from a discontinuous two-arm spiral  structure that can be identified in the map of Fig.~\ref{COmaps-full}  at larger radii ($r\geq300$~pc). On these larger scales the CO distribution  can be explained in the framework of the  
 expected response of the gas inside the corotation of the $\sim4$kpc-size stellar bar detected by \citet{Gas19} along $PA\sim46^{\circ}$. The morphology of the inner region shows 
a compact bright $r\leq30$~pc CO disk, which is responsible for the obscuration of the Seyfert 2 nucleus. The disk is embedded in a smoother gas distribution out to $r\leq150$~pc. Like in NGC~5643, this compact CO disk shows an excellent correspondence in size and orientation with the 351~GHz continuum disk imaged by ALMA.  Furthermore, together with NGC~5506 and NGC~5643, it is one of three GATOS sources where a  similar HCO$^{+}$(4--3) torus 
 is detected  (see Sect.~\ref{faces}).  The strong CO emission detected at 
the AGN implies a molecular gas mass $M_{\rm gas}[r\leq3-4$~pc]~$\sim(2.5-3.6)\times10^5 M_{\sun}$ and a column density $N$(H$_{\rm 2})\sim(2.3)\times10^{23}$cm$^{-2}$ 
 The orientation of the dusty molecular torus ($PA_{\rm Gauss}\sim85^{\circ}$; Table~\ref{Tab4}) is mostly equatorial relative  to the ionized outflow axis:   \citet{Sch16} observed a biconical outflow extending out to $\sim 1.6\,$kpc from  the AGN along $PA_{\rm out}\sim15^{\circ}-20^{\circ}$. Although the outflow is mostly edge-on, the $V-H$ color map shows that the southern redshifted lobe is obscured by the ISM of the host. 
 
 \subsubsection{NGC~6814}
Molecular gas in this  Seyfert 1.5 galaxy  shows weak and clumpy CO(3--2) emission  in the radial range 200~pc~$> r >$~70~pc, i.e, well inside the 2.8~kpc-size NIR stellar bar located along  $PA=21-25^{\circ}$ \citep{Mul97, Mar99}. The CO clumps  tend to concentrate along the  two-arm spiral structure identified in the 
$V-H$ map, where it stands out by its redder colors. At smaller radii, $r<70$~pc, the spiral is connected to the
northeast to an asymmetric molecular bar-like structure, which is oriented along $PA\sim45^{\rm \circ}$. The bar-like structure shows a CO deficiency 
 towards the AGN. Emission of CO is barely detected at the AGN locus at a 3$\sigma$ level implying upper limits on the molecular gas mass and the H$_{\rm 2}$ column densities of  $M_{\rm gas}[r\leq5-6$~pc]~$\leq(1.9-2.3)\times10^4 M_{\sun}$ and $N$(H$_{\rm 2})\leq(0.6-1.2)\times10^{22}$cm$^{-2}$, respectively.  The region of the CO bar-like structure is characterized by its overall redder $V-H$ colors, as similarly reported by \citet{Marq03}. In particular, a conspicuous dusty ring of $\sim$20~pc-radius located  around the AGN can be identified in the {\it HST} image.  The eastern hemisphere of the dusty ring has a weak yet significant CO counterpart.  
The orientation of the ionized outflow in this source is controversial: values of $PA_{\rm out}$ go from $\sim30^{\circ}-35^{\circ}$ \citep{Mue11} to $150^{\circ}$ \citep{Sch96}.   As in NGC~4388, the extended component of the 351~GHz continuum shows a mixed  geometry if we assume $PA_{\rm out}\sim33^{\circ}$ but with a mostly undefined relative angle ($\Delta$=[57$^{\circ}$, 28$^{\circ}$]; Table~\ref{Tab4}). Of particular note, the orientation of the CO bar-like structure is within the range of values derived for the extended component of the 351~GHz 
continuum. 

 \subsubsection{NGC~7213}
 This nearly face-on ($i=35^{\rm o}$; Appendix~\ref{kinemetry}) Seyfert 1.5 galaxy shows the weakest CO(3--2) emission among the GATOS sample. On the CND scales displayed in Fig.~\ref{CO-colors}, the bulk of molecular gas is concentrated in a one-arm spiral structure that unfolds in the plane of the galaxy  from ($PA$, $r$)~$\sim$~($145^{\circ}$, $200$~pc)  to  $\sim$~($270^{\circ}$, $140$~pc). The CO spiral shows an excellent correspondence with the dusty arm detected in the $V-H$ color image of the galaxy.
Leaving aside a number of isolated clumps located north of the AGN, there is a dearth of molecular gas probed by the 3--2 line of CO at $r<100$~pc. In particular, the CO line is undetected at the position of the AGN. This non-detection implies 3$\sigma$ upper limits on the molecular gas mass and the H$_{\rm 2}$ column densities of  $M_{\rm gas}[r\leq4$~pc]~$\leq(1.8-2.2)\times10^4 M_{\sun}$ and $N$(H$_{\rm 2})\leq(1.1-1.8)\times10^{22}$cm$^{-2}$, respectively.  Farther out ($r>200$~pc) the spiral structure traced by CO becomes filamentary and multi-arm (Fig.~\ref{COmaps-full}). The recent lower resolution ($\sim0.6\arcsec$) ALMA  CO(2--1) maps of the galaxy  published by \citet{Ram19}  \citep[see also][]{Alo20,Sal20} showed a similar morphology to that seen in the CO(3--2) line, in particular regarding the deficit of molecular gas at $r<100$~pc. This is in stark contrast with the picture drawn from the $V-H$ map, which nevertheless shows significant dust  extinction in the nuclear region.

Previous  lower resolution images of the  galaxy showed an unresolved  centimeter and millimeter radio continuum source at the AGN interpreted as  either  synchrotron  or  free-free  emission \citep{Bra98, The00, Bla05, Murphy10, Sal20}. However, the improved resolution of the 351~GHz continuum image shows spatially-resolved emission at nuclear scales stemming from a $\sim30$~pc-size disk oriented along $PA_{\rm Gauss}\sim44^{\circ}-71^{\circ}$ (Table~\ref{Tab4}) and a network of filaments. 
The overall spectral index derived from the 230~GHz and 351~GHz fluxes measured at a common aperture of $\sim0.6\arcsec$ is $\sim$1.5, which suggests that the contribution of dust emission is significant.
Although the [OIII] emission of the galaxy imaged by {\it HST} is described as mostly compact and halo-like by \citet{Sch03}, there is tantalizing evidence of an extension of the [OIII] emission around $PA_{\rm out}=-15^{\circ}$ to $-30^{\circ}$. If this component is taken as evidence of an ionized outflow, the orientation of the 351~GHz extended component is close to equatorial.

  \begin{figure}
   \centering
    \includegraphics[width=1\textwidth]{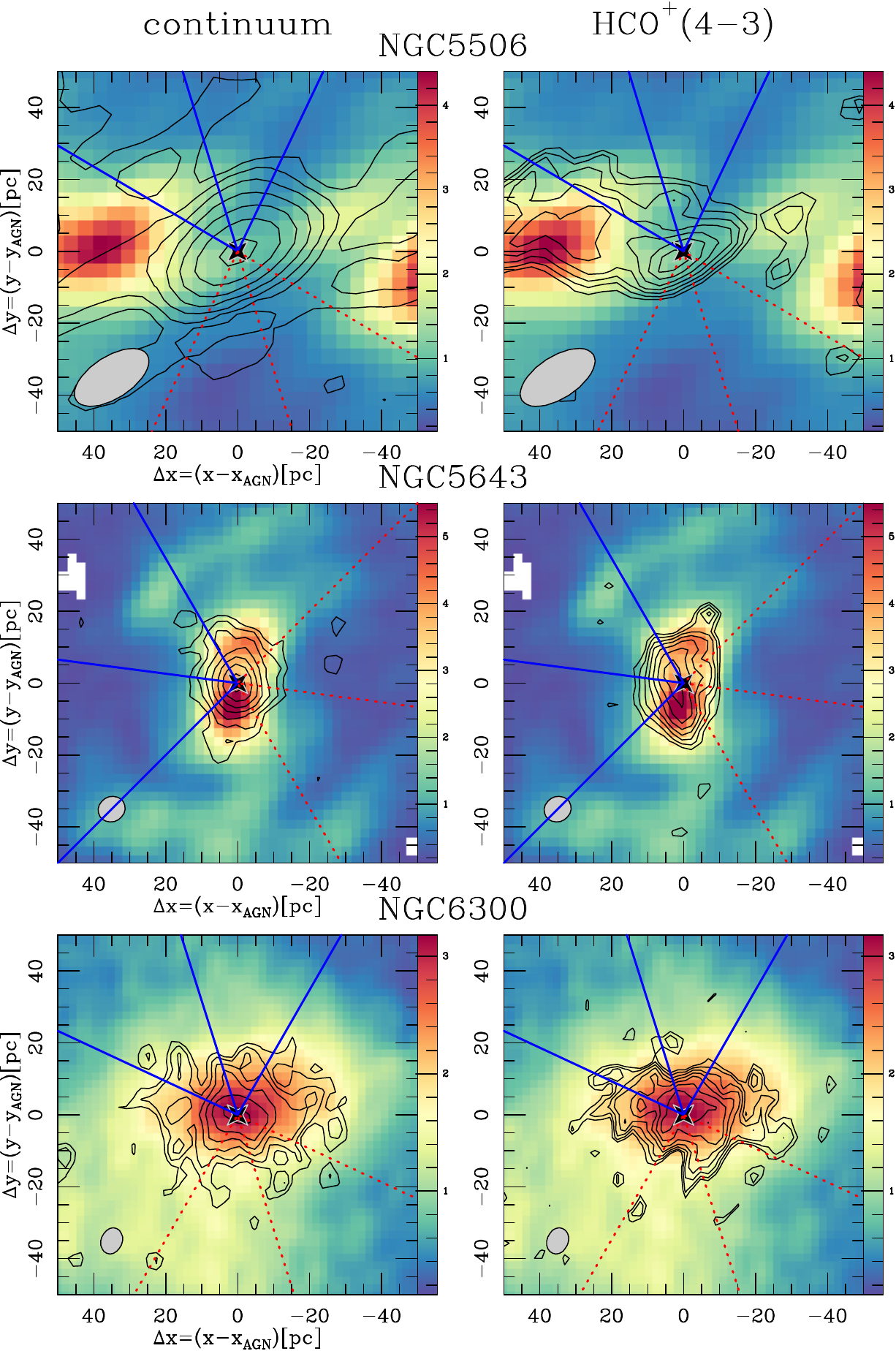}
   \caption{{\it Left panels:}~Overlay of the continuum emission ($I_{\rm cont}$; contours) on the velocity-integrated CO(3--2) maps ($I_{\rm CO}$; linear color scale) of the central $\Delta x\times\Delta y$=100~pc~$\times$~100~pc regions of \object{NGC~5506},  \object{NGC~5643}, and \object{NGC~6300}, derived from their ALMA HSR datasets. Color scale spans the range [$3\sigma_{\rm CO}$, $I_{\rm CO}^{\rm max}$] in units of Jy~km~s$^{-1}$. Contour levels have a logarithmic spacing from 3$\sigma_{\rm co}$ to 90$\%$ of the peak continuum inside the displayed field-of-view ($I_{\rm cont}^{\rm max}$)  in steps of $\sim$0.18~dex on average. {\it Right panels:}~Same as {\it left panels} but showing the overlay of the velocity-integrated HCO$^{+}$(4--3) contours ($I_{\rm HCO^+}$) on the  CO(3--2) maps. Contour levels have a logarithmic spacing from 3$\sigma_{\rm HCO^+}$ to 90$\%$ of the peak HCO$^+$ intensity  ($I_{\rm HCO^+}^{\rm max}$)  in steps of $\sim$0.08~dex on average. Markers and lines are as defined in Fig~\ref{cont-CO}.}  
   \label{cont-hcop-co}
    \end{figure}
 
  \begin{figure*}
   \centering
    \includegraphics[width=0.85\textwidth]{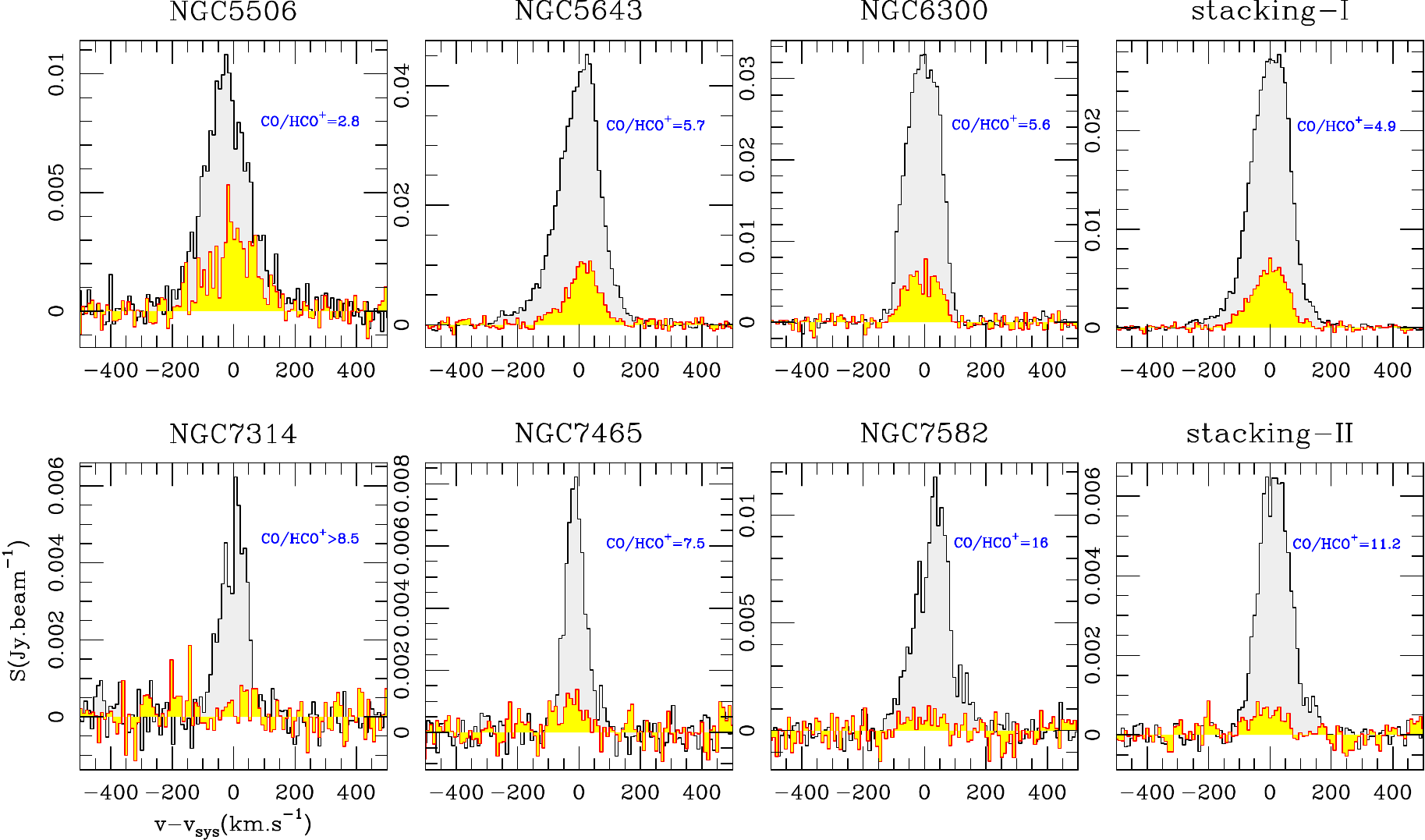}
   \caption{A comparison of the molecular gas emission towards the AGN  probed by the HCO$^+$(4--3) (yellow-filled histograms) and CO(3--2) (gray-filled histograms) lines in two subsets of GATOS galaxies. The {\it upper panels} illustrate this comparison for the galaxies shown in Fig.~\ref{cont-hcop-co} where strong ($>15\sigma$) and/or spatially  resolved emission is detected in HCO$^+$(4-3): NGC~5506, NGC~5643 and NGC~6300.  The spectra stacked for these sources is  shown under the label {\it stacking-I}. The {\it lower panels} compare the AGN spectra obtained in NGC~7314, NGC~7465 and NGC~7582. For these galaxies, either the detection of HCO$^+$ or the lower limit on the CO/HCO$^+$ ratio are statistically significant. The spectra stacked for these sources is  shown under the label {\it stacking-II}.}   
 \label{cont-hcop-AGN}
 \end{figure*}

 \subsubsection{NGC~7314}
 
 This intermediate barred galaxy is classified as a Sy 1.9 nucleus. Although NED gives a high inclination for the galaxy ($i=70^{\circ}$; based on photometry) we
 attribute to $i$ a somewhat lower value based on the kinematics of the CO(3--2) line ($i=55^{\rm o}$, Appendix~\ref{kinemetry}). The bulk of the CO emission detected inside  the ALMA FOV resides in an elongated 
 300~pc-size  source oriented along  $PA\sim20^{\circ}$. In particular, the distribution of CO around the AGN shows a twin-peak morphology with two CO peaks located symmetrically around the AGN at a radial distance $r\sim20$~pc
 along $PA\sim20^{\circ}$. Like in NGC~4941, the two CO peaks show a conspicuous butterfly shape. Although the CO emission does not peak at the very center the line is nevertheless detected at the AGN, 
 implying values for the molecular gas mass and H$_{\rm 2}$ column densities of  $M_{\rm gas}[r\leq5-7$~pc]~$\sim(4.3-6.5)\times10^4 M_{\sun}$ and $N$(H$_{\rm 2})\sim1.4\times10^{22}$cm$^{-2}$, respectively. 
 At larger radii ($200$~pc~$>r>20$~pc), molecular gas shows hints of a nuclear spiral-like structure.
  
  Da Silva et al. (in prep) detected a wide opening angle ($\sim90^{\circ}-100^{\circ}$) fan-shape [OIII]  structure that extends mostly west of the nucleus and interpreted this as the signature of an ionized outflow with a mean axis $PA_{\rm out}\sim100^{\circ}-120^{\circ}$.  The eastern lobe of the
  outflow seems to lie below the galaxy plane and therefore suffers the extinction of the galaxy ISM, as expected if the eastern side of the disk is the near side. 
    The southern (unresolved) component of the double radio source detected by \citet{The00} at 8.4~GHz coincides with the  
 AGN locus defined by the position of the ALMA 351~GHz continuum point source. The extended component of the 351~GHz 
 continuum extends for 50-60~pc along $PA_{\rm Gauss}\sim20^{\circ}-30^{\circ}$, i.e., in agreement with the overall orientation of the 
 CO elongated source. Therefore the dusty molecular disk structure has an equatorial (torus-like) geometry, based on their projected orientation perpendicular to the outflow axis.

  \subsubsection{NGC~7465}
Molecular gas in this  relatively inclined  ($i=55^{\rm o}$; Appendix~\ref{kinemetry}) Seyfert~2 galaxy is oriented along $PA\sim30^{\circ}$ at $r<200$~pc. In particular, the distribution of molecular gas shows a pair of spiral-like
gas lanes connected closer to the center to an elongated disk of $\sim70$~pc-radius. The CO spiral becomes filamentary and multi-arm at larger radii (Fig.~\ref{COmaps-full}). The overall distribution of molecular gas  at CND scales
shows a good correspondence with the network of dust lanes identified in the $V-H$ color image of the galaxy, which are oriented  perpendicular to the photometric major axis of the system ($PA=120^{\circ}$). As in NGC~6814, the $V-H$ image
shows a dusty ring of $\sim$40~pc-radius located  around the AGN. Furthermore, the 
orientation of the elongated CO disk is roughly perpendicular to the [OIII] bi-cone structure identified by \citet{Fer00} ($PA_{\rm out}\sim100^{\circ}-115^{\circ}$). The latter can be interpreted as the signature of an ionized outflow stemming from the Seyfert~2 nucleus of the galaxy.
Although the CO emission does not peak at the AGN, the detection of the line at the nucleus implies values for the molecular gas mass and H$_{\rm 2}$ column densities of  $M_{\rm gas}[r\leq5-7$~pc]~$\sim(1.4-1.9)\times10^5 M_{\sun}$ and $N$(H$_{\rm 2})\sim(4.5-5.4)\times10^{22}$cm$^{-2}$, respectively.  The weak extended component of the 351~GHz continuum shows hints of equatorial (torus-like)
geometry with an angle ($PA_{\rm Gauss}\sim4^{\circ}$; Table~\ref{Tab4}) close to that of the CO disk at small radii ($PA\sim20^{\circ}$). NOEMA CO(2--1) observations of the
nuclear regions show the presence of two emitting clumps separated by $\sim 1\arcsec$ on
both sides of the AGN oriented at approximately  $30^{\circ}$ \citep{Dom20}. These molecular clumps appear further 
resolved in the ALMA CO(3--2) observations. 

 \subsubsection{NGC~7582}

 On the CND scales displayed in Fig.~\ref{CO-colors}, the CO(3--2) emission in this highly inclined ($i=59^{\rm o}$; Appendix~\ref{kinemetry}) barred Seyfert~2 galaxy
 is detected at every single position inside $r<200$~pc. However, a sizeable fraction of the molecular gas appears concentrated in a nuclear ring of $\sim200$~pc (deprojected) radius.
 Molecular gas in this nuclear ring is feeding  an active  star formation episode detected at optical as well as near and mid IR wavelengths. The molecular gas ring is the likely signature of the gas response to the $\sim140\arcsec$ (16~kpc) long 
 stellar bar at its ILR region. The bar, detected in the NIR by \citet{Qui97} and \citet{But09}, shows a prominent boxy-shape morphology. At the larger radii imaged inside the ALMA FOV (Fig.~\ref{COmaps-full}), molecular gas shows a two-arm spiral structure that is connected to the nuclear ring. 
 Closer to the Seyfert~2 nucleus ($r<50$pc), molecular gas probed by CO is concentrated in an asymmetric ringed disk of $\sim30-40$~pc radius located around the AGN and oriented along $\sim160^{\circ}$. The molecular disk shows a similar orientation to the extended component of the 351~GHz continuum emission detected by ALMA ($PA_{\rm Gauss}\sim160^{\circ}-162^{\circ}$; Table~\ref{Tab4}). The CO disk appears to be partly incomplete: its southwest hemisphere is on average a factor of 3 weaker than its northeast counterpart. The $V-H$ map shows a dusty ring feature in excellent correspondence with the morphology of the CO ringed disk. There is nevertheless significant molecular gas inside the ring: CO emission is detected towards the position of the AGN defined by the position of the ALMA 351~GHz continuum point source. This implies values for the molecular gas mass and H$_{\rm 2}$ column densities of  $M_{\rm gas}[r\leq7-9$~pc]~$\sim(1.5-2.9)\times10^5 M_{\sun}$ and $N$(H$_{\rm 2})\sim(3.5-3.6)\times10^{22}$cm$^{-2}$, respectively.   
 The galaxy shows a bright ionization cone southwest of the nucleus oriented along $PA_{\rm out}\sim235^{\circ}-245^{\circ}$ and characterized by blueshifted velocities, which betray an outflow \citep{Mor85,Sto91,Dav16,Ric18}.  The northeast ionization cone appears to be obscured by the host galaxy. The orientation of the $70-90$~pc-size disk detected by ALMA in continuum and CO is therefore equatorial (torus-like).  

\section{The many faces of dusty molecular tori: dust, CO and HCO$^+$ emission}\label{faces}

We have shown in Sect.~\ref{CND} that the CO images obtained by ALMA reveal the presence of significant amounts of molecular gas on nuclear scales in all the galaxies of our sample excluding NGC\,7213. In particular, molecular gas emission in 8 sources is associated with the detection of continuum emission from conspicuous dusty disks. These disks are characterized by an equatorial projected geometry ($\Delta\geq60^{\circ}$). In addition, there are 4 sources where significant CO emission is associated with a continuum disk that shows either a polar ($\Delta\leq30^{\circ}$) or a mixed projected geometry ($30^{\circ}<\Delta<60^{\circ}$).  While the CO emission does not generally peak at the AGN, the 3--2 line  is detected close to the central engines (at radii $r\leq3-10$~pc) in all the galaxies, with the exception of NGC~6814 and NGC~7213. However, the 3--2 line of CO, sensitive to moderately dense molecular gas ($\sim$a few 10$^{4}$cm$^{-3}$) is not optimized to isolate a morphologically distinct molecular torus feature, in particular in those sources where molecular tori are seen to be connected to the gas reservoir of the CND, which can also be a bright CO emitter. As shown in Sect.~\ref{continuum}, the identification of dusty tori probed by their submillimeter continuum emission is comparatively easier, since dust emission is proportional to the dust column density and to the expected high dust temperatures of AGN environments. Alternatively, the use of molecular line tracers that are more specific to the high densities and temperatures, which are likely prevalent in the torus, can facilitate this identification.

The emission of HCO$^+$(4--3) was observed simultaneously with the CO(3--2) line during our observations. The  critical densities for the HCO$^+$(4--3) line,  $n_{\rm crit}\sim2-4\times10^{6}$~cm$^{-3}$ for T$_{\rm k}\sim$~10-100~K \citep{Shi15}, are about two orders of magnitude higher than the corresponding values for CO(3--2) within the same range of temperatures. Although the excitation of HCO$^+$ lines can be significant at effective densities well below the critical densities of the transitions due to the effects of radiative trapping at high opacities,  compared to the 3--2 line of CO,  the 4--3  line of HCO$^+$ traces denser molecular gas in AGN \citep[n(H$_{\rm2}$)~$\geq10^5-10^6$~cm$^{-3}$;][]{Ima13, Ima14, Ion14, GB14,Vit14, Izu15, Izu16, Martin16, GB19}.  We obtained clear detections of the HCO$^+$ line in four of our sources.

 Figure~\ref{cont-hcop-co} compares the CO(3--2), HCO$^+$(4--3) and dust continuum emission from the dusty molecular  tori  of NGC~5506, NGC~5643, and NGC~6300. In these galaxies HCO$^+$(4--3) emission is detected at nuclear scales at high significance ($\simeq15\sigma-35\sigma$ in velocity-integrated intensities). The emission of HCO$^+$ is weaker yet statistically significant in NGC~7465 ($\geq5\sigma$). Furthermore, the emission of HCO$^+$  is spatially resolved in NGC~5643 and NGC~6300. In NGC~5506, the  HCO$^+$ AGN disk is connected to more extended emission stemming from the disk of the host. Figure~\ref{cont-hcop-co} shows that  the HCO$^+$ tori  show an excellent  correspondence in their position, size and orientation with the CO and dusty disks identified in these sources.

  \begin{figure}
   \centering
    \includegraphics[width=0.98\textwidth]{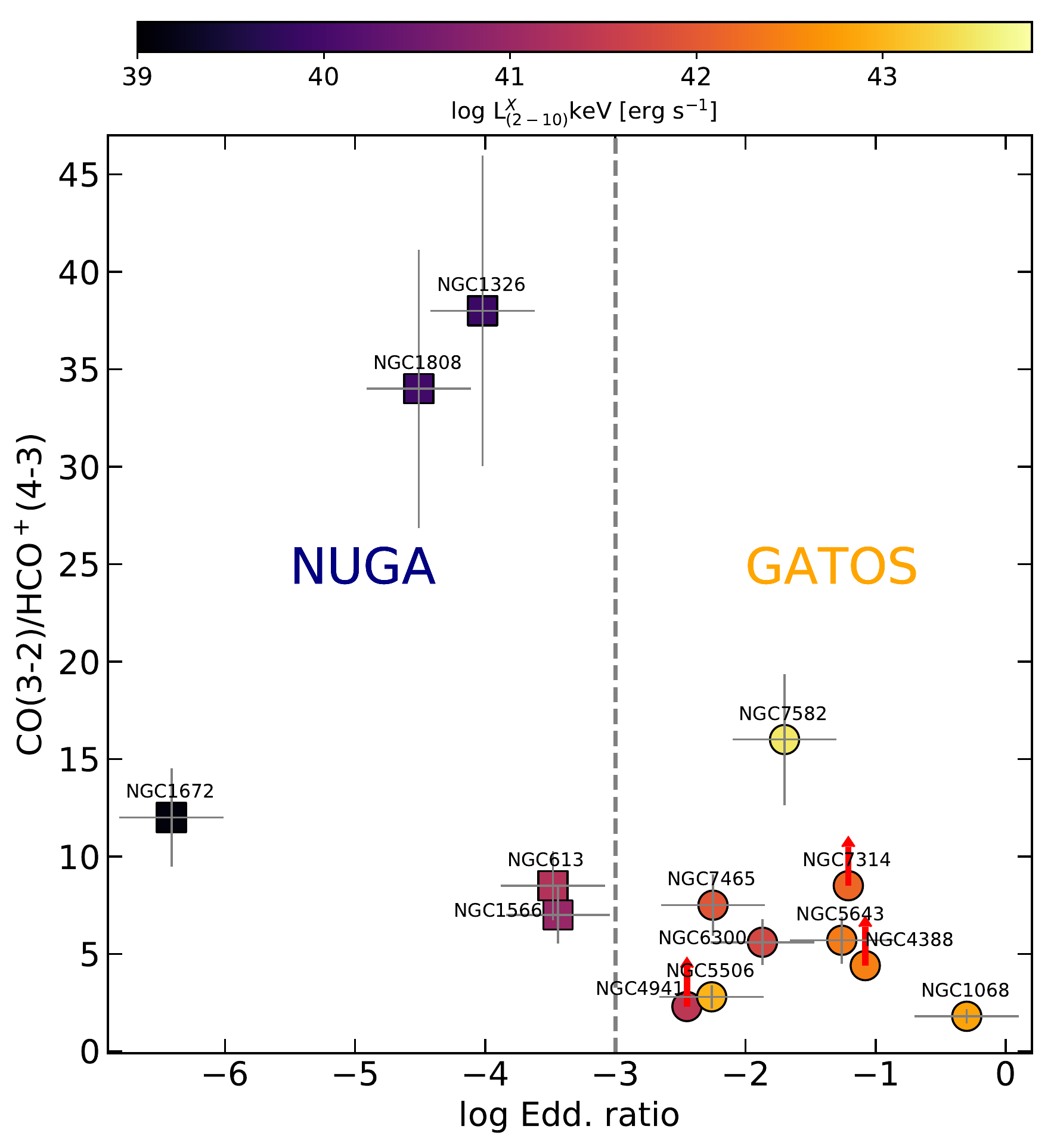}
   \caption{A comparison of the CO(3--2)/HCO$^{+}$(4--3) ratios measured at the AGN locus in the galaxies of the NUGA (square markers) and GATOS (circle markers)  samples as a function of the Eddington ratio. Symbols are color-coded as a function of the  AGN luminosities measured in the 2--10~keV band for  the sample galaxies. Lower limits on the ratios are identified by the (red) arrows. Uncertainties on the ratios ($\sim21\%$) are driven by the absolute flux calibration errors on the lines ($\sim15\%$). Eddington ratios have a $\sim\pm$0.4~dex uncertainty. }   
 \label{AGN-ratios}
 \end{figure}

Figure~\ref{cont-hcop-AGN} compares the HCO$^+$(4--3)  and CO(3--2) spectra and the CO/HCO$^+$ intensity ratios derived at the AGN positions in 6 galaxies of the GATOS core sample. The upper panels illustrate this comparison for the 3 galaxies showing the brightest HCO$^+$ emission in their tori, shown in Fig.~\ref{cont-hcop-co}. We also show the CO and HCO$^+$ spectra stacked for these sources, obtained after subtraction of the $v_{sys}$ values derived in Appendix~\ref{kinemetry}. The fairly low CO/HCO$^+$ velocity-integrated intensity ratios measured towards the AGN in these sources, $\sim2.8-5.7$ ($\sim4.9$ from the stacked spectra), suggest the presence of high-density molecular gas at small radii $\leq3-10$~pc from the central engines. These ratios are comparable to the value measured by \citet{GB19} towards the central offset of the NGC~1068 torus, CO/HCO$^+$~$\sim2.0$, where \citet{Vit14} estimated that the average molecular gas density is n(H$_2$)~$\simeq$~a few~$10^6$~cm$^{-3}$.
The  lower panels of Fig.~\ref{cont-hcop-AGN} compare the AGN spectra obtained in 3 GATOS sources characterized by significantly weaker HCO$^+$ emission and correspondingly higher CO/HCO$^+$ intensity ratios $\geq7.5-16$ ($\sim11.2$ from the stacked spectra): NGC~7314, NGC~7465 and NGC~7582 \footnote{Lower limits on the CO/HCO$^+$ ratios obtained  in NGC~4388 and NGC~4941, $\geq2.3-4.4$, are less statistically significant. In NGC~7213 and NGC~6814, the CO line is either not detected (NGC~7213) or barely detected at 
a  $\simeq3\sigma$ level (NGC~6814).}.

 Figure~\ref{AGN-ratios} shows the CO(3--2)/HCO$^{+}$(4--3) ratios measured towards the AGN  of the  GATOS and NUGA galaxies as a function of the Eddington ratio of the sources. Compared to the GATOS Seyferts, the CO/HCO$^+$ ratios measured in 5 of the  lower luminosity and the lower Eddington ratio  sources of NUGA  show a larger span of values: $\sim8.5-38$. In particular, NGC~1808 and NGC~1326 show very high  CO/HCO$^+$ ratios ($\geq34-38$), an indication that the average gas densities of their molecular tori are the lowest among our sample. Conversely, the high molecular gas densities hinted at by the low  CO/HCO$^+$ ratios measured in GATOS Seyferts, might indicate that diffuse molecular gas has been preferentially driven out by outflows, while higher density clumps have remained in the molecular tori of these sources. 
 
 Overall, the order of magnitude differences in the CO/HCO$^+$ ratios measured within our sample are indicative of a very different density radial stratification inside the molecular tori of these Seyferts.

 \section{Dusty molecular tori and  the processing of X-rays}\label{Xrays}

  \begin{figure*}
   \centering
    \includegraphics[width=0.45\textwidth]{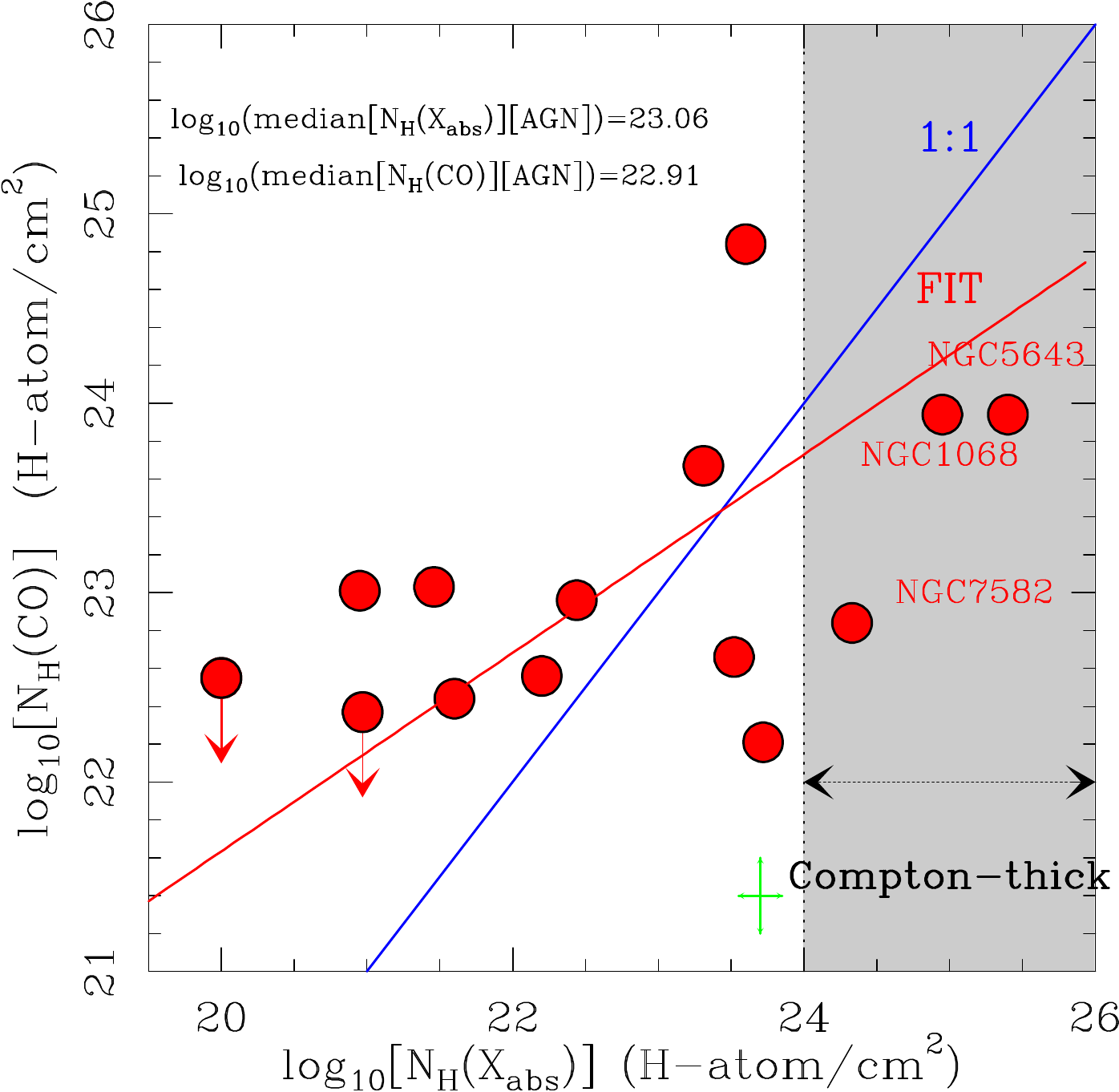}
    \includegraphics[width=0.45\textwidth]{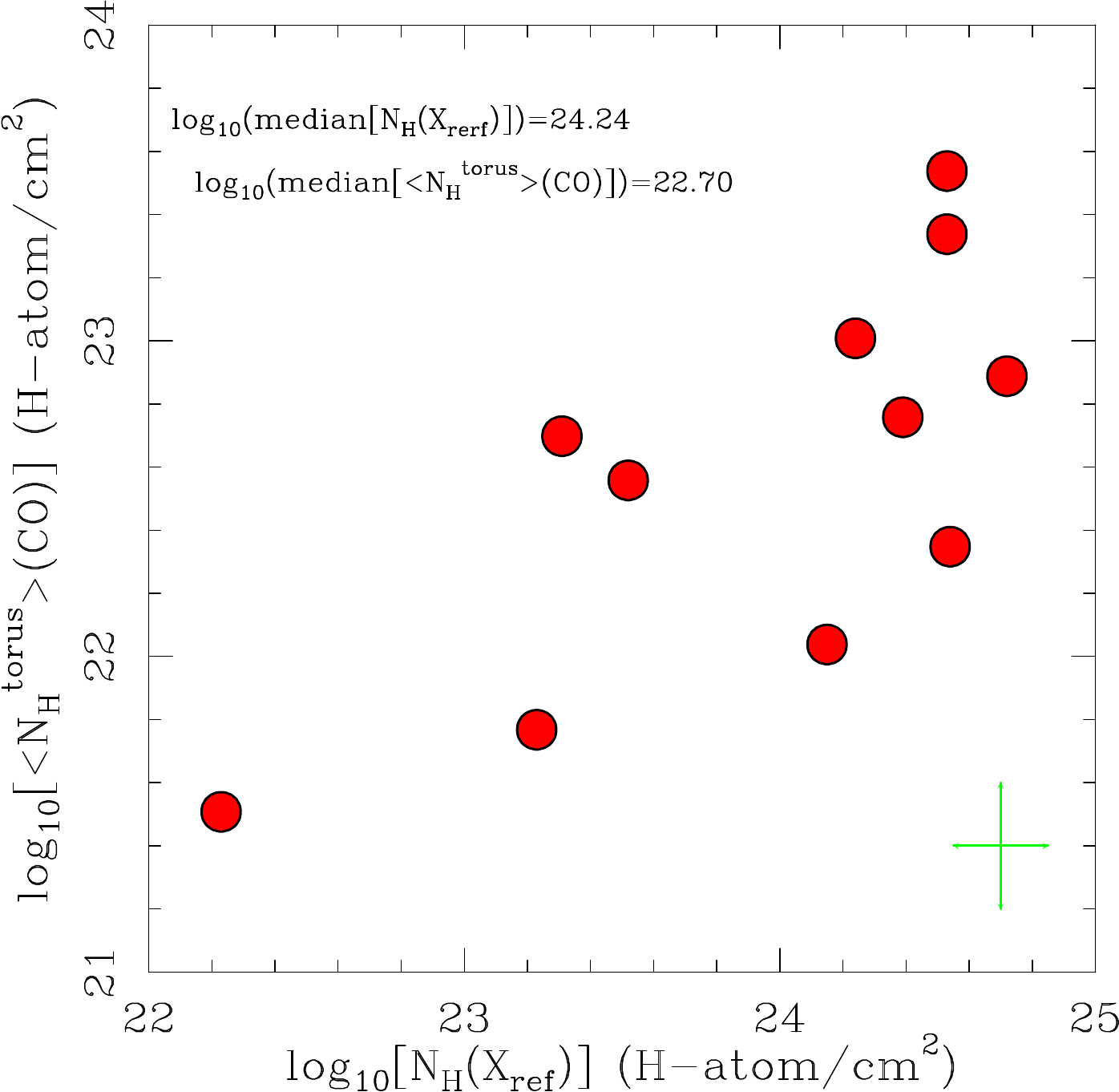}    
   \caption{{\it Left panel:}~A comparison of the column densities of neutral gas measured from CO towards the AGN ($N_{\rm H}$(CO); in units of H atoms per cm$^2$) with the  column densities derived by modelling the absorption of X-rays \citep[$N_{\rm H}$(X$_{\rm abs}$);][]{Ric17a}  in the galaxies listed in Table~\ref{Tab3} and in NGC~613 . In this comparison we use the HSR datasets of ALMA for the core sample of GATOS. The blue color is used to draw the 1:1 line, while the red color identifies the linear bisector fit to te data. {\it Right panel:}~Comparison of the average surface density of molecular gas  measured for the molecular tori  of the galaxies listed in Table~\ref{Tab3} ($<N_{\rm H}^{\rm torus}>$(CO), in units of H atoms per cm$^2$) with the corresponding column density derived in Sect.~\ref{reflection} by modelling the reflection of hard X-rays ($N_{\rm H}$(X$_{\rm ref}$)). Errorbars at the lower right corners in both panels account for the total uncertainties on the molecular gas mass estimates due to  the assumed conversion factors ($\sim\pm$0.2~dex), and for a $\pm0.15$~dex uncertainty on the gas column densities derived from the modelling of X-rays.} 
   \label{Xray-scaling}
    \end{figure*}

\subsection{Dusty molecular tori and X-ray absorption}\label{obscuration}

Many previous studies  tried  to find a connection between  the line-of-sight gas column densities responsible for the absorption of X-rays ($N_{\rm H}$(X$_{\rm abs}$)) and those derived from the obscuration measurements obtained at other wavelengths ($N_{\rm H}$) in different populations of AGN. The early work  of \citet{Mai01} failed to find a connection using the gas-to-dust ratio ($\rm{A_{V}/N_{H}}$), where the dust reddening was estimated using optical/near-infrared emission lines. \citet{Mai01} attributed their results to the different properties exhibited by dust in the circumnuclear regions of AGN. This kind of analysis has been recently revisited by \citet{Oga21}, who explain the results of \citet{Mai01} and their own measurements with an updated unified picture  of AGN structure that comprises a dusty torus, a dusty polar component, and a dust-free gaseous disk component. Other works have also used the optical extinction from spectral fitting of the dust continuum at mid-infrared wavelengths \citep[][Esparza-Arredondo et al. 2021, submitted]{Esp19}. In this section we explore for the first time the connection between the  molecular gas column densities derived from CO   ($N_{\rm H}$(CO)) and $N_{\rm H}$(X$_{\rm abs}$) using our sample of AGN.

Figure~\ref{Xray-scaling} (left panel) illustrates  the comparison between $N_{\rm H}$(CO), in units of H atoms per cm$^2$,  and  $N_{\rm H}$(X$_{\rm abs}$),  taken from Table~5 of \citet{Ric17a} in the 13 galaxies of Table~\ref{Tab3} and the NUGA source NGC~613. \citet{Ric17a} used different models to fit the broadband X-ray spectra for these sources, accounting for the complexity and nature of each particular object. As for the reflection component, in the four AGN included in this analysis that lie in the Compton-thick regime, namely NGC~1068, NGC~4941, NGC~5643, and NGC~7582, \citet{Ric17a} used a physically motivated torus 
model \citep[{\tt BNtorus};][]{Bri11}, different from the screen geometry used by the same authors in the less obscured objects \citep[{\tt pexrav};][]{Mag95} \footnote{See Sect.~\ref{reflection} for an alternative estimate of the X-ray reflected component in our sources using data from the X-ray satellite {\tt NuSTAR}.}. 

The scatter plot of Fig.~\ref{Xray-scaling} (left panel) shows an admittedly large dispersion of values around the 1:1 line. However, the median values of the distributions of log($N_{\rm H}$(CO)) and log($N_{\rm H}$(X$_{\rm abs}$)) differ only by  $\simeq0.15$~dex. Furthermore, there is a statistically significant correlation between the two variables as confirmed by the estimated Pearson's correlation coefficient, $r=+0.57$, and its associated one-sided $p$ value $\sim0.02$, derived from the physically-motivated alternative hypothesis that  log($N_{\rm H}$(CO)) and log($N_{\rm H}$(X$_{\rm abs}$)) show a positive correlation coefficient. Similarly, the Spearman and Kendall tests show that both column densities have positive rank parameters, $\rho_{\rm Sp}=+0.46$ and $\rho_{\rm Ke}=+0.35$, respectively, with one-sided $p$ values $\sim$0.05.

While the estimates of $N_{\rm H}$(CO) are uncertain to the extent that they depend on the assumed CO--to--H$_2$ and 3--2/1--0 conversion factors (see Sect.~\ref{gas-CND-nuclear} for a detailed discussion), the existence of a positive correlation between $N_{\rm H}$(CO) and 
$N_{\rm H}$(X$_{\rm abs}$) and the similar median values of their distributions suggest that the neutral gas line-of-sight column densities sampled by ALMA at scales $\simeq7-10$~pc (the spatial resolution of our observations) make an important contribution to the obscuration of X-rays in the galaxies of Fig.~\ref{Xray-scaling}. 

A closer look at Fig.~\ref{Xray-scaling} reveals that less absorbed galaxies ($N_{\rm H}$(X$_{\rm abs}$)~$<10^{22}$cm$^{-2}$) tend to lie systematically above the 1:1 line. In contrast, galaxies with high X-ray obscuration ($N_{\rm H}$(X$_{\rm abs}$)~$>10^{22}$cm$^{-2}$) are closer to or lie below the 1:1 line. This result is expected if the X-ray absorption in type 1 objects is dominated by a $\leq$pc-scale dust-free gas disk component, while $N_{\rm H}$(CO) traces the comparatively higher molecular gas column density of the torus probed by ALMA on scales $\simeq7-10$~pc (the spatial resolution of our observations). Conversely, in type 2 objects, the agreement between $N_{\rm H}$(CO) and $N_{\rm H}$(X$_{\rm abs}$)
is expected to be better, in agreement with the scenario discussed by \citet{Oga21}. We can foresee that higher spatial resolution ALMA observations would find an even better correlation for type 2 AGN.

\subsection{Dusty molecular tori and  X-ray reflection}\label{reflection}

In this section we adopt a torus model different from the one used by \citet{Ric17a} to fit single epoch  {\tt NuSTAR} X-ray observations obtained in a subset of galaxies of our sample with the aim of obtaining an independent and self-consistent estimate of the X-ray reflected component in 11 galaxies of our sample. 

To accomplish this, we fitted  all the spectra obtained in the galaxies listed in Table~\ref{Tab6} to a combination of an intrinsic continuum (modelled with a cutoff power-law model, {\tt cutoffpl} in {\tt XSPEC}), a distant and neutral reflection component, and two emission lines (fitted with Gaussians, {\tt zgauss$_{\rm i}$, i=1,2} in {\tt XSPEC}) centered at the 6.7 and 6.97 keV Fe emission lines, for which we assumed a 0.1~keV width, smaller than the spectral resolution of {\tt NuSTAR}. The intrinsic continuum is absorbed by a partial covering screen along the line of sight ({\tt zpcfabs} in {\tt XSPEC}). For the reflection component we used the smooth model {\tt borus02} provided by \citet{Bal18}. This can be written in {\tt XSPEC} format as:

\begin{equation}
{\tt zpcfabs} \times {\tt cutoffpl} + {\tt borus02} + {\tt zgauss_{\rm 1}} + {\tt zgauss_{\rm 2}}
\end{equation}

\noindent where the free parameters are the line-of-sight column density associated with the partial covering ($N_{\rm H}^{\rm LOS}$),  the line-of-sight covering factor of the source ($f_{\rm cov}$), the torus column density associated with the {\tt borus02} component  ($N_{\rm H}$(X$_{\rm ref}$)), and the photon index of the cutoff power-law component ($\Gamma$). Note that we also let  vary the viewing angle toward the observer and the half opening angle of the torus for the reflection component. However, these two values are fully unconstrained for most of the sources analyzed. While this model does not consider line-of-sight Compton scattering, which can be important to properly reproduce the X-ray continuum in heavily obscured sources, it is included for the reflection component within the {\tt borus02} model. 
We also stress that, at least in some cases (e.g. NGC\,6814), the addition of neutral partial covering could lead to values of the line-of-sight column density rather different from those reported in the literature. We therefore preferred to use the line-of-sight column densities obtained by \citet{Ric17a}. Furthermore, some objects are known to require the inclusion of  two reflection components to fit their spectra \citep[such as in NGC~1068;][]{Bau15}. The column density of the reflector component  derived here for NGC~1068 nevertheless matches that accounting for most of the FeK${\alpha}$ line in \citet{Bau15}. Finally, some type 1 AGN might require a disk reflection component.

Table~\ref{Tab6} lists the best-fit $\rm{\chi^2}$ statistics together with the resulting parameters. We also explored if the line-of-sight obscuration could be the same as the reflecting material. For that purpose we linked together these two values ($N_{\rm H}^{\rm linked}$) when fitting and perform an f-test to see in which cases the linked scenario was enough to reproduce the data. In seven out of the 11 objects the absorptions must be unlinked. The exceptions are NGC\,4388, NGC\,4941, NGC\,5643, and NGC\,7213. Table~\ref{Tab6} also lists the resulting column densities when the linking of both absorptions was preferred as solution.

 Figure~\ref{Xray-scaling} (right panel) compares the average gas surface density of the dusty molecular tori of the galaxies listed in Table~\ref{Tab6} ($<N_{\rm H}^{\rm torus}>$(CO), in units of H atoms per cm$^2$) with the column density responsible for the reflection of hard X-rays derived above ($N_{\rm H}$(X$_{\rm ref}$)).  Within the limits of our small sample statistics, we see a positive correlation between the two column densities, similar to the one shown by  log($N_{\rm H}$(CO)) and log($N_{\rm H}$(X$_{\rm abs}$)) in Sect.~\ref{obscuration} (Pearson's correlation coefficient $r=+0.64$ and  associated  two-sided and one-sided  $p$ values  
 $\sim0.03$ and $\sim0.02$, respectively). However, in stark contrast with the comparison of column densities discussed in Sect.~\ref{obscuration}, the median values of the distributions of log($<N_{\rm H}^{\rm torus}>$(CO))  and  log$(N_{\rm H}$(X$_{\rm ref}$)) differ now by $\simeq1.54$~dex.  This flagrant discrepancy can be explained if the bulk of the neutral gas producing the reflection of X-rays is preferentially located at the inner walls of the tori, therefore at very small radii  $<1$~pc, as argued by \citet{Liu15}. Resolved studies of the characteristic Fe K X-ray fluorescence line in nearby AGN  \citep[e.g.][]{Gan15, Min15} suggest that the bulk of the X-ray reflection arises at or within the infrared dust sublimation radii on sub-pc scales \citep[see, however,][]{Bau15, Mar17, Fab17, Fab18, Yi21}. These regions would typically have the highest gas densities, as suggested by the density radial stratification revealed by the 2--3~pc spatial resolution ALMA images of the torus of NGC~1068 \citep[e.g.][]{GB19}. 
 
 Despite the discrepancy in their mean values, the linear correlation between log($<N_{\rm H}^{\rm torus}>$(CO)) and log($N_{\rm H}$(X$_{\rm ref}$)) reported above suggests that both quantities are linked to the same region. Whether this scenario applies for dusty molecular tori in general remains to be proved by new ALMA and   {\tt NuSTAR} observations using larger samples of nearby AGN.

  \begin{table}[bth]
\caption{H$_{\rm 2}$ surface densities derived at different spatial scales for the combined sample of GATOS and NUGA galaxies shown in Fig.~\ref{masses-radii}.}
\centering
\resizebox{0.8\textwidth}{!}{ 
\begin{tabular}{lccc} 
\hline
\noalign{\smallskip} 
  Name      & log$_{\rm 10}\Sigma_{\rm H_2}^{\rm torus}$  &  log$_{\rm 10}\Sigma_{\rm H_2}^{\rm 50pc}$  & log$_{\rm 10}\Sigma_{\rm H_2}^{\rm 200pc}$  \\
             	  &  M$_{\sun}$~pc$^{-2}$  &  M$_{\sun}$~pc$^{-2}$ &  M$_{\sun}$~pc$^{-2}$ \\  
	   \hline
	   \hline
 \noalign{\smallskip} 	   
{\it NGC613}	& 4.03  &     3.36   &    	2.54  \\     
NGC1068	 &	  2.60  &     2.72       &		2.97 	 \\    
{\it NGC1326}	&   3.08   &    2.54       &	1.81   \\    
NGC1365	&  2.44   &    2.25       &		1.55   \\   
{\it NGC1566}	&  2.92   &    2.62       &	1.92  \\    
{\it NGC1672}	&   3.28   &    3.19       &		2.36   \\   
{\it NGC1808}	 &  4.15    &   2.71       &		2.23   \\   
NGC3227	  &     2.91   &    2.91     & 		2.77  \\    
NGC4388  &	 2.26   &    2.00     &		2.16  \\   
NGC4941	&	 1.75   &    2.24     &  		1.45  \\    
NGC5506	&	 2.72   &    2.45     &  		2.37  \\    
NGC5643	&       3.42   &    3.20     &  		2.65  \\    
NGC6300	&	3.18   &    3.13     &  		2.45  \\    
NGC6814	&	1.80   &    1.70     &   		0.99  \\    
NGC7213	&      1.61   &    1.21     &  		0.66  \\    
NGC7314	&	 2.17   &    2.25     &  		1.56  \\    
NGC7465	&	 2.58   &    2.68     &  		2.10  \\   
NGC7582	&	 2.62   &    2.80     &  		2.96  \\   
\hline 
\hline
\end{tabular}}
\tablefoot{Galaxies of the NUGA sample are highlighted in italics.} \label{radii} 
\end{table}


\begin{figure*}
    \includegraphics[width=8.50cm]{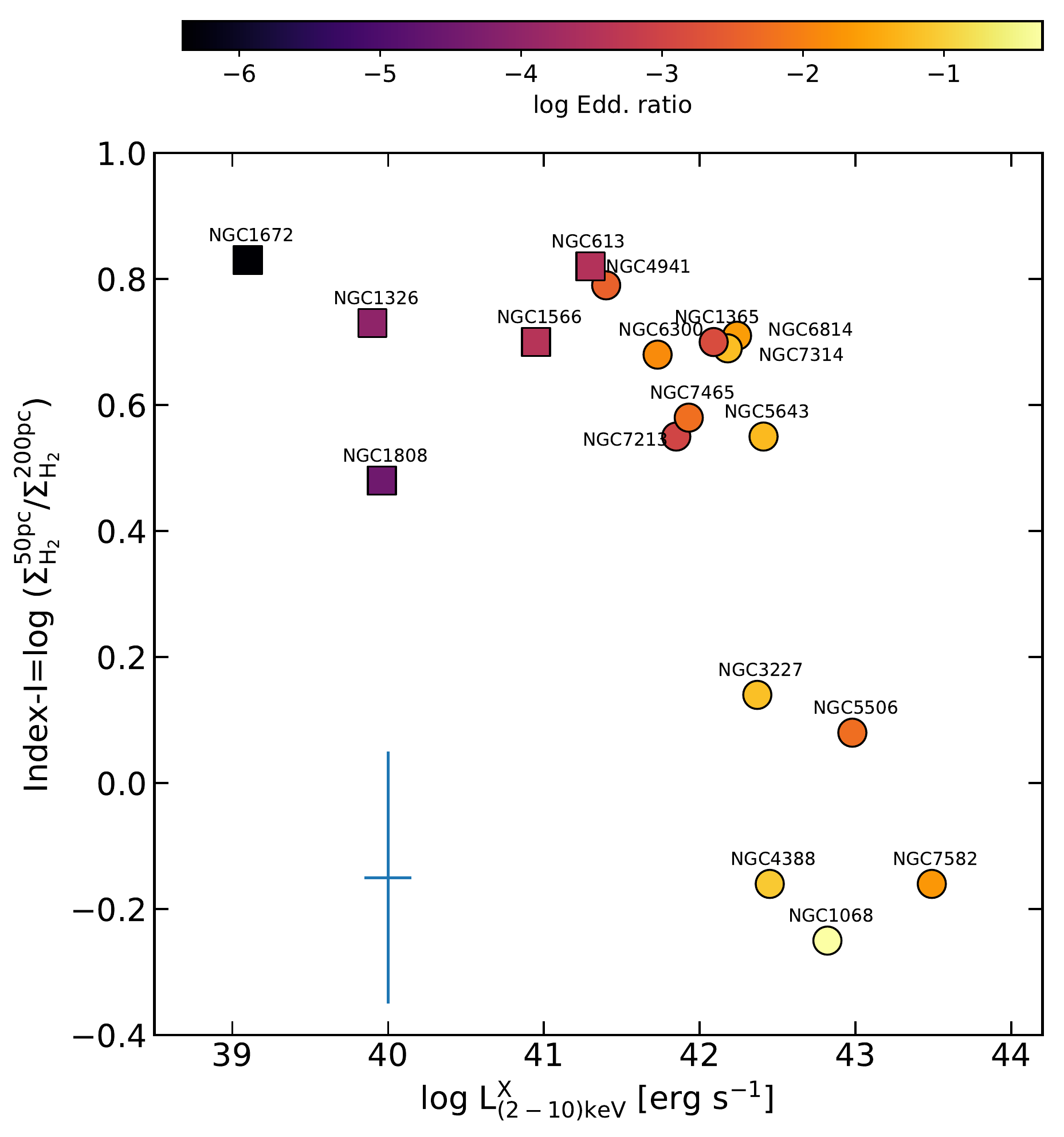}
    \includegraphics[width=8.65cm]{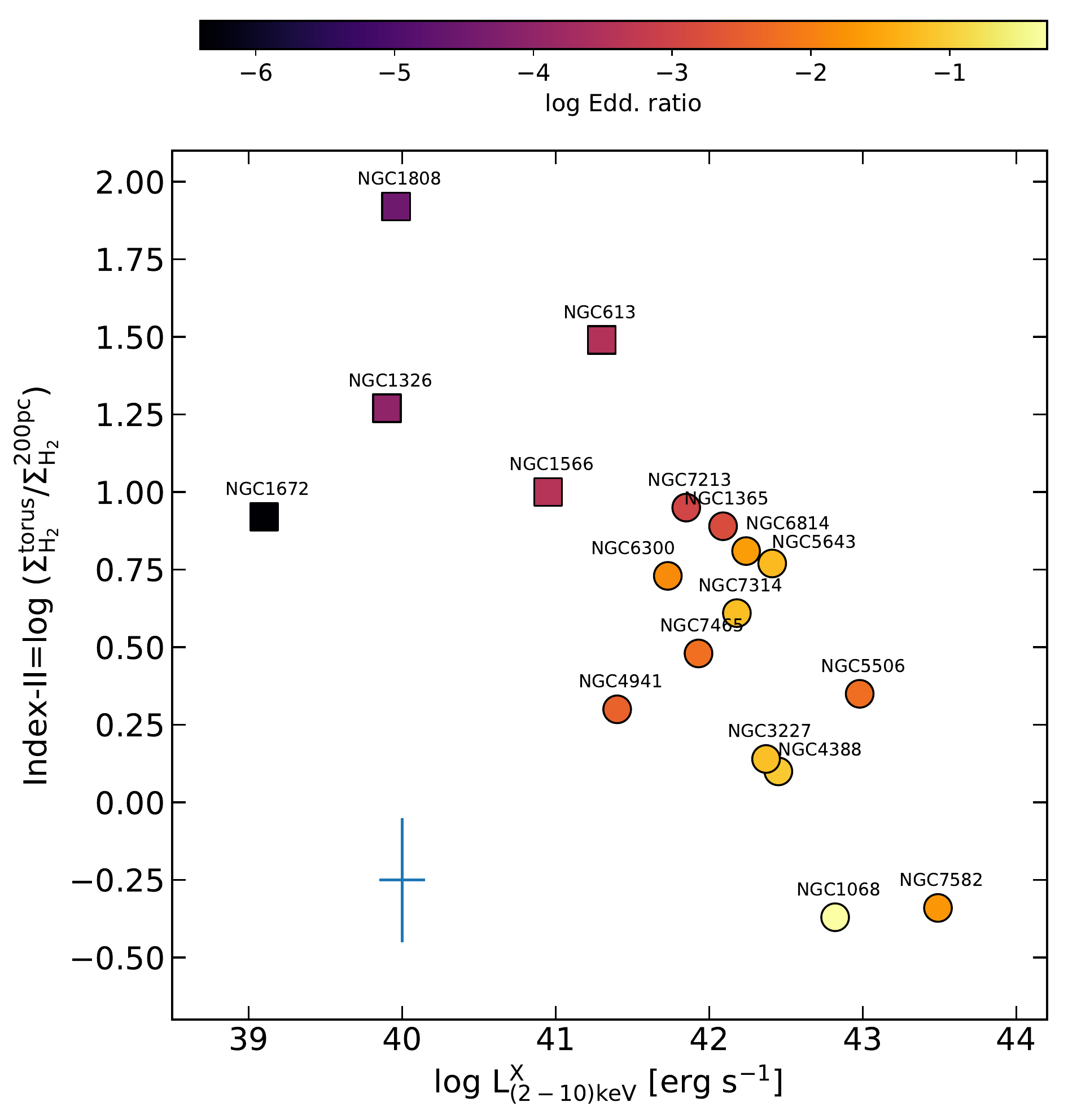}    
  \caption{{\it Left panel:}~An estimate of the concentration of molecular gas in the central regions of the galaxies of GATOS  and NUGA  samples derived from the ratio of the average H$_{\rm 2}$ surface densities measured at two spatial scales:   $r\leq50$~pc ($\Sigma_{\rm H_ 2}^{\rm 50pc}$) and  $r\leq200$~pc ($\Sigma_{\rm H_ 2}^{\rm 200pc}$); these scales are characteristic of the nuclear and circumnuclear regions, respectively. The ratio (index-I) is represented as a function of the AGN luminosities measured in the 2--10~keV band in log units; symbols  are color-coded as a function of the Eddington ratios for the sample galaxies: NUGA (square markers) and GATOS (circle markers). {\it Right panel:}~Same as {\it left panels} but replacing $\Sigma_{\rm H_ 2}^{\rm 50pc}$ by the average H$_{\rm 2}$ surface densities measured  inside the torus (or polar) regions identified in Sect.~\ref{continuum-fits} for GATOS sources (index-II). For NUGA galaxies we adopted the torus sizes derived by \citet{Com19}.  Errorbars account for the range of uncertainties on the molecular gas mass ratio estimates due to the assumed conversion factors, discussed in Sect.~\ref{gas-CND-nuclear} ($\sim\pm$0.2~dex), and  on the AGN luminosities ($\sim\pm$0.15~dex).}  
   \label{masses-radii}
\end{figure*}   

 

\begin{table*}[tbh!]
\caption{Statistical estimates of rank correlations.}
\centering
\resizebox{1\textwidth}{!}{ 
\begin{tabular}{ccccccc} 
\hline
\hline
\noalign{\smallskip}     	
 &	$\log (\Sigma_{\rm H_2}^{\rm 50pc}/\Sigma_{\rm H_2}^{\rm 200pc})$ 	 &	$\log (\Sigma_{\rm H_2}^{\rm torus}/\Sigma_{\rm H_2}^{\rm 200pc})$		&	$\log (\Sigma_{\rm H_2}^{\rm torus})$&  $\log (N_{\rm H_2}^{\rm AGN})$  &	$D_{\rm torus}$		\\
\noalign{\smallskip}   
 \hline  
 \noalign{\smallskip}  
$\log L^{\rm X}_{\rm (2-10)keV}$	&	 $\rho_{\rm Sp}$={\bf --0.75} ,  $p$={\bf 4}${\bf\times10^{-4}}$	&	$\rho_{\rm Sp}$={\bf --0.79} ,  $p$=${\bf 10^{-4}}$		&	$\rho_{\rm Sp}$=--0.36 ,  $p$=0.14	&	$\rho_{\rm Sp}$=--0.33 ,  $p$=0.18 	&	$\rho_{\rm Sp}$=+0.14 ,  $p$=0.59				\\	
	
&	$\rho_{\rm Ke}$={\bf --0.57} ,  $p$={\bf 6}${\bf\times10^{-3}}$  	&	$\rho_{\rm Ke}$={\bf --0.60} ,  $p$={\bf 3}${\bf \times10^{-4}}$		&	$\rho_{\rm Ke}$=--0.22 ,  $p$=0.22	&	$\rho_{\rm Ke}$=--0.19 ,  $p$=0.29 	&	$\rho_{\rm Ke}$=+0.06 ,  $p$=0.76 \\

&	$r$={\bf --0.64} ,  $p$={\bf 4}${\bf\times10^{-3}}$  	&	$r$={\bf --0.73} ,  $p$={\bf 6}${\bf \times10^{-4}}$		&	$r$={\bf --0.44} ,  $p$={\bf 0.05}	&	$r$=--0.41 ,  $p$=0.08 	&	$r$=+0.34 ,  $p$=0.16 \\

\noalign{\smallskip}	
\hline
\noalign{\smallskip}							
$\log {\rm Edd.ratio}$	&	$\rho_{\rm Sp}$={\bf --0.62} ,  $p$={\bf 7}${\bf \times10^{-3}}$  	&	$\rho_{\rm Sp}$={\bf --0.81},  $p$={\bf 5}${\bf\times10^{-5}}$		&	$\rho_{\rm Sp}$=--0.40 ,  $p$=0.10	&	$\rho_{\rm Sp}$=--0.30 ,  $p$=0.22 	&	$\rho_{\rm Sp}$=+0.02 ,  $p$=0.90				\\
&	$\rho_{\rm Ke}$={\bf --0.48} ,  $p$={\bf 5}${\bf \times10^{-3}}$  	&	$\rho_{\rm Ke}$={\bf --0.63} ,  $p$=${\bf10^{-4}}$		&	$\rho_{\rm Ke}$=--0.25 ,  $p$=0.15	&	$\rho_{\rm Ke}$=--0.18 ,  $p$=0.32 	&	$\rho_{\rm Ke}$=--0.03 ,  $p$=0.90  \\  
&	$r$={\bf --0.57} ,  $p$={\bf 0.01}  	&	$r$={\bf --0.69} ,  $p$=${\bf10^{-3}}$		&	$r$=--0.43 ,  $p$=0.08	&	$r$=--0.36 ,  $p$=0.14 	&	$r$=+0.10 ,  $p$=0.68 \\

\noalign{\smallskip} 
\hline 
\hline
\end{tabular}}
\tablefoot{A list of the rank and correlation parameters, as well as their associated two-sided  $p$ values, according to the Spearman's, Kendall's and Pearson's tests for the pairs of variables tabulated above. The pairs of variables that show statistically significant low $p$ values $\leq$0.05 are highlighted in boldface.} \label{stat} 
\end{table*}


\section{The radial distribution of molecular gas: the imprint of feedback?}\label{gas-radial}
 
 \subsection{Circumnuclear  versus nuclear scales}\label{gas-CND-nuclear} 

In Sect~\ref{CO-global} we concluded that  the molecular gas content measured on $\sim$~kpc scales  shows no dependence on the AGN luminosity or the Eddington ratio of the galaxies of our sample. The absence of a clear trend can be explained by the very different spatial scales and timescales relevant to the feeding and feedback cycle of Seyfert galaxies on the one hand and  to the assembly of the kpc-scale gas reservoirs on the other.

On smaller scales AGN winds and jets have nevertheless been seen to leave an imprint on the radial distribution of molecular gas in a number of Seyfert galaxies. High-resolution ALMA images obtained in 3 galaxies of our 
sample show that AGN winds and jets can push outward significant amounts of ISM material inside the disk and create deficits of molecular gas at scales $r\leq30-100$~pc in NGC~1068 \citep{GB14, GB19}  and NGC~3227 \citep{Alo19}, and even on larger scales in  NGC~5643 \citep{Alo18, Gar21}. Similar {\it holes} in CO emission identified in the circumnuclear disk of the Seyfert~2 galaxy NGC~5728 were attributed to molecular gas removal by the 
AGN wind \citep{Shi19}.

  \begin{figure*}[bth!]
   \centering
    \includegraphics[width=8.3cm]{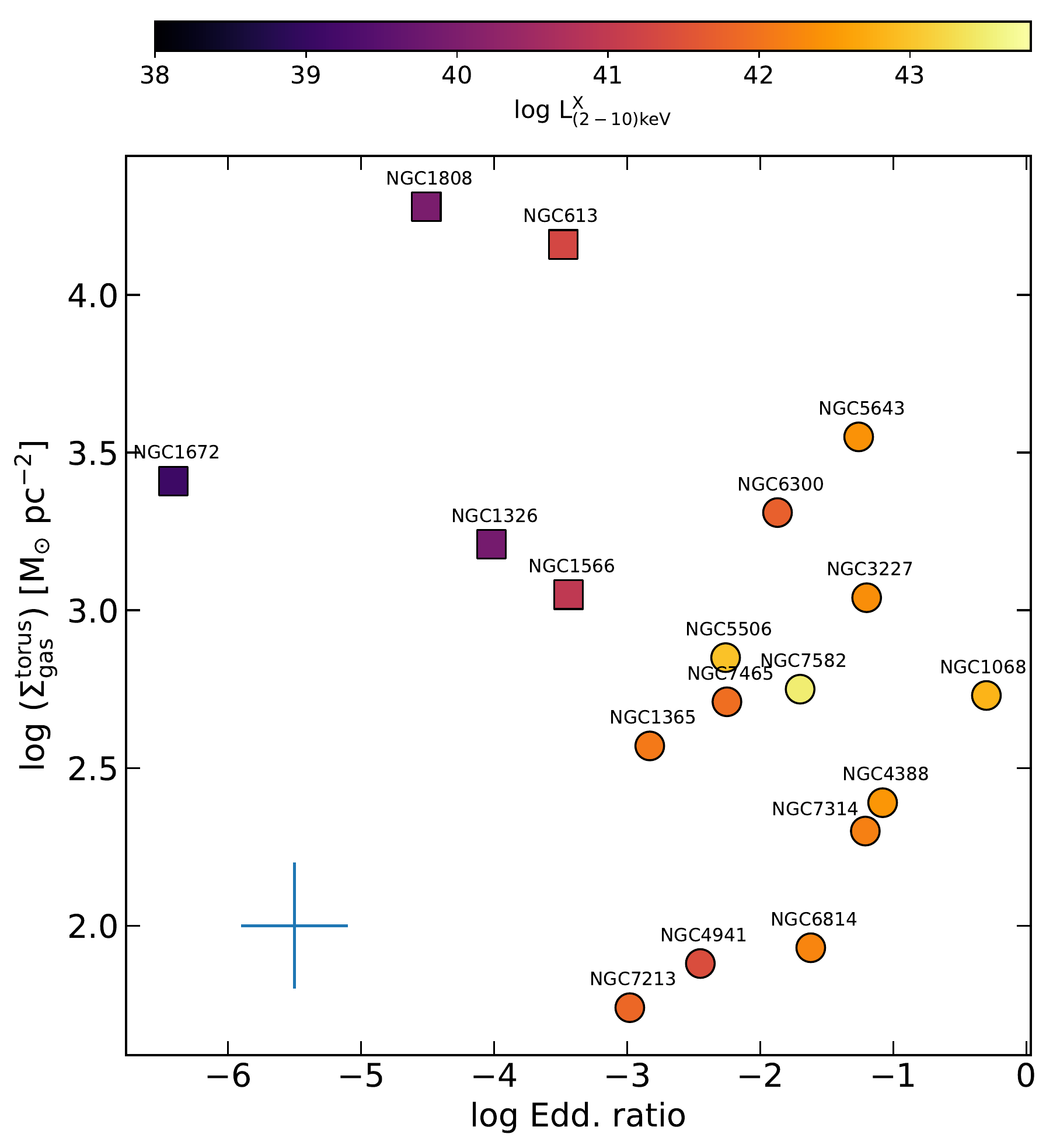}   
     \includegraphics[width=8.45cm]{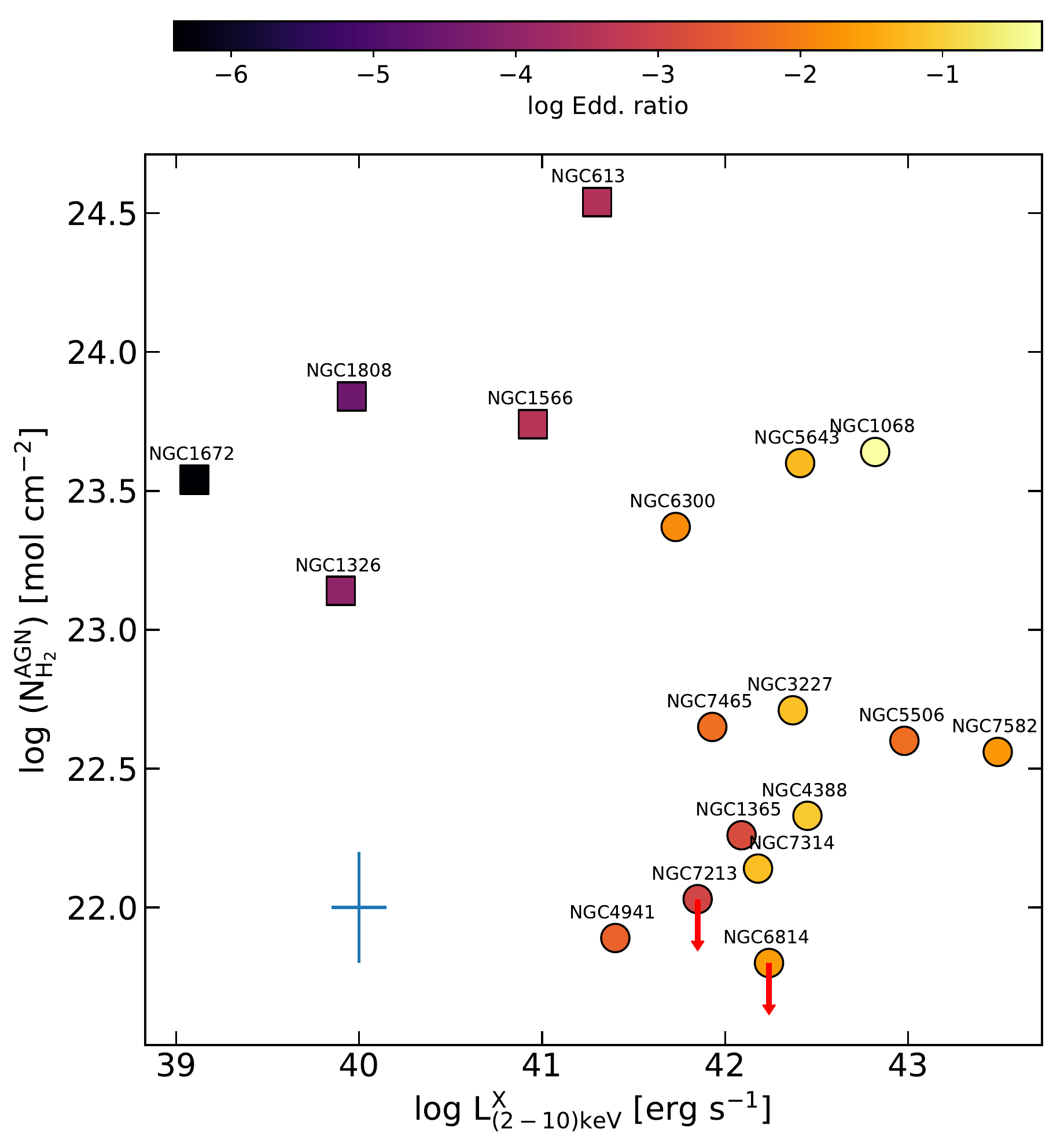} 
       \caption{A comparison of physical properties of the molecular  tori in GATOS (circle markers) and NUGA (square markers).  Average molecular gas surface densities  ($\Sigma_{\rm gas}^{\rm torus}$; {\it left panel}) and  H$_{\rm 2}$ column densities measured towards the AGN ($N_{\rm H_2}^{\rm AGN}$; {\it right panel}) are shown as a function of the  Eddington ratio ({\it left panel}) or the 2--10~keV band AGN luminosities  ({\it right panel}) of our sources. Errorbars account for the total uncertainties on the molecular gas mass estimates due to  the assumed conversion factors ($\sim\pm$0.2~dex), as well as on the AGN luminosities ($\sim\pm$0.15~dex) and Eddington ratios ($\sim\pm$0.4~dex).}  
   \label{torus-props}
    \end{figure*}

We explore below if a simple parameterization of the radial distribution of molecular gas in the galaxies of our combined sample is able to capture the effects of AGN feedback.
Based on the results obtained in NGC~1068, NGC~3227 and NGC~5643, we purposely focus on the central $r\leq200$~pc regions where molecular gas reservoirs are expected to build up and potentially reflect the effects of AGN-driven outflows. In particular we investigate if the concentration of molecular gas inside typical `circumnuclear' disk scales shows any dependence on $L_{\rm AGN}$ and/or on the Eddington ratio in our sample galaxies.  To accomplish this, we  first derived  the average H$_{\rm 2}$ surface densities measured using two alternative definitions of the `nuclear' spatial scales: 
\begin{itemize}
\item A region common for all galaxies defined by $r\leq 50$~pc: $\Sigma_{\rm H_2}^{\rm 50pc}$. Fig.~\ref{histo-sizes} shows that the typical sizes of dusty molecular tori in our sample galaxies are a factor of $\sim2$ smaller. This scale 
therefore probes the region occupied by the tori and their immediate surroundings  \footnote{CO images were deprojected using values of $PA$ and $i$ derived in the kinematic fit of Appendix~\ref{kinemetry} or taken from LEDA for 
the NUGA galaxies.}. 

\item  A region defined on a case-by-case basis by the extended emission components  identified at 870~$\mu$m (Sect.~\ref{continuum-fits}) for the GATOS sources: $\Sigma_{\rm H_2}^{\rm torus}$.  For sources of the NUGA sample we adopted the torus sizes published by \citet{Com19}.  According to Fig.~\ref{histo-sizes}, this definition of the nuclear scales typically encompasses  regions $r\leq 20-30$~pc for most of our sources.   

\end{itemize}

 We normalized the nuclear surface densities $\Sigma_{\rm H_2}^{\rm 50pc}$ and $\Sigma_{\rm H_2}^{\rm torus}$ to the  H$_{\rm 2}$ surface densities measured on 
 'circumnuclear' scales ($r\leq200$~pc), hereafter referred to as $\Sigma_{\rm H_2}^{\rm 200pc}$. The adopted  normalization is intended to neutralize any potential bias that may result from the different overall 
 molecular gas content of the galaxies of our sample and it therefore  serves as a measurement of the gas concentration.  Furthermore, as our gas concentration index is derived from surface 
 densities averaged inside the regions defined above we minimize the variance that could be introduced by the analysis of individual radial profiles. To derive the mass estimates on 
circumnuclear and nuclear scales we assumed slightly different 3--2/1--0 brightness temperature ratios of $\sim2.0$ and  $\sim2.9$, respectively, based on the values  measured on similar spatial scales in NGC~1068 \citep{GB14}. The H$_{\rm 2}$ surface densities derived at these different spatial scales for the galaxies of our combined sample are listed in Table~\ref{radii}.

Figure~\ref{masses-radii} represents the molecular gas concentration indices, estimated by $\log (\Sigma_{\rm H_2}^{\rm 50pc}/\Sigma_{\rm H_2}^{\rm 200pc})$ (hereafter, index-I)
 and $\log (\Sigma_{\rm H_2}^{\rm torus}/\Sigma_{\rm H_2}^{\rm 200pc})$ (hereafter, index-II), as a function of the intrinsic (absorption-corrected) AGN luminosities measured in the 2--10~keV band ($\log L^{\rm X}_{\rm (2-10)keV}$). Symbols are color-coded to explore in parallel any potential dependence of the molecular gas concentration  indices on the Eddington ratio. A visual inspection of Fig.~\ref{masses-radii} shows that there is a significant trend pointing to lower molecular gas concentration indices in the higher luminosity and the higher Eddington ratio sources of our sample. The trend is statistically significant for both indices as confirmed by the estimated Spearman's and Kendall's rank coefficients ($\rho_{\rm Sp}$ and $\rho_{\rm Ke}$), as well as by the Pearson's correlation coefficient ($r$), and their associated two-sided $p$ values, listed in Table~\ref{stat}. In particular, the gas 
concentration measured by index-I changes by $\sim1$~dex across the sample. The corresponding decrease is of $\sim$2~dex for index-II.

The GATOS sources with the highest AGN luminosities (NGC~1068, NGC~3227, NGC~4388, NGC~5506, and NGC~7582) show a flat distribution of surface densities, characterized by an index-I 
$\simeq$ --0.2 to +0.2. This is in stark contrast with the more centrally peaked distributions shown by the lower AGN luminosity sources among the GATOS sample and NUGA sources, which show a common range for index-I $\simeq$+0.5 to +0.8. Furthermore, while both indices show a similar spread of values for GATOS sources, index-II is comparatively boosted for  NUGA sources (index-II $\simeq$+1.0 to +2.0).  

The estimates of the molecular gas masses derived  from the 3--2 line of CO on nuclear scales are subject to uncertainties related, first,  to the assumed 3--2/1--0 ratio, which is here taken from the value measured in the molecular torus of the GATOS source NGC~1068 \citep{GB14, Vit14, GB19}, and second, to the adopted CO--to--H$_2$ conversion factor, which we assumed to be the canonical value of the Milky Way.  We can give a  conservative estimate of the overall uncertainties accounting for the systematic errors associated with the two conversion factors mentioned above. We estimate that the latter may contribute up to $\sim$0.12-0.13~dex each if we allow them to explore the full range of values seen in the observations of different populations of galaxies (on all ranges of spatial scales) and, also, if we assume the unlikely worst case scenario where these uncertainties are fully uncorrelated. Under these extreme hypotheses the total uncertainty amounts to $\pm$0.20~dex \footnote{These estimates are in line with those typically assumed in the literature: $\sim\pm0.23$~dex \citep[e.g.][]{Gen10}}.  While we do not have quantitative estimates for these conversion factors for the galaxies of our combined sample,  we nevertheless expect that, compared to GATOS sources, molecular gas will be less excited on nuclear scales for NUGA sources, which are characterized by lower AGN luminosities, and therefore lower X-ray and UV irradiation of molecular gas.  The trend shown by the CO(3--2)/HCO$^+$(4--3) ratio in Fig.~\ref{AGN-ratios} seems to independently corroborate this picture. Adopting a lower  3--2/1--0 ratio would therefore increase the nuclear-scale mass estimates for NUGA Seyferts. In addition, if at all, the  CO--to--H$_2$ conversion factor would have to be lowered relative to the assumed Milky Way value for the nuclear scales of GATOS Seyferts, where molecular gas is expected to live in a more heavily irradiated and dynamically perturbed environment. This habitat is more akin to the disturbed medium of ULIRGs and nuclear starbursts where this factor is assumed to be 1/4 of the Milky Way value. In summary, any foreseeable change of these conversion factors in our sample galaxies would rather combine to further reinforce the trends shown in Fig.~\ref{masses-radii} well beyond the uncertainties estimated above.

\subsection{The dusty molecular tori scales}\label{gas-tori} 

We have shown in Sect~\ref{gas-CND-nuclear} that NUGA galaxies show a notably higher degree of concentration of their molecular gas on circumnuclear scales compared to that of GATOS sources. These differences are amplified most singularly if we estimate the gas concentration by the  $\Sigma_{\rm H_2}^{\rm torus}/\Sigma_{\rm H_2}^{\rm 200pc}$ ratio, namely when nuclear scales are set by the tori.  In this section we study the variance in a set of physical properties of the dusty molecular tori between GATOS and NUGA sources. Figure~\ref{torus-props} compares  the average gas surface densities of the  tori, $\Sigma_{\rm gas}^{\rm torus}\equiv1.36\times\Sigma_{\rm H_2}^{\rm torus}$, and the H$_{\rm 2}$ column densities measured from the velocity-integrated CO intensities towards the AGN, $N_{\rm H_2}^{\rm AGN}$ (listed in Table~\ref{Tab3})\footnote{MSR and HSR datasets give virtually identical values for $N_{\rm H_2}^{\rm AGN}$ except for the two upper limit estimates in NGC~6814 and NGC~7213 where we adopt the more compelling value derived from the MSR dataset.}, in our sample galaxies.

The distribution of $\Sigma_{\rm gas}^{\rm torus}$  for GATOS sources shows a large scatter within an overall wide range of values 
$\simeq5.0\times10^{1}$--$4.0\times10^{3}~M_{\sun}$pc$^{-2}$, which show no trend with the Eddington ratio within the range covered by this sample. Molecular tori in NUGA sources nevertheless show comparatively higher 
 $\Sigma_{\rm gas}^{\rm torus}$ values $\simeq1.3\times10^{3}$--$2.0\times10^{4}~ M_{\sun}$pc$^{-2}$. Within the limits imposed by the limited number of very low Eddington ratio sources, the combined NUGA and GATOS samples do suggest a tantalizing monotonic decrease of  $\Sigma_{\rm gas}^{\rm torus}$ with the Eddington ratio or the AGN luminosity.

 The trend is on the borderline of statistical significance as estimated by the Pearson's correlation coefficient ($r\sim-0.4$) and its associated two-sided $p$ values=0.05-0.08, listed in Table~\ref{stat}.  The significance of this anticorrelation increases under the physically-motivated alternative hypothesis that  log($\Sigma_{\rm gas}^{\rm torus}$) and log($L_{\rm AGN}$) have a negative Pearson's correlation coefficient,  in which case the associated one-sided test  $p$ value decreases to $\sim$0.03. The one-sided test using the Eddington ratio as a substitute for  the AGN luminosity gives a $p$ value$\sim$0.04. This trend could explain the different behaviour of the gas concentration indices  discussed in Sect.~\ref{gas-CND-nuclear}. 
 
 Furthermore, the deconvolved diameters of  tori ($D_{\rm torus}$) show no significant trend with the AGN luminosity or the Eddington ratio. In particular, GATOS and NUGA sources show a common median value for $D_{\rm torus}\sim$42~pc. The value estimated for $D_{\rm torus}$ in NGC~5506, which stands out as an outlier in the distribution, is likely overestimated due to the contribution of dust emission stemming from the highly inclined disk of the host (see AH21).

 As expected, the variance in $N_{\rm H_2}^{\rm AGN}$ values between GATOS and NUGA sources is similar to that shown by $\Sigma_{\rm gas}^{\rm torus}$ and so are their associated rank and correlation parameters listed in Table~\ref{stat}. Molecular tori in NUGA sources tend to show higher gas column densities compared to GATOS. The existence of a monotonic decrease of $N_{\rm H_2}^{\rm AGN}$ with the AGN luminosity or the Eddington ratio for the combined sample is nevertheless on the border of statistical significance. In particular, the physically-motivated alternative hypothesis that  log($N_{\rm H_2}^{\rm AGN}$) and log($L^{\rm X}_{\rm (2-10)keV}$) have a negative Pearson's correlation coefficient has a one-sided $p$ value $\sim$0.04. The same test using the Eddington ratio as a substitute for  the AGN luminosity gives a $p$ value$\sim$0.07.

  \begin{figure*}[th!]
   \centering
       \includegraphics[width=18cm, angle=0]{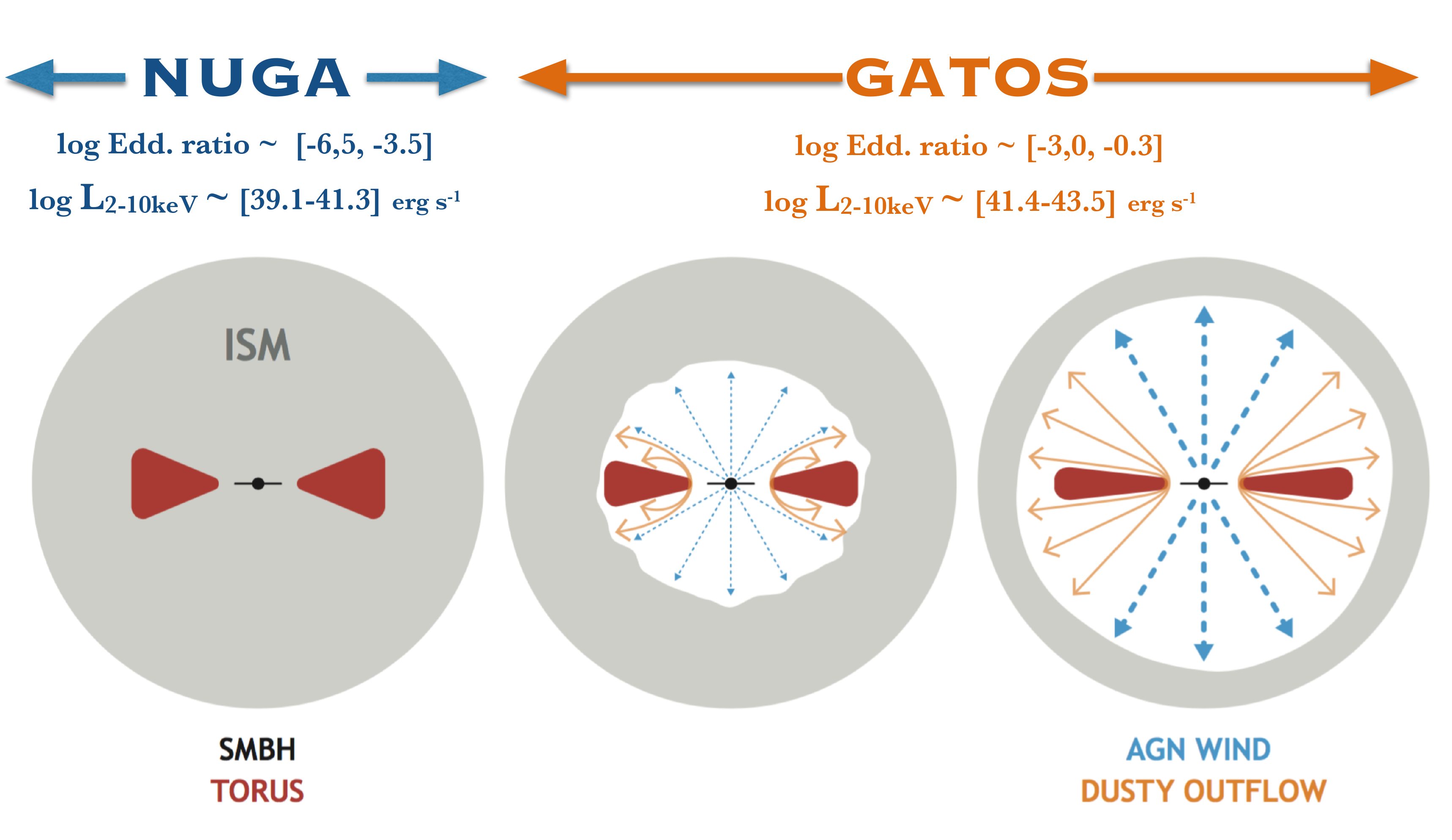}
       \vspace{-0.5cm}
               \caption{A cartoon illustrating the different nuclear and circumnuclear environments of the NUGA and GATOS sources analyzed in this paper. The figure is a (not to scale)  simplified sketch 
       of the CND  around the SMBH of a  Seyfert galaxy representative of the NUGA ({\it left panel}) and GATOS samples ({\it middle} and {\it right panels}). These samples are characterized by different Eddington ratios and AGN luminosities (NUGA: Eddington ratios$\sim$10$^{-6.5}$--$10^{-3.5}$ and $L_{\rm AGN}$(2-10~keV)$~\sim10^{39.1-41.3}$erg~s$^{-1}$; GATOS: Eddington ratios$\sim$10$^{-3.0}$--$10^{-0.3}$ and $L_{\rm AGN}$(2-10~keV)$~\sim10^{41.4-43.5}$erg~s$^{-1}$). Based on the monotonic trends of Fig.~\ref{masses-radii}, we distinguish 
       between the typical environment of the highest luminosity and highest Eddington ratio GATOS Seyferts ({\it right panel}, well represented by sources like NGC~1068,  and  an intermediate case between NUGA       
      and the most extreme GATOS Seyferts  ({\it middle panel}), well represented by sources like NGC~5643. The imprint left by AGN winds on the  
       distribution of molecular gas in the torus and its surroundings through the launching of dusty winds, and by the impact of AGN winds on the ISM at larger radii (ISM in the figure), is seen to be more extreme in the higher luminosity and  higher Eddington ratio sources.}  
   \label{cartoon}
    \end{figure*}
 
 \subsection{A scenario describing the imprint of AGN feedback in Seyfert galaxies}\label{scenario}

 Figs.~\ref{CO-colors} and \ref{cont-CO} visualize the existence of a depression  in the radial distribution of molecular gas, and in some extreme cases (e.g. NGC~1068) of a distinct {\it hole} at nuclear scales  for a significant fraction of GATOS sources.  As expected, galaxies like NGC~1068 and NGC~3227, where  the presence of AGN-driven molecular outflows was established \citep{GB19, Alo19}, are among the galaxies showing the highest gas deficiencies on nuclear scales according to Fig.~\ref{masses-radii}.

 NGC~1068 is the most extreme case exemplifying the effect of AGN feedback on the radial redistribution of molecular gas in the disk.  The CND of NGC~1068  shows  a highly-contrasted ring structure with a steep radial profile on its inner edge. This morphology reflects the accumulation of  molecular gas at the working  surface  of  the  AGN  wind and jet, which is currently situated at radii $\sim50-80$~pc. This distinct signature is found in all the high-resolution images of the CND obtained in a large variety of molecular gas tracers that probe several orders of magnitude in volume densities and  kinetic temperatures \citep{Mue09, Sto12, GB14, Ima18, GB19}. The molecular torus of NGC~1068 is connected to the $r\sim200$~pc CND ring through a network of gas streamers, which are detected at $r\leq50$~pc, i.e., inside the distinct cavity excavated by the AGN wind and the radio jet \citep{GB19}. The kinematics of molecular gas show unambiguous signs of outflowing motions in the torus itself, but also, in the gas streamers and the CND ring. Most of the mass, momentum and energy of the molecular outflow  is contained by the CND, where the AGN wind and the radio jet are currently sweeping the gas of the disk at a high rate \citep[$\dot{M}\simeq60~M_{\sun}$yr$^{-1}$;][]{GB19}. 
 
  The molecular outflow of NGC~3227 studied by \citet{Alo19} has also left an imprint on the distribution of molecular gas on nuclear scales. The NGC~3227 outflow nevertheless appears as less extreme or less evolved compared 
  to that of NGC~1068: it extends up to a factor of 10 smaller radial distances   ($\sim30$~pc) when compared to NGC1068 and is pushing the ISM in the disk of the galaxy with moderate outflow rates that are 
  an order of magnitude lower ($\dot{M}\simeq5~M_{\sun}$yr$^{-1}$).  
 
Like in NGC~1068 and NGC~3227, ALMA detected the presence of molecular outflowing components at different spatial scales in  NGC~5643 \citep{Alo18, Gar21}.  The CO(2-1) profiles comprise several velocity components
 that result from the interaction of the AGN outflow with molecular gas in the disk $\sim500$~pc from the AGN. This interaction created a large-scale molecular outflow with a high mass outflow rate \citep[$\dot{M}\simeq51~M_{\sun}$yr$^{-1}$;][]{Gar21}. On nuclear scales, \citet{Alo18} detected non-circular motions in the $\sim30-40$~pc-diameter molecular torus that could be explained as radial outflows, as similarly found in the outflowing NGC~1068 torus \citep{Gal16, GB16, GB19, Imp19, Ima20}. 
  
As discussed in  Appendix~\ref{App1} and Sect.~\ref{CND}, AGN winds are present in the majority of the GATOS sources. Although in a smaller number of objects, the existence of radio jets has also been reported.  AGN winds and radio jets could therefore launch and promote molecular outflows in other GATOS sources similar to those of NGC~1068, NGC~3227, and NGC~5643. 

In order to search for signatures of  molecular outflows in other sources, we used the software package {\tt kinemetry}, developed by \citet{Kra06}, to fit the velocity fields derived from the CO(3--2) data of the GATOS core sample. 
The outcome of this analysis, presented in Appendix~\ref{kinemetry}, shows evidence of molecular outflows in the CND of three  sources: NGC~4388, NGC~5506 and  NGC~5643.  In particular, the bulk of the non-circular motions 
identified can be interpreted as due to the presence of coplanar molecular outflow components. For NGC~4388 and NGC~5643 these results confirm the findings of \citet{Dom20}, \cite{Alo18}, and \citet{Gar21}, who found evidence of molecular outflows in these sources based on a study of the gas kinematics  derived from lower resolution CO(2--1) images obtained with  NOEMA and ALMA.  The newly detected outflows in NGC~4388 and 
NGC~5506 are seen to extend  out to deprojected radii $\sim100-200$~pc from the AGN and have moderate radial velocities $\sim50-100$~km~s$^{-1}$.  Furthermore, we find only tentative evidence of moderate $\sim50-100$~km~s$^{-1}$ molecular outflows in NGC~4941 and NGC~7465. The location of NGC~5643, NGC~4941, and NGC~7465 in Fig.~\ref{masses-radii} is intermediate between the two extremes represented by NGC~1068 and the NUGA sources. 

In NGC~7582, the other galaxy  showing a  gas deficit on nuclear scales illustrated by Fig.~\ref{masses-radii}, we only find marginal evidence of moderate outflow motions ($<50$~km~s$^{-1}$) up to 100~pc from the AGN. We nevertheless find evidence of inflows triggered by the stellar bar down to the deprojected radii of the likely location of the ILR region of the bar ($r\sim250-300$~pc). Molecular gas is distributed in a two-arm nuclear spiral that ends
up in a ring. This morphology can be interpreted as the gas response to the 16~kpc-long bar near its ILR region.  The nuclear-scale gas deficit detected in NGC~7582 may have been caused by a previous episodic molecular outflow event or, alternatively, by the gravity torques exerted by the stellar bar over longer time scales on the gas. Inside the ILR ring torques are expected to be positive and be able to drive gas outward. Both AGN feedback and bar-driven secular evolution can combine their efforts to quench AGN fueling at nuclear scales.

To summarize, we find clear evidence of molecular outflows in four of the five GATOS Seyferts  showing the most extreme nuclear-scale gas deficits in our sample, namely  NGC~1068, NGC~3227, NGC~4388, and NGC~5506. We nevertheless find marginal evidence of outflow in NGC~7582.

The existence of a molecular outflow is not always accompanied by a conspicuous nuclear-scale molecular gas deficit. The case of the NUGA galaxy NGC\,613 is a good illustration of this paradox. \citet{Aud19} detected
a high-speed ($v_{\rm out}\sim300$~kms$^{-1}$) compact molecular outflow in the central $r<20-25$~pc region around the Seyfert  nucleus of the galaxy. In the absence of a powerful AGN wind in this source,  \citet{Aud19} concluded that the outflow is likely triggered by the highly-collimated 200~pc-size radio jet imaged by the VLA. The low Eddington ratio of NGC~613 and the high value of $N_{\rm H_2}^{\rm AGN}$ place the galaxy in 
a region of the parameter space that is not conducive to  launch dusty outflows from the central engine, according to the predictions of \citet{Ven20} (see discussion in AH21).  This would suggest that the jet is the only agent able to launch a molecular outflow in this source. 
 The picture drawn from the ALMA image of NGC\,613 indicates that although the radio jet has triggered a small molecular outflow at its very base, it has left virtually intact the distribution of molecular gas outside $r\sim20-25$~pc: NGC\,613 shows high values of the molecular gas concentration indices (see Fig.~\ref{masses-radii}).  The 3D geometry of the  molecular outflow and the jet remain unconstrained. However, the low coupling  efficiency of the jet with the ISM  can be explained in a scenario where the jet lies at a large angle relative to the disk of NGC\,613. Alternatively, given the very short timescales associated with the outflow and the radio jet \citep[$\sim10^{4}$~yr;][]{Aud19}, we can speculate that the nascent jet will be able to inflate a cocoon of hot gas and eventually leave a more distinct imprint on the gas distribution in the future.

 Furthermore, the trends discussed in Sect.~\ref{gas-tori} suggest the existence of a tentative  decline of the tori-related parameters, $\Sigma_{\rm gas}^{\rm torus}$ and $N_{\rm H_2}^{\rm AGN}$, with the increasing Eddington ratio and/or the AGN luminosity in the galaxies of our
 combined sample (Fig~\ref{torus-props}).  It is tempting  to consider $\Sigma_{\rm gas}^{\rm torus}$ or  $N_{\rm H_2}^{\rm AGN}$ as proxies for the dust covering fraction (CF) of the torus, defined as the fraction of the sky that is blocked by dust, as seen from the central engine.  \citet{Ric17b} recently found that the X-ray CF declines with the Eddington ratio based on a survey of 731 AGN in the local Universe \citep[see also][]{Ezh17, Zhu18, Tob19, Tob21}. 
 In particular, \citet{Ric17b} found evidence of higher  CF ($\sim85\%$) at Eddington ratios $<10^{-1.5}$  compared to those measured at  Eddington ratios $>10^{-1.5}$  ($\sim40\%$).  Although other works that use different proxies for the CF found no correlation with the Eddington ratio \citep[e.g.][]{Cao05, Gar19}, the results of  \citet{Ric17b} indicate that  radiative feedback on dusty gas is a key mechanism able to regulate the internal distribution of material inside the torus and its surroundings.
 While the link  between  $\Sigma_{\rm gas}^{\rm torus}$ or  $N_{\rm H_2}^{\rm AGN}$ and CF is admittedly indirect, the tentative trends of  Fig~\ref{torus-props} add supporting evidence to this scenario.

 Overall, the trends discussed in Sect.~\ref{gas-CND-nuclear} suggest that  AGN feedback has significantly modified  the radial distribution of the circumnuclear molecular gas preferentially in those sources characterized by having  higher luminosities and higher Eddington ratios within our combined sample of NUGA and GATOS galaxies. This is reflected in the  $\sim1$~dex monotonic decrease of index-I, sensitive to the scales of the tori and their immediate surroundings out to $r\sim50$~pc, but more clearly in the $\sim$2~dex decrease of index-II, tailored to capture the normalized concentration of molecular gas on torus scales.  
 The differences between the nuclear and circumnuclear environments of  NUGA and GATOS sources are illustrated by the cartoon shown in Fig.~\ref{cartoon}.  
 The different degrees of nuclear-scale gas deficits illustrated by Fig.~\ref{masses-radii} allow us to distinguish the typical CND environment of sources like NGC~1068 from that observed in sources like NGC~5643, a representative case intermediate between the low luminosity NUGA and the highest luminosity GATOS targets. In agreement with the picture drawn from Fig.~\ref{cartoon}, \citet{Oma19} found that the percentage of good fits to the infrared SED of active galaxies increases significantly for higher luminosity AGN when the two-component models of \citet{Hoe17}, which include both a clumpy disk and a polar clumpy wind, are confronted with observations. In contrast, one-component torus models, like those of \citet{Nen08a, Nen08b}, provide better fits to the SED of lower luminosity AGN.  
  
  The search for molecular outflow signatures in our Seyfert galaxies was aimed at identifying unambiguous `smoking-gun' evidence of AGN feedback. The properties of the sources where we have identified the existence of molecular outflows  suggest that  AGN feedback can be  `caught in the act' more frequently among the higher luminosity and /or higher Eddington ratio sources of the GATOS sample.   AH21 show evidence of extended mid-IR polar emission likely associated with AGN-driven dusty outflows  in some of the GATOS sources where we also find evidence of molecular outflows on larger scales. Modelers usually account for the mid-IR polar emission detected in Seyfert galaxies as due to a clumpy dusty wind launched `locally' by radiation pressure from the inner region of the tori \citep{Hoe17}. These dusty winds, which adopt a hollow cone geometry,  can  exist for a certain range of  gas column densities and Eddington ratios \citep[e.g.][see also AH21]{Ven20}.  The efficiency of AGN feedback is maximized if the  AGN wind and/or the radio jet is strongly coupled with the ISM of the host disk on large spatial scales, i.e., only if the angular momentum of the accretion disk and the torus are decoupled from that of the disk. In this scenario AGN winds and/or jets can eventually intersect a large fraction of the disk. This favourable configuration can explain the launching of mostly coplanar massive molecular outflows across the wide range of spatial scales identified in the disks of NGC~1068, NGC~5643 or NGC~3227. Furthermore, this result is in agreement with the findings of \citet{Dav14}, who reported the detection of molecular outflows using the  1-0 S(1) H$_{\rm2}$ line in five nearby AGN; similarly, these outflows lie in a geometry configuration  close to coplanar with the host galaxy disk.

\section{Summary and outlook}\label{summary}

 We used ALMA to image the emission of molecular gas and dust using the CO(3--2) and HCO$^+$(4--3) lines as well as their underlying continuum emission at 870~$\mu$m with a set of spatial resolutions ($0.1\arcsec\simeq7-13$~pc) in the circumnuclear disks  of 10 nearby ($D<28$~Mpc) Seyfert galaxies. The selected targets are part of the GATOS core sample, which was drawn from an ultra-hard X-ray survey. This is the first paper in a series aimed at understanding the properties of the dusty molecular tori and their connection to the host galaxy in nearby Seyferts. In  paper II of this series AH21 study the relation between the torus and polar dust emissions  and compare them with predictions from radiative transfer disk+wind models. Our project expands the range of AGN luminosities and Eddington ratios covered by other ALMA surveys of Seyferts such as NUGA and allows us to study the gas feeding and feedback cycle  in a combined sample of 19 Seyferts.  The main results of this paper are summarized as follows:
  
  \begin{itemize}
 
 \item
 
 ALMA detected   870~$\mu$m continuum emission  stemming from spatially-resolved disks  located around the AGN in all the galaxies of the GATOS core sample.  We performed a morphology-wise decomposition to fit the continuum images with a set of point sources and extended components to minimize the contamination from free-free and synchrotron emission, which can contribute to the 870~$\mu$m emission on spatial scales unresolved by ALMA. After correction, the bulk of the continuum flux of the extended components can be accounted for by thermal emission from dust in all of our sources with the exception of NGC\,7213. This result is confirmed by the spectral index maps of the submillimeter continuum emission obtained after correction in a subset of our targets.  
 
 \item

 Our data support that the orientation of the dusty disks relative to the AGN winds, which are present in the majority of our sources, shows a statistically significant preference.  
 In particular, out of the 13 galaxies analysed in our combined sample, a group of 9 galaxies tend to show  an equatorial orientation  of their extended components, i.e. close to perpendicular relative to the AGN wind axes, as expected for a torus geometry for the dusty disks. A smaller subset of four galaxies display a mixed geometry intermediate between polar and equatorial.   
 
 \item
 
  Significant amounts of molecular gas are associated with the dusty disks, as confirmed  by the detection of bright CO emission in all of our sources, with the exception of NGC\,7213. 
  In particular, while CO emission does not generally peak at the nuclei, the 3--2 line is detected coincident with the AGN in all the galaxies with the exception of NGC~6814 and NGC~7213.  The range of CO fluxes measured in front of the AGN in our sources corresponds to a range of molecular gas column densities $N(H_{\rm 2})\sim8\times10^{21}-4\times10^{23}$~cm$^{-2}$.
   
 \item
 
 The dusty molecular  tori imaged by ALMA in the GATOS Seyferts have full-sizes (diameters) ranging from $\sim25$~pc to $\sim130$~pc, with a median value of $\sim42$~pc, and molecular gas masses ranging from $\simeq5\times10^{4}$ to $1\times10^{7}$~M$_{\sun}$, with a median value of $\sim6\times10^{5}$~M$_{\sun}$. While the median value of the molecular tori in NUGA sources is virtually identical to that measured in GATOS, the distribution of masses of NUGA tori, from $\simeq2\times10^{6}$ to $1\times10^{7}$~M$_{\sun}$, is skewed toward the highest values within our combined sample.   

\item

We detected the emission of the 4--3 line of HCO$^+$,  a tracer  of dense molecular gas (n(H$_{\rm2}$)~$\geq10^5-10^6$~cm$^{-3}$) in four GATOS Seyferts. The HCO$^+$ emission arises from spatially-resolved disks  that   show an excellent  correspondence in their position, size and orientation with the CO and dusty disks identified in two sources: NGC~5643 and NGC~6300. 
 We measure differences of up to an order of magnitude in the CO/HCO$^+$ ratios within our combined sample of NUGA and GATOS: CO/HCO$^{+}\sim$2.8--38. This wide range of values point to a very different density radial stratification inside the dusty molecular tori in our sample galaxies. 

\item

 We find a positive correlation between the line-of-sight gas column densities responsible for the absorption of X-rays and  the  molecular gas column densities derived from CO   towards the AGN in our sources. Furthermore, the median values of the distributions of both  column densities are similar.
This suggests that the neutral gas line-of-sight column densities of the dusty molecular tori imaged by ALMA with $\simeq7-10$~pc spatial resolution make a significant contribution to the obscuration of X-rays.
We also find a positive correlation between the average gas surface densities of the dusty molecular tori and the gas column densities responsible for the reflection of hard X-rays. However the median values of these two column densities differ by more than an order of magnitude. This discrepancy could be accounted for if the bulk of the neutral gas producing the reflection of X-rays is located at the inner walls of the tori.

\item

We analyzed the normalized radial distributions of molecular gas in the CND of our Seyfert galaxies to look for signs of nuclear-scale molecular gas deficits. 
We find that the imprint of AGN feedback is seen to be more extreme in higher luminosity and/or higher Eddington ratio sources, as captured by the nuclear gas deficit trends  and the scaling laws of the dusty molecular tori found in this work.  We find clear evidence of molecular outflows in the sources that show the most extreme nuclear-scale gas deficits in our sample.  This suggests that  AGN feedback can be  `caught in the act' more frequently among the higher luminosity and/or higher Eddington ratio sources of the GATOS sample. In agreement with this picture, AH21 find  extended mid-IR polar emission likely associated with AGN-driven dusty outflows launched from the tori in the five GATOS sources where we detect large-scale molecular outflows and the most extreme nuclear-scale gas deficits. 
  
 \end{itemize}

  It is tempting to claim the existence of  an evolutionary link connecting the Seyferts along the sequence depicted in Fig.~\ref{cartoon}.  In this idealized scenario, a nascent AGN would start its activity cycle as a low-luminosity galaxy akin to a NUGA source. At this stage the galaxy would have a massive dusty molecular torus, which  could have been assembled following an AGN fueling episode.  During the growing phase of the activity cycle,  a wide-angle ionized wind launched from the accretion disk could impact a sizeable fraction of gas `locally' in the torus and its immediate surroundings ($\sim$1--to-tens of pc). These `local' dusty outflows could help regulate the fueling of the central engine and the growth of the  dusty molecular torus. If the geometry helps AGN winds and/or radio jets to be strongly coupled with the ISM of the host disk,  molecular outflows can propagate in the disk and clear the gas on CND scales ($\sim$hundreds  of pc) as the galaxy reaches its maximum level of nuclear activity (modulo the travel time).  AGN fueling would be therefore thwarted on nuclear scales during the phase of peak activity.  The gas reservoir in the torus would be eventually exhausted, and as a consequence, both the AGN luminosity  and the impact of AGN feedback would be reduced.  This cycle could be restarted after the completion of another AGN fueling phase.

 The misalignment between the inner molecular torus and the large-scale disk can help enhance the impact of AGN feedback on the ISM.  Several mechanisms have been described in the literature to explain the origin of this misalignment as well as the lifetime and recurrence of this particular configuration in active galaxies \citep[e.g. see][and references therein]{Com21}. However, it is still unclear if the orientation of the accretion disks and molecular tori is more easily `randomized' relative to that of the large-scale disks during the low activity phase or if these processes are expected to be at work throughout the entire activity cycle. 
 
 Furthermore, the evolutionary scenario described in Sect.~\ref{scenario} needs to be confirmed  by adding high-resolution ALMA observations of a sample of active galaxies that could significantly expand the range of AGN luminosities and/or Eddington ratios covered by the present work. We might foresee finding a more dramatic imprint of AGN feedback in higher luminosity objects. Similarly, high-resolution ALMA observations  of a comparison sample of non-AGN galaxies, properly matched in mass and Hubble type with our current sample,  would be a key to validating this picture.

\begin{acknowledgements}
       This paper
makes use of the following ALMA data: ADS/JAO.ALMA$\#$2017.1.00082.S, $\#$2018.1.00113.S, and $\#$2016.1.00254.S.
ALMA is a partnership of ESO (representing its member states), NSF (USA)
and NINS (Japan), together with NRC (Canada) and NSC and ASIAA (Taiwan),
in cooperation with the Republic of Chile. The Joint ALMA Observatory is
operated by ESO, AUI/NRAO and NAOJ. The National Radio Astronomy
Observatory is a facility of the National Science Foundation operated under cooperative
agreement by Associated Universities, Inc. 
We used observations made
with the NASA/ESA Hubble Space Telescope, and obtained from the Hubble
Legacy Archive, which is a collaboration between the Space Telescope Science
Institute (STScI/NASA), the Space Telescope European Coordinating Facility
(ST-ECF/ESA), and the Canadian Astronomy Data center (CADC/NRC/CSA).     
SGB and MQ acknowledge support from the  research project PID2019-106027GA-C44 of the Spanish Ministerio de Ciencia e Innovaci\'on. 
SGB and CRA acknowledge support from the Spanish MINECO and FEDER funding grant AYA2016-76682-C3-2-P.
CRA acknowledges the Ram\'on y Cajal Program of the Spanish Ministry of Economy and Competitiveness through project RYC-2014-15779, from the European Union's Horizon 2020 research and innovation programme under Marie Sk\l odowska-Curie grant agreement No 860744 (BiD4BESt), from the State Research Agency (AEI-MCINN) of the Spanish MCIU under grants "Feeding and feedback in active galaxies" with reference PID2019-106027GB-C42, and "Quantifying the impact of quasar feedback on galaxy evolution (QSOFEED)" with reference EUR2020-112266
AAH, SGB and AU acknowledge support through grant PGC2018-094671-B-I00 (MCIU/AEI/FEDER,UE). AAH's  work was done under project No. MDM-2017-0737 Unidad de Excelencia "Mar\'{\i}a de Maeztu"- Centro de Astrobiolog\'{\i}a (INTA-CSIC). SFH acknowledges support by the EU Horizon 2020 framework programme via the ERC Starting Grant DUST-IN-THE-WIND (ERC-2015-StG-677117). AB acknowledges support from a Royal Society International Exchange Grant.
BG-L acknowledges support from the Spanish Agencia Estatal de Investigaci\'on del Ministerio de Ciencia e Innovaci\'on (AEI-MCINN) under grant with reference PID2019-107010GB-I00.
AB has received funding from the European Research Council (ERC) under the European Union's Horizon 2020 Advanced Grant 789056 ``First Galaxies''. CR acknowledges support from the Fondecyt Iniciaci\'on grant 11190831. IGB acknowledges support from STFC through grant ST/S000488/1. AL acknowledges the support from Comunidad de Madrid through the Atracci\'on de Talento Investigador Grant 2017-T1/TIC-5213, and PID2019-106280GB-I00 (MCIU/AEI/FEDER,UE). MPS acknowledges support from the Comunidad de Madrid through the Atracci\'on de Talento Investigador Grant 2018-T1/TIC-11035 and PID2019-105423GA-I00 (MCIU/AEI/FEDER,UE). PB acknowledges financial support from the Czech Science Foundation project No. 19-05599Y. The authors thank G. P\'erez D\' iaz (SMM, IAC) for his help in designing Figure 20.
\end{acknowledgements}

\bibliographystyle{aa}
\bibliography{aa-2}

\begin{appendix}

\section{Notes on individual galaxies}\label{App1}
\smallskip
\noindent  {\em NGC~4388} -- The host galaxy of this Seyfert 1.9 is classified  as an
SA(s)b and is highly-inclined  \citep[$i=78^{\circ}$;][$i=90^{\circ}$; HyperLeda]{Vei99a}. The galaxy is part of the Virgo cluster and it is
a candidate for suffering strong ram pressure stripping as betrayed by the detection a rich complex of ionized gas that extends 
~up to 35~kpc above the disk \citep{Vei99b, Yos02}.
NGC~4388 has a prominent ionization cone/NLR
detected with {\it HST} [OIII] $\lambda$5007\AA~narrow-band imaging \citep{Sch03} 
 and  near-infrared  IFU observations of the [SiVI]1.96$\mu$m
and Br$\gamma$ emission lines \citep{Gre14}. The ionization cone  has an 
approximate $PA$ of $190^{\circ}-200^{\circ}$, extends for  $\sim 1\,$kpc and  
presents a large opening angle ($\sim 95^{\circ}-105^{\circ}$). In the optical the northeast 
side of the ionization cone is mostly obscured by the host galaxy
whereas it appears nearly symmetric in the near-IR, as expected if the north side  of the galaxy is the near side.
 We see the south ionization cone coming  towards us and the north cone directed away from us.
Radio emission extends out of the galaxy plane and consists of
 two point-like sources and a diffuse lobe located north \citep{Sto88, Hum91}.
The northern knot coincides with the nucleus and the southern knot coincides with a bright region
of ionized gas seen in the [OIII] line, which suggests a jet--ISM
interaction \citep{Fal98, Rod17}. The column densities toward the central engine 
 of NGC~4388 derived from X-ray observations show significant variations: $N(H)\sim (2.1-6.5) \times10^{23}$cm$^{-2}$ \citep{Yi21}. The most recent 
 determinations give $N(H)\sim(2.6-2.7)\times10^{23}$cm$^{-2}$ \citep{Mil19}. 

\smallskip

\noindent  {\em NGC~4941} -- This nearly face-on ($i= 37^{\circ}$; HyperLeda), ring galaxy
(morphological type  (R)SAB(r)ab) is classified as a Seyfert 2 and has,
together with NGC~7465, the lowest 14-150\,keV luminosity in our
sample. Recent VLT/MUSE optical IFU observations showed that the inner 2\,kpc
are dominated by AGN/shock emission in a spiral-like structure
 \citep{Err19}.  Using optical IFU [SIII]~$\lambda$9069\AA~observations,  \citet{Bar09} proposed the presence of  a compact outflow
associated with the emission of a radio jet. The gas emission is enhanced in the regions
where significant radio emission is detected. The average orientation of the
outflow is  $PA\sim-40^{\circ}$ to $-60^{\circ}$, which roughly coincides with the direction of the
large-scale high excitation emission detected with MUSE.  The nuclear
radio emission appears to be partially resolved, and presents
a double source morphology extended by $\sim$15~pc along
$PA\sim-25^{\circ}$ \citep{The00, Sch01}.  

\smallskip

\noindent  {\em NGC~5506 }-- This is a highly-inclined  ($i\sim90^{\circ}$, HyperLeda) galaxy (morphological
type Sa peculiar) within the Virgo supercluster classified as a Seyfert 1.9. {\it HST} narrow-band
[OIII]$\lambda$5007\AA \, imaging shows a one-sided ionization cone/NLR
perpendicular to the disk of the host galaxy at an approximate
$PA=15^{\circ}-20^{\circ}$ and a 5$\arcsec$ extension to the northeast \citep{Fis13}. The $V-H$ colour map also reveals 
a blue fan-shaped structure to the N of the nucleus \citep[][see also this work]{Mar03}.
The northern [OIII] ionization cone
is  inclined slightly towards our line-of-sight  and avoids the high extinction of the galaxy disk. Both sides of 
the ionization cone appear well defined in the velocity dispersion maps
of the near-IR [FeII]$\lambda$1.257$\mu$m and Br$\gamma$ emission lines
 \citep{Sch19}. From the modeling of the optical NLR kinematics,
\citet{Fis13} derived an outer semi-opening angle of the cone of $40^{\circ}-45^{\circ}$. 
The VLA 4.9~GHz map of \citet{Col96b} shows that besides the central source there is elongated radio 
emission out to several kpc from the nucleus in the direction of the minor axis. Overall, there is multiple
observational evidence implying the presence of
a minor-axis wind. The nuclear X-ray source of NGC~5506 is absorbed by  a column density $N(H)\sim3\times10^{22}$cm$^{-2}$, i.e. in the 
Compton-thin regime. However, the obscuring matter covering the nucleus has an average  integrated column 
density about two orders of magnitude larger \citep{Sun18}.

\smallskip

\noindent  {\em NGC~5643} -- The host of this Seyfert 2 is a spiral galaxy with a
 morphological type of SAB(rs)c, seen close to a face-on  orientation
 ($i=30^{\circ}$, HyperLeda). \citet{Jun97}  and  \citet{Mul97} identified the presence of a
large-scale stellar bar in the near-IR with a diameter $D\sim67\arcsec$(5.6~kpc) and oriented along $PA\sim85^{\circ}$.  A large ionization cone/NLR/outflow
 is identified using  [OIII]$\lambda$5007\AA \, \citep{Sim97, Cre15, Min19, Gar21}.  
 VLA radio observations show emission corresponding to a radio jet  of this radio-quiet galaxy \citep{Mor85, Lei06}.The outflow and radio
jet extend for approximately 2\,kpc  close to the east-west direction ($PA\sim80^{\circ}-85^{\circ}$). The NLR modelling of
\citet{Fis13} inferred an outer semi-opening angle of $50^{\circ}-55^{\circ}$ for
the ionization cone.  \citet{Dav14} found evidence of non-circular motions  in their study of the kinematics of hot molecular gas traced by the H$_{\rm 2}$ 1-0 S(1) line in a region located $\sim2\arcsec$ northeast of the nucleus. NGC~5643 is classified as a compton-thick galaxy with an obscuring column density $N(H)\sim(2.5)\times10^{25}$cm$^{-2}$ \citep{Gua04, Ann15, Ric15}. It has a moderate X-ray luminosity ($L_{\rm 2-10keV}\sim10^{42}$erg~s$^{-1}$). 
\citet{Ram16} did not detect optical broad lines in polarized light  in NGC~5643. This non-detection  has been explained by \citet{Ram16} as due to  
its Compton-thick nature and  low AGN luminosity or, alternatively, as due to the different properties of the material responsible for the scattering.

\smallskip

\noindent  {\em NGC~6300} -- This ring barred spiral galaxy (morphological type SB(rs)b) is observed
at an intermediate inclination ($i=53^{\circ}$, HyperLeda) and is classified as a
Seyfert 2. \citet{Ram16} detected a broad H$\alpha$ component identifying it as hidden BLR candidate.
The large-scale stellar bar detected in the Ks band by \citet{Gas19} has a projected length $\sim1\arcmin$ and is oriented
along $PA\sim46^{\circ}$.  \citet{Mul97} discussed the presence of a secondary nuclear bar at scales $\leq4\arcsec$.
Two spiral dust lanes connect inside the bar the outer ring with the nucleus.
 \citet{Gas19}  found evidence of the presence of large amounts of hot dust  in the nuclear region itself but also in a circumnuclear disk up 
 to $r\sim27$~ pc, and attributed to this compact disk part of the obscuration in this Seyfert 2 nucleus.
 \citet{Dav14} modelled the kinematics of the near-IR
H$_2$ line at $2.12\,\mu$m and inferred the presence of an edge-on
large opening angle outflow ($\sim90^{\circ}-100^{\circ}$) oriented in an approximate north-south direction ($PA\sim
15^{\circ}-20^{\circ}$)  superimposed on a rotating disk.
Using optical IFU observations,  \citet{Sch16} 
observed that this biconical structure extends out to approximately
20$\arcsec$ ($\sim 1.6\,$kpc) from  the AGN. It presents enhanced
[NII]/H$\alpha$ ratios and relatively high  velocity dispersion when 
compared to the HII regions observed in the 
circumnuclear ring. Both the obscuration of the nuclear region and the intrinsic AGN luminosity of NGC~6300 were found to evolve over the years 
based on X-ray observations: $N(H)\sim(1.1-5.8)\times10^{23}$cm$^{-2}$ and $L_{\rm bol}\sim(0.7-3.8)\times10^{42}$erg~s$^{-1}$  \citep{Jan20}.

\smallskip

\noindent  {\em NGC~6814} -- This galaxy at moderate inclination ($i=52^{\circ}$, HyperLeda) has a morphological type
of SAB(rs)bc,  although a large-scale stellar bar is clearly observed in the near-IR. The bar has  a diameter of 24$\arcsec$ and it extends along $PA=21-25^{\circ}$
 \citep{Mul97, Mar99}.  The {\it HST} sharp-divided image of NGC 6814 published by \citet{Marq03}  shows a two-armed spiral  structure  inside $r<1\arcsec$=120~pc. This dusty feature
  transforms into a multi-arm filamentary feature at larger radii up to $r\sim4-5\arcsec$=500~pc. 
  On smaller scales, the nucleus is absorbed by a dusty gas disk that lies along $PA\sim45^{\circ}$ within $r<0.5\arcsec$=60~pc.  
Spectroscopically the galaxy is classified as a Seyfert 1.5.  Using {\it HST} narrow-band [OIII]$\lambda$5007\AA,  \citet{Sch96}
detected a compact core and extended emission along $PA\sim150^{\circ}$ over a total extent of $1\arcsec$. 
The radio emission imaged by \citet{Ulv84}  shows a compact core and an elongation up to $2\arcsec$ to the W of the nucleus at $PA\sim105^{\circ}$, 
which has no counterpart in the [OIII] image. \citet{Mue11} modelled the kinematics of the near-IR Br$\gamma$ and 
[SiVI]1.96$\mu$m emission lines with a rotating disk oriented along $PA\sim145^{\circ}$ and a biconical
outflow oriented at $PA\sim30^{\circ}-35^{\circ}$ with an opening angle $\sim100^{\circ}-120^{\circ}$. This Seyfert 1.5 galaxy shows a X-ray variability of a factor of 10 over time-scales of years
 \citep{Muk03}. \citet{Tor18} estimated an X-ray obscuring column density towards the nucleus $N(H)\sim 3.5\times10^{23}$cm$^{-2}$.

\smallskip

\noindent  {\em NGC~7213} -- This nearly face-on ($i=39^{\circ}$, HyperLeda) SA(s)a galaxy is classified
as a Seyfert 1.5 galaxy. The {\it HST} [OIII]$\lambda$5007\AA \,
emission has a relatively compact halo-like emission extending for
approximately  $1\arcsec$  \citep{Sch03}. However, there is evidence of an extension of the [OIII] emission 
around $PA=-15^{\circ}$ to $-40^{\circ}$.  Using optical IFU
observations, \citet{Sch14} found evidence of
inflowing gas along a nuclear two-arm spiral structure that extends  from the nucleus out to $r\sim4\arcsec$ (440~pc)
to the NW and SE. This spiral structure seen in ionized gas is roughly co-spatial with a dusty spiral  feature revealed by the structure map of the
galaxy published by \citet{Sch14}. The line of nodes is derived from the stellar kinematics is oriented along  $PA\sim356^{\circ}$ \citep{Sch14}.
At centimeter and millimeter wavelengths  the  galaxy shows a compact and mostly unresolved structure  in the radio continuum \citep{Bra98, The00, Bla05, Murphy10, Sal20}. 
\citet{Bra98} interpreted the radio emission at 3~cm  as  either  nuclear  synchrotron  or  free-free  emission.  The  radio  power of 
the galaxy, $\sim 3\times10^{29}$~erg~s$^{-1}$Hz$^{-1}$ at 20~cm, is a factor of 10 higher than usually found in radio-quiet Seyfert galaxies.
Among the galaxies in our sample, it shows
the lowest X-ray column density and it is a relatively faint and unobscured AGN with evidence of significant variability \citep{Yan18}.
\citet{Rus14} fitted a clumpy torus model to the MIR SED finding a low $i=21^{\circ}$ and a very low gas column density $N(H)\sim10^{18}$~cm$^{-2}$,
which seems to correspond to its  Seyfert~1.5 type classification.

\smallskip

\noindent  {\em NGC~7314} --  This highly inclined galaxy ($i=70^{\circ}$, HyperLeda) is morphologically
classified as an SAB(rs)bc and contains a Sy1.9 nucleus.  \citet{The00} detected a double radio structure oriented in an almost
north-south direction and extending for approximately $4\arcsec$ (350~pc). An {\it HST}/STIS  acquisition image at 7230\AA \, shows a bright compact
nuclear source and a dust lane  east of the AGN and also oriented
north-south \citep{Hug03}. NGC 7314 experiences extreme variability events in X-ray flux \citep{Yaq96, Ebr11, Emm16}. \citet{Ebr11} found evidence that the source is 
seen through a warm absorber situated within a clumpy torus. The reported 
changes in the neutral column density were interpreted by \citet{Ebr11} as due to the crossing of 
a cloud of neutral gas that is grazing the edge of the torus.
 Da Silva et al. (in prep) detected a wide opening angle ($\sim90^{\circ}-100^{\circ}$) [OIII]  structure that extends mostly west of the nucleus, interpreted 
 as the signature of an ionized outflow with a mean axis $PA_{\rm out}\sim100^{\circ}-120^{\circ}$.

\smallskip

\noindent  {\em NGC~7465 }-- This relatively inclined ($i=64^{\circ}$, HyperLeda)  barred galaxy
(morphological type (R')SB(s)0) hosts a Seyfert 2 nucleus.  However, \citet{Ram09} classified the galaxy as a type-1 LINER
through its near-IR spectrum, which shows broad Pa$\beta$ and Br$\gamma$.
The galaxy belongs to
the NGC~7448 group and it forms
an interacting pair with NGC~7464. The optical continuum {\it HST} images  of this galaxy, published by \citet{Fer00}, show an
inverted S-shaped morphology on large scales and tend
to a photometric major axis along $PA=120^{\circ}$ in the inner 4$\arcsec$ (540 pc).
The morphology of the central regions is heavily obscured by a set of dust lanes 
that are oriented  perpendicular to the photometric major axis of the galaxy.
 {\it HST} narrow-band [OIII]$\lambda$5007\AA \, imaging
reveals extended emission (approximately $4\arcsec$) oriented
at $PA\sim100^{\circ}-120^{\circ}$ \citep{Fer00}. The [OIII] morphology resembles a
bi-cone  with the brighter emission coming from the northwest of the
nucleus with an opening angle $\sim90^{\circ}-100^{\circ}$. \citet{Kra11} determined a kinematic major axis along $PA\sim167^{\circ}$, based on 
 the stellar kinematics of the galaxy. They also found evidence of a an inner decoupled kinematic core
 with a $PA$ tilted by $\sim100^{\circ}$ relative to the large-scale disk.
 \citet{Nyl16} detected spatially resolved radio continuum emission at 5~GHz stemming from 
 an elongated  source of size $0.4\arcsec\times0.3\arcsec$ at $PA\sim20^{\circ}$. 
 \citet{Bau13} detected NGC~7465 in hard X-rays with Swift. 
 
\smallskip

\noindent  {\em NGC~7582} --  This is a highly inclined ($i=68^{\circ}$, HyperLeda) barred galaxy
(morphological type (R')SB)(s)ab) optically classified as a Seyfert 2
with a bright circumnuclear ring of star formation. It shows a
spectacular ionization cone of wide opening angle ($\sim90^{\circ}-100^{\circ}$), which was first identified at optical wavelengths
\citep{Mor85, Sto91, Dav16} with an
approximate  $PA=235^{\circ}-245^{\circ}$. The southwest part of the cone shows the
brightest emission whereas the northeast part appears to be obscured
by the host galaxy, at least in the optical. 
The kinematics of the [OIII] $\lambda$5007\AA~ line differs significantly from the kinematics of H$\alpha$ \citep{RicT18}. 
The structure seen in the radial velocity map of  [OIII] has a $PA\sim230^{\circ}$, with  the blueshifted structure being associated with the inner section of the ionization cone.
\citet{Rif09} analyzed the kinematics of the hot molecular gas, traced by the NIR H$_{\rm 2}$ line emission and found little evidence of outflowing motions for this component.
Soft X-ray observations also revealed the 
ionization cone of the galaxy \citep{Bia07, Bra17}.
NGC 7582 is a X-ray bright source with a significant obscuration: $N(H)\sim10^{23-24}$~cm$^{-2}$. The source shows
 very rapid  changes of the column density of an inner absorber, but it nevertheless requires the existence of additional constant components \citep{Bia09}. It behaves like a changing-look AGN with 
highly variable absorption and strong reflection features \citep{Riv15, Bra17}.
The radio emission of this galaxy at 8.4~GHz appears to be extended over
several arcseconds \citep{The00} and is likely spatially
coincident with the ring of star formation with no clear association with the AGN source.

\section{{\it HST} observations}\label{App2}

Table~\ref{tab_hst} lists the details of the {\it HST} observations used in this paper. We list the instruments, filters, exposure times, and pixel sizes of the observations, as well as 
the proposal identifications and their principal investigators. 

\begin{table*}[bth!]
\centering

\resizebox{0.7\textwidth}{!}{
\begin{tabular}{lcccccccc}
\hline
\hline
ID & Instrument	& Filter & Exptime (s) &	Proposal   &    PI & DATEOBS & Provenance & Pixel size (arcsec) \\
\hline
NGC1068  &    NICMOS/NIC2   	&   F160W  &   384   &    7215	 &  Thompson  &     1998-02-21     &   MAST/CALNIC 	& 0.075	 \\
	 &    WFPC2/PC 		&   F547M  &   440   &    5754	 &  Ford      &     1995-01-17     &   HLA 		& 0.05	 \\
NGC3227  &    NICMOS/NIC2   	&   F160W  &   192   &    7172	 &  Rieke     &     1998-04-06     &   MAST/CALNIC	& 0.075  \\	  
	 &    WFPC2/PC 		&   F606W  &   500   &    5479	 &  Malkan    &     1995-02-23     &   HLA 		& 0.05	 \\
NGC4388  &    WFC3     		&   F160W  &   422   &    12185  &  Greene    &     2011-06-08     &   HLA		& 0.09   \\	  
	 &    WFPC2/PC 		&   F606W  &   560   &    8597	 &  Regan     &     2001-02-08     &   HLA 		& 0.05	 \\
NGC4941  &    WFC3     		&   F160W  &   597   &    15133  &  Erwin     &     2018-02-24     &   MAST/CALWF3	& 0.128	 \\	  
	 &    WFPC2/PC 		&   F606W  &   560   &    8597	 &  Regan     &     2001-06-02     &   HLA 		& 0.05	 \\
NGC5506  &    NICMOS/NIC2   	&   F160W  &   640   &    7330	 &  Mulchaey  &     1998-04-25     &   MAST/CALNIC	& 0.075	 \\	  
	 &    WFPC2/PC 		&   F606W  &   500   &    5479	 &  Malkan    &     1994-07-21     &   HLA 		& 0.05	 \\
NGC5643  &    WFC3     		&   F160W  &   1006  &    15145  &  Riess     &     2018-01-16     &   MAST/CALWF3	& 0.128	 \\	  
	 &    WFPC2/PC 		&   F606W  &   560   &    8597	 &  Regan     &     2001-05-29     &   HLA 		& 0.05	 \\
NGC6300  &    NICMOS/NIC2   	&   F160W  &   640   &    7330	 &  Mulchaey  &     1998-05-06     &   MAST/CALNIC   	& 0.075	 \\
	 &    WFPC2/PC 		&   F606W  &   500   &    5479	 &  Malkan    &     1995-04-02     &   HLA 		& 0.05	 \\
NGC6814  &    WFC3     		&   F160W  &   5041  &    12961  &  Bentz     &     2013-08-14     &   HLA 		& 0.09 	 \\
	 &    WFPC2/PC 		&   F606W  &   500   &    5479	 &  Malkan    &     1994-05-18     &   HLA 		& 0.05	 \\
NGC7213  &    WFC3     		&   F160W  &   1612  &    15181  &  Rosario   &     2018-05-20     &   MAST/CALWF3 	& 0.128	 \\
	 &    WFPC2/PC 		&   F606W  &   500   &    5479	 &  Malkan    &     1994-10-05     &   HLA 		& 0.05	 \\
NGC7314  &    WFPC2/PC 		&   F450W  &   460   &    9042	 &  Smartt    &     2001-07-03     &   HLA 		& 0.05	 \\
	 &    WFPC2/PC 		&   F606W  &   500   &    5479	 &  Malkan    &     1994-08-24     &   HLA 		& 0.05	 \\
NGC7465  &    NICMOS/NIC2   	&   F160W  &   1152  &    11219  &  Capetti   &     2007-08-17     &   HLA 		& 0.05	 \\
	 &    WFPC2/PC 		&   F606W  &   500   &    5479	 &  Malkan    &     1994-11-30     &   HLA 		& 0.05	 \\
NGC7582  &    NICMOS/NIC2   	&   F160W  &   640   &    7330	 &  Mulchaey  &     1997-09-16     &   HLA 		& 0.05	 \\  
	 &    WFPC2/PC 		&   F606W  &   560   &    8597	 &  Regan     &     2001-07-24     &   HLA 		& 0.05	 \\
\hline	   					 			    					 			      
\end{tabular}}	
\caption{Details of the {\it HST} observations.}
\label{tab_hst}
\end{table*}

\section{{\tt NuSTAR} X-ray observations}\label{App3}

In Table~\ref{Tab6} we list the details of the  {\tt NuSTAR} X-ray observations used in this paper, as well as the main parameters obtained in the fitting procedure described in Sect.~\ref{reflection}. 
 
\begin{table*}[tbh!]
\caption{Parameters of the fit to   {\tt NuSTAR} X-ray observations.}
\centering
\resizebox{.9\textwidth}{!}{ 
\begin{tabular}{lccccccc} 
\hline
\hline
\noalign{\smallskip}
    source	   	&	OBSID	 	&	exposure time		&	log$N_{\rm H}^{\rm LOS}$		&    log$N_{\rm H}$(X$_{\rm ref}$)				&			f$_{\rm cov}$	&	chi$^2$/dof		&	log$N_{\rm H}^{\rm linked}$ \\
\noalign{\smallskip}   
 \hline  
 \noalign{\smallskip}   
NGC1068 	&	60002030002  	&	57.8			&	24.75 (24.72 - 24.76) 	&	23.31 (23.22 - 23.35) 	&	1* 				&	570.8/488   	&			\\
NGC3227 	&	60202002002  	&	49.8			&	23.16  (22.88 - 23.33) 	&	24.24 (24.15 - 24.33 ) 	&	0.30 (0.26 - 0.37) 	&	1063.7/995  	&			\\
NGC4388 	&	60061228002  	&	21.4			&	23.60 (23.54 - 23.64) 	&	23.52 (23.37 - 23.64) 	&	0.88 (0.87 - 0.90) 	&	403.3/375   	&	23.57 (23.54 - 23.61)		\\
NGC4941 	&	60061236002  	&	20.7			&	23.62 (23.49 - 23.73) 	&	24.15 (24.06 - 24.22) 	&	1* 				&	67.5/57   		&	23.92 (23.90 - 23.95)		\\
NGC5506 	&	60061323002  	&	56.4			&	23.18 (23.11 - 23.25) 	&	24.72 (24.65 - 24.90)		&	0.52 (0.50 - 0.55) 	&	1269.9/1196	&			\\
NGC5643 	&	60061362002  	&	22.4			&	24.3                	        		&	> 24.4 			  	&	< 0.78 			&	81.3/67   		&	24.53 (24.41 - 24.92)		\\
NGC6300 	&	60261001004  	&	23.5			&	23.37 (23.33 - 23.42) 	&	24.53 (24.43 - 24.68) 	&	0.91 (0.90 - 0.93) 	&	533.2/644		&			\\
NGC6814 	&	60201028002  	&     148.1			&	24.33 (24.30 - 24.38) 	&	23.23 (23.21 - 23.31) 	&	0.19 (0.18 - 0.20) 	&	1418.8/1334	&			\\
NGC7213 	&	60001031002  	&     101.5			&	 < 22.1                		&	22.23 (22.08 - 22.34) 	&	> 0.75 			&	854.2/836  	&	22.08 (22.04 -22.11)		\\
NGC7314 	&	60201031002  	&     100.4			&	23.29 (23.18 - 23.37)		&	24.54 (24.47 - 24.60) 	&	0.31 (0.28 - 0.34) 	&	1179.6/1129  	&			\\
NGC7582 	&	60201003002  	&	48.3			&	23.53 (23.52 - 23.55) 	&	24.39 (24.32 - 24.45) 	&	0.95 (0.94 - 0.96) 	&	934.5/889		&			\\
  \noalign{\smallskip} 
\hline 
\hline
\end{tabular}}
\tablefoot{We list the identification of the  {\tt NuSTAR} X-ray observations, their exposure time, and  the  $\rm{\chi^2}$ statistics of the best-fits  together with their resulting parameters:  $N_{\rm H}^{\rm LOS}$, $N_{\rm H}$(X$_{\rm ref}$), $f_{\rm cov}$, $N_{\rm H}^{\rm linked}$, and associated errors as 1$\sigma$ deviations. Asterisks are shown when the parameter was frozen to a certain value along the fit.} \label{Tab6} 
\end{table*}

\section{Fits of mean-velocity fields}\label{vels} \label{kinemetry}

Figure~\ref{COvels-full} shows the mean-velocity fields derived from the CO(3--2)  line  ($<v_{\rm CO}>$) in the GATOS core sample galaxies inside the 17$\arcsec$ ALMA FOV. To maximize the reliability of the maps, isovelocities have been obtained by integrating the emission with a threshold of 3$\sigma$. Figure~\ref{CO-vels} zooms in  on the central  $\Delta x\times\Delta y$=100~pc~$\times$~100~pc regions to show the overlay of the velocity-integrated emission on $<v_{\rm CO}>$ for  the core sample of GATOS galaxies as well as NGC~1068 and NGC~3227\citep{GB19, Alo19}. 

The velocity field of the gas shows the expected rotation pattern on $\sim$~kpc scales (Fig.~\ref{COvels-full}) but also $\sim$50~pc from the AGN (Fig.~\ref{CO-vels}) in all the galaxies. The presence of non-circular motions, which distort the rotation pattern, is nevertheless identified across a range of spatial scales for a significant fraction of the sources. Some of the distortions in the large-scale velocity fields visualized in Fig.~\ref{COvels-full} can be easily associated with gas inflow motions driven by bars and spiral arms, which have left  their imprint in the disks of NGC~4941, NGC~5643, NGC~6300, NGC~7314, NGC~7465 and NGC~7582. However, other distortions of the velocity fields identified closer to the nucleus in NGC~4388, NGC~5506, NGC~5643, NGC~3227, and NGC~1068 are rather indicative of outflow motions \footnote{The outflows of NGC~5643, NGC~3227, and NGC~1068 have already been amply studied in previous works referred to in Sect.~\ref{gas-radial}.}. We use below the software package {\tt kinemetry} developed by \citet{Kra06} to fit the CO(3--2) mean-velocity fields of Fig.~\ref{COvels-full} using an iterative process and a minimum number of free parameters. Our objective is to identify the signature of outflows that can be launched by AGN winds and/or radio jets, which are present in a fraction of our sources.

\subsection{Kinemetry fits}

The general description of the two-dimensional line-of-sight velocity field of a galaxy disk can be expressed as:
\begin{equation}
v_{\rm los} (x,y) = v_{\rm sys} + v_\theta (x,y) \cos\psi\sin i + v_R (x,y) \sin\psi\sin i\;,
\label{eq-1}
\end{equation}
where $(v_R,v_\theta)$ is the velocity in polar
coordinates, $\psi$ is the phase angle measured in the galaxy plane from the receding side
of the line of nodes, and $i$ is the inclination angle restricted to the range $0<i<\pi/2$.

Alternatively, the observed mean-velocity field of a galaxy disk, $v_{\rm mean}(r, \psi)$,  expressed as a function of the radius ($r$) and the phase angle ($\psi$, measured from the receding side
of the line of nodes), can be divided into a number of elliptical ring profiles with a geometry defined by the position angle and inclination ($PA$, $i$). We can decompose $v_{\rm mean}(r, \psi)$ as a Fourier series with harmonic coefficients $c_j(r)$ and $s_j(r)$, where 
\begin{equation}
v_{\rm mean} (r, \psi)= c_0 + \sum_{j=1}^n [c_j(r) \cos (j\psi) + s_j(r) \sin (j\psi)]
\label{eq-2}
.\end{equation}

The term $c_1$ accounts for  the contribution of circular rotation and the other terms of the series contain the contributions of noncircular 
motions \citep{Sch97, Sch99}. Expanding the series of Eq.~\ref{eq-2} out to $n=3$ 
provides a fairly complete description of $v_{\rm los}$ in most models \citep{Tra08}. In a disk with axisymmetric circular rotation 
($v_c$), $c_0=v_{\rm sys}$ and $c_1=v_c \sin i$ and all remaining terms can be neglected. In the case
of a pure axisymmetric radial flow ($v_R$), $c_0=v_{\rm sys}$ and $s_1=v_R \sin i$, with the rest of coefficients being 0.

\begin{figure*}
   \centering
    \includegraphics[width=1\textwidth]{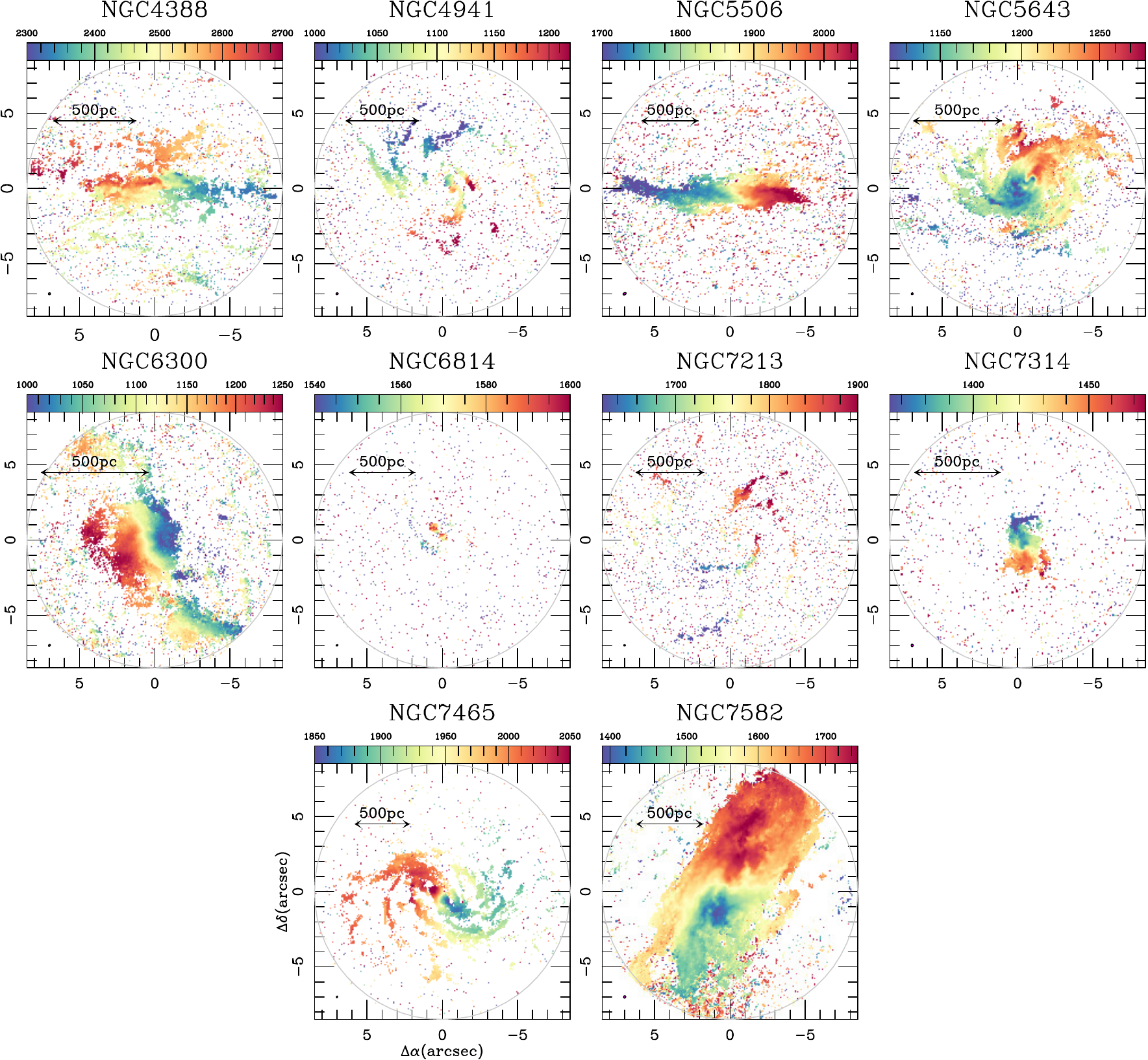}
    \caption{Same as Fig.~\ref{COmaps-full} but showing the mean-velocity field  of the 3--2 line of CO ($<v_{\rm CO}>$ in km~s$^{-1}$-units) inside the 17$\arcsec$ ALMA field-of-view for the galaxies of the core sample of GATOS. Mean-velocities have been derived using a 3$\sigma$ clipping on the emission.}      
 \label{COvels-full}
   \end{figure*}

\begin{figure*}
   \centering
    \includegraphics[width=0.92\textwidth]{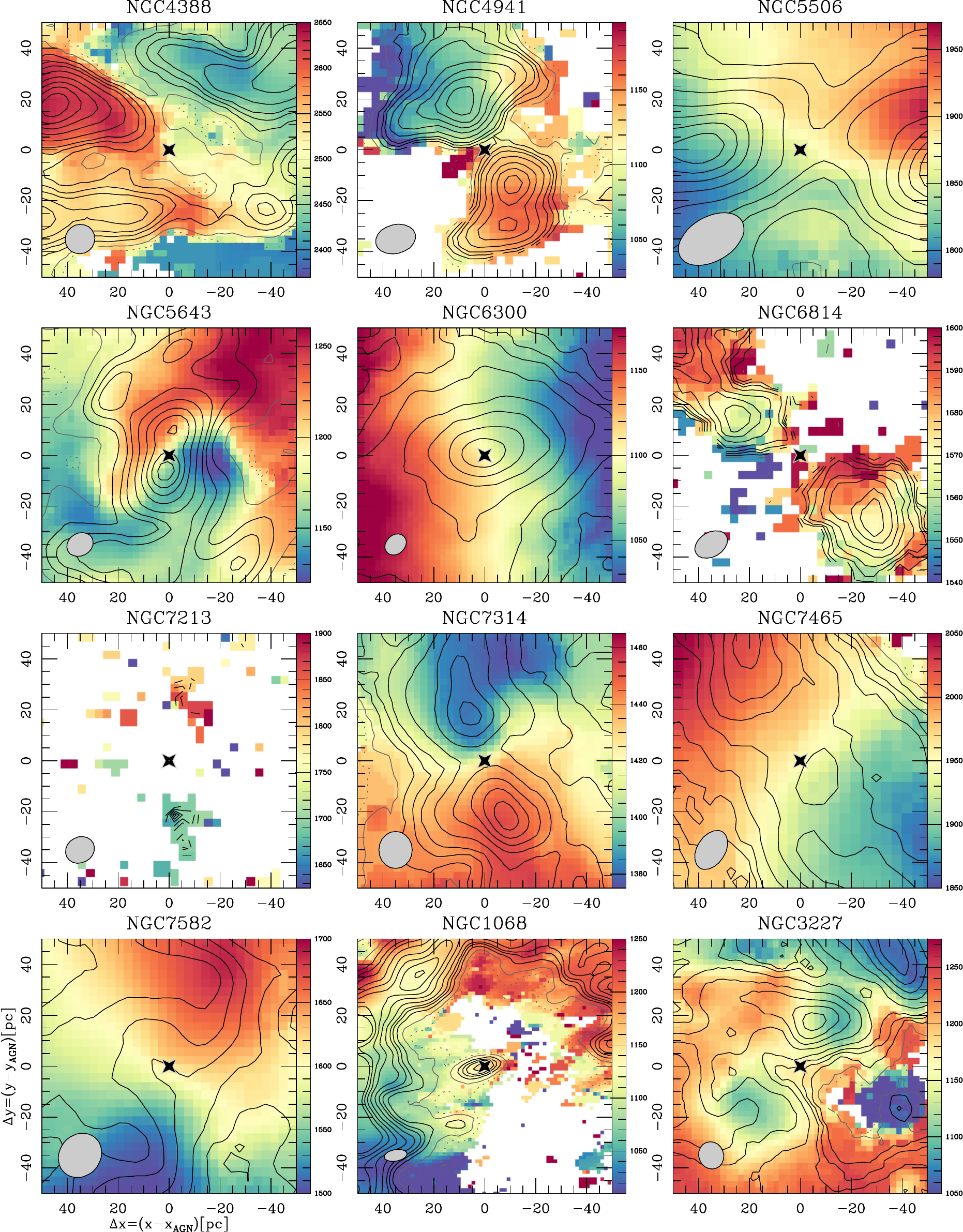}
   \caption{Overlay of the velocity-integrated emission ($I_{\rm CO}$; contours) on the mean-velocity field ($<v_{\rm CO}>$; linear color scale in km~s$^{-1}$-units) of molecular gas derived from the 3--2 line of CO  for  the central  
  $\Delta x\times\Delta y$=100~pc~$\times$~100~pc regions of the core sample of GATOS galaxies. We also include the images of 
   NGC~1068 \citep{GB19} and NGC~3227 \citep{Alo19}. We used a 3$\sigma$ clipping on the MSR data to simultaneously derive  $I_{\rm CO}$ and $<v_{\rm CO}>$. Contours: 2.5$\%$ (dashed gray), 5$\%$ (gray),
    10$\%$, 15$\%$ 20$\%$, 30$\%$ to 90$\%$ of $I_{\rm CO}^{max}$ in steps of 15$\%$ of $I_{\rm CO}^{max}$ (black). Symbols as in Fig.~\ref{cont-CO}.}  
   \label{CO-vels}
    \end{figure*}

We used the software package {\tt kinemetry}  to fit the CO(3--2) mean-velocity fields of Fig.~\ref{COvels-full}, following the approach adopted by \citet{GB14} and \citet{GB19} to fit the velocity field of NGC~1068. We used in the fit  32 ellipses with semi major axes covering the disks from $r=0\farcs25$ to $r=8\farcs25$, with 
 a spacing $\Delta r=0\farcs25$.  At each step, {\tt kinemetry} finds the best fitting ellipses by 
 maximizing the contribution of circular motions and by minimizing the contribution of noncircular motions ($v_{\rm nonc}$), evaluated as:
  
 \begin{equation}
v_{\rm nonc} (r)= \sqrt{s_1^2(r)+s_2^2(r)+c_2(r)^2+s_3^2(r)+c_3^2(r)}  
\label{eq-3}
\end{equation}
 
 By construction  {\tt kinemetry}  provides a good estimate of the $c_1$ term.

We  assumed that the dynamical centers for our sources coincide with the AGN positions derived in Sect.~\ref{continuum-fits}. 
In a first step we left the position angles $PA(r)$ and the inclinations $i(r)$  as free parameters in the fit. From this first iteration we obtained  the best fit for the systemic velocities. We then subtract $v_{\rm sys}$ from 
$v_{\rm mean}$ and re-derive the Fourier decompositions of Eq.~\ref{eq-1}. We then derived the average values of $PA(r)$ and $i(r)$:   
 $\langle PA \rangle$ and $\langle i \rangle$. In a second step, we fixed  $i$ to $\langle i \rangle$ and re-determined the  
 $PA(r)$ profile. From this second iteration, we derived $\langle PA \rangle$. Finally, we derived the Fourier decomposition of the 
 velocity  fields fixing  $PA$ and $i$ at all radii to the mean values derived in the two previous iterations.

The approach followed in the present analysis implicitly assumes that the gas kinematics  in our sources can be modeled roughly by orbits in thin disks as a way to identify deviations in the velocity fields that can be attributed to mostly coplanar molecular inflows or outflows.  Although this approach might appear simplistic at first sight, we note that there is mounting observational evidence suggesting that most of the molecular outflows launched in nearby AGN lie in a geometry configuration  close to coplanar with the host galaxy disk \citep{GB14,Dav14,Mor15,Alo18,GB19,Alo19,Gar21}.  More realistic 3D scenarios including vertical components for the velocity fields will nevertheless be explored in paper III for the GATOS core sample galaxies.
 
\begin{figure*}
   \centering
    \includegraphics[width=0.49\textwidth]{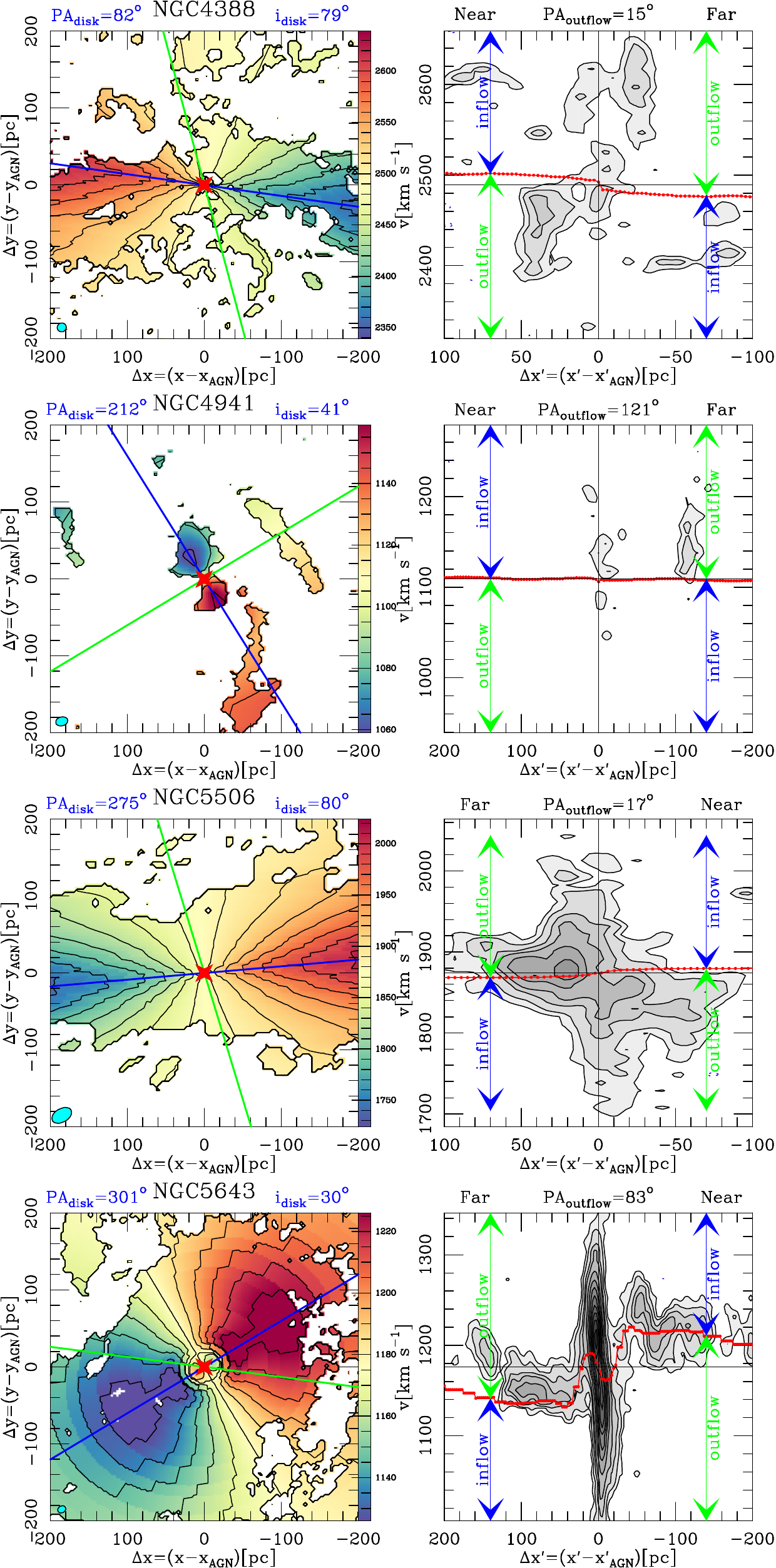}
     \includegraphics[width=0.49\textwidth]{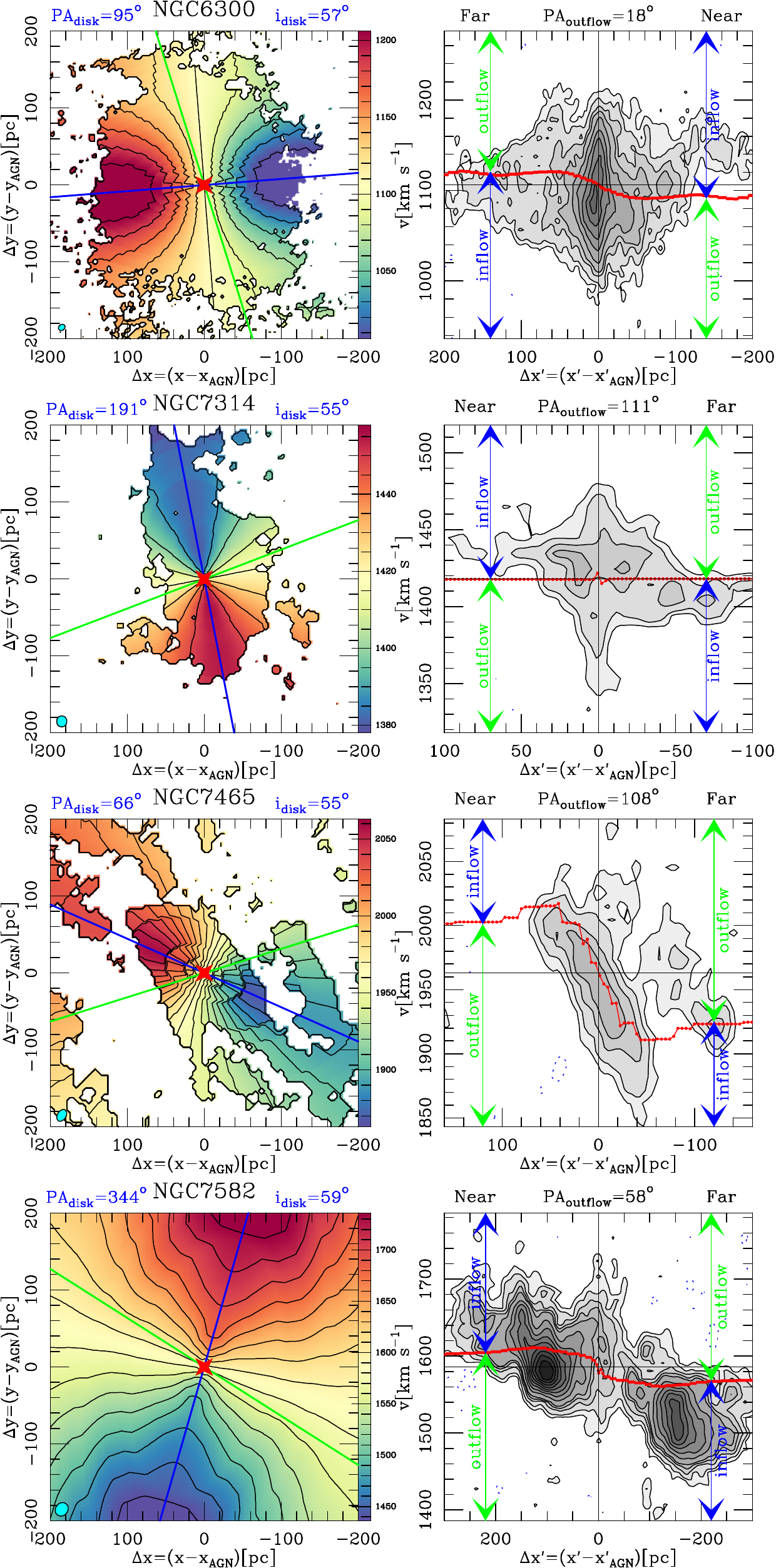}    
   \caption{Rotational motions inside the central $\Delta x\times\Delta y$=400~pc~$\times$~400~pc regions of eight galaxies of the core sample of galaxies, derived from the fit  to the  CO velocity fields obtained by {\tt kinemetry}, as described in Appendix~\ref{kinemetry}. We indicate the best-fit values estimated for  the  position angle (PA$_{\rm disk}$; blue line) and the inclination of the disk (i$_{\rm disk}$). The green line shows the orientation of the outflow axes. The CO(3--2) position-velocity (p-v) diagrams derived along the outflow  axes are  shown on the right hand side beside the velocity field panels for each galaxy.  The p-v diagrams are centered around the AGN locus and $v_{sys}$.  We identify the near and far sides on the upper x' axes of the outflow p-v diagrams, based on the morphology of the $V-H$ color maps of Fig~\ref{CO-colors} and on the assumption 
  that the spiral structures seen in some galaxies are trailing. Contour spacing: -2.5$\sigma$ (dashed blue), 2.5$\sigma$, 5$\sigma$, 10$\sigma$, 15$\sigma$ to the peak intensity in steps of 10$\sigma$. The red curves show the  velocities along the outflow axes that can be attributed to rotational motions derived from the model. We identify in the p-v diagrams the regions corresponding to coplanar inflow or outflow radial motions.}  
   \label{kinemetry-pvs}
    \end{figure*}

Figure~\ref{kinemetry-pvs}  shows the rotational motions inside the central $\Delta x\times\Delta y$=400~pc~$\times$~400~pc regions derived from the fit  to the  CO velocity fields obtained by {\tt kinemetry} in eight of the GATOS core sample galaxies. These targets are purposely selected based on the availability of good estimates for the orientation of their AGN wind outflows. Based on the  {\tt kinemetry} fit, we can therefore estimate the velocity gradients that we can attribute to purely rotational motions predicted by the model. We overlay in Fig.~\ref{kinemetry-pvs} the  expected velocities due to rotational motions predicted by the model  on the CO(3--2) position-velocity (p-v) diagrams taken along the (projected) direction of the ionized outflow axes. In the majority of the sources shown in Fig.~\ref{kinemetry-pvs} the projected outflow axes lie at a large angle relative to the kinematic major axes of the disks derived by the model. The contribution of tangential components of both circular and non-circular motions in the p-v plots of Fig.~\ref{kinemetry-pvs} is therefore expected to be totally negligible. Under the hypothesis that molecular gas lies mostly in a coplanar geometry, and taking into account the known orientation of the galaxy disks relative to the plane of the sky (as identified on each side of the p-v plots), we can therefore attribute any significant ($>\pm30$~km~s$^{-1}$) deviation from the modelled circular rotation to either inflow or outflow bulk motions. We highlight the regions where inflow/outflow motions can be identified in the p-v plot panels of Fig.~\ref{kinemetry-pvs}.

We find evidence of outflow motions affecting a sizeable fraction of molecular gas on CND scales in three of the sources of Fig.~\ref{kinemetry-pvs}: NGC~4388, NGC~5506 and  NGC~5643.  
The non-circular motions detected in the CO(3--2) velocity  field of NGC~5643 are very similar to those revealed by the previous ALMA CO(2--1) data of \citet{Alo18} and \citet{Gar21}, who modelled them in terms of a molecular outflow. In NGC~4388, we confirm the results of \citet{Dom20}, who claimed the existence of a molecular outflow based on a fit to the gas velocity field derived from a lower resolution ($\sim0\farcs6$) CO(2--1) NOEMA image of the galaxy. In the case of NGC~5506, the outflows identified in Fig.~\ref{kinemetry-pvs} extend in the disk out to deprojected radii $\sim100-200$~pc from the AGN and are characterized by moderate radial deprojected velocities $\sim50-100$~km~s$^{-1}$, similar to those of the NGC~4388 outflow.

In NGC~7582 a perturbed kinematics and marginal evidence of moderate  ($<50$~km~s$^{-1}$) outflow motions are found $\sim100$~pc from the AGN. We nevertheless detect the clear signature of inflow motions of up to $\sim120$~km~s$^{-1}$ down to  $r\sim250-300$~pc. The latter is plausibly linked to the canonical  gas response to the stellar bar outside the ILR region. Similar yet more moderate inflow motions are detected in the p-v diagrams  linked to the gas response to the stellar bar in NGC~6300 and to the crossing of the spiral arms in NGC~7314. Finally, we find only tentative evidence of outflowing components in the p-v plots of NGC~4941 and NGC~7465. In NGC~7465, velocities redshifted by $\sim50-100$~km~s$^{-1}$ around 70~pc northwest of the AGN are similar to the [OIII] velocities associated to the outflow  in this region \citep{Fer00, You21}.

%

\end{appendix}

\end{document}